\documentclass[twocolumn,aps,prc,showpacs,superscriptaddress,nofootinbib]{revtex4}
\usepackage{epsfig}
\newcommand {\snn}	{\sqrt{s_{_{\rm NN}}}}
\newcommand {\pt}	{p_{T}}
\newcommand {\psiRP}	{\psi_{\rm RP}}
\newcommand {\psiEP}	{\psi_{\rm EP}}
\newcommand {\EPres}	{\epsilon_{\rm EP}}
\newcommand {\Npart}	{N_{\rm part}}

\newcommand {\etal}	{{\it et al.}}
\newcommand {\lpv}	{LPV}
\newcommand {\cme}	{CME}
\newcommand {\ep}	{EP}
\newcommand {\dca}	{{\sc dca}}
\newcommand {\ud}	{UD}
\newcommand {\lr}	{LR}

\newcommand {\udlr}	{\ud$-$\lr}

\newcommand {\be}	{\begin{equation}}
\newcommand {\ee}	{\end{equation}}
\newcommand {\bea}	{\begin{eqnarray}}
\newcommand {\eea}	{\end{eqnarray}}

\newcommand {\NpU}	{N_{+,{\rm U}}}
\newcommand {\NpD}	{N_{+,{\rm D}}}
\newcommand {\NmU}	{N_{-,{\rm U}}}
\newcommand {\NmD}	{N_{-,{\rm D}}}
\newcommand {\NpL}	{N_{+,{\rm L}}}
\newcommand {\NpR}	{N_{+,{\rm R}}}
\newcommand {\NmL}	{N_{-,{\rm L}}}
\newcommand {\NmR}	{N_{-,{\rm R}}}
\newcommand {\Npm}	{N_{\pm}}
\newcommand {\Nu}	{N_{\pm,{\rm U}}}
\newcommand {\Nd}	{N_{\pm,{\rm D}}}
\newcommand {\Nud}	{N_{\pm}}

\newcommand {\vhigh}	{v^{\rm obs}_{2,\pt>2{\rm\,GeV}/c}}
\newcommand {\vlow}	{v^{\rm obs}_{2}}

\newcommand {\Ap}	{A_{+}}
\newcommand {\Am}	{A_{-}}

\newcommand {\ApUD}	{A_{+,{\rm UD}}}
\newcommand {\ApLR}	{A_{+,{\rm LR}}}
\newcommand {\AmUD}	{A_{-,{\rm UD}}}
\newcommand {\AmLR}	{A_{-,{\rm LR}}}
\newcommand {\ApmUD}	{A_{\pm,{\rm UD}}}
\newcommand {\ApmLR}	{A_{\pm,{\rm LR}}}

\newcommand {\phiw}	{\phi_{\rm w}}
\newcommand {\dphiw}	{\Delta\phi_{\rm w}}

\newcommand {\APM}[2]	{A_{\pm,#1^{o}\pm#2^{o}}}

\newcommand {\ApmWD}	{A_{\pm,\phiw\pm\dphiw}}

\newcommand {\mean}[1]	{\langle #1\rangle}
\newcommand {\cosijk}   {\cos(\phi_{\alpha}+\phi_{\beta}-2\psi_{\rm RP})}
\newcommand {\sinijk}   {\sin(\phi_{\alpha}+\phi_{\beta}-2\psi_{\rm RP})}
\newcommand {\mcosijk}  {\mean{\cosijk}}
\newcommand {\mcosabc}  {\mean{\cos(\phi_{\alpha}+\phi_{\beta}-2\phi_{c})}}
\newcommand {\mcosabcv} {\mcosabc/v_{2,c}}

\newcommand {\Apsq}	{\langle A^{2}_{+}\rangle}
\newcommand {\Amsq}	{\langle A^{2}_{-}\rangle}
\newcommand {\Apmsq}	{\langle A^{2}_{\pm}\rangle}
\newcommand {\AsqUD}	{\langle A^{2}_{\rm UD}\rangle}
\newcommand {\AsqLR}	{\langle A^{2}_{\rm LR}\rangle}
\newcommand {\Asq}	{\langle A^{2}\rangle}
\newcommand {\ApsqUD}	{\langle A^{2}_{+,{\rm UD}}\rangle}
\newcommand {\AmsqUD}	{\langle A^{2}_{-,{\rm UD}}\rangle}
\newcommand {\ApsqLR}	{\langle A^{2}_{+,{\rm LR}}\rangle}
\newcommand {\AmsqLR}	{\langle A^{2}_{-,{\rm LR}}\rangle}
\newcommand {\ApmsqUD}	{\langle A^{2}_{\pm,{\rm UD}}\rangle}
\newcommand {\ApmsqLR}	{\langle A^{2}_{\pm,{\rm LR}}\rangle}
\newcommand {\ApAm}	{\langle A_{+}A_{-}\rangle}
\newcommand {\AAUD}	{\langle A_{+}A_{-}\rangle_{\rm UD}}
\newcommand {\AALR}	{\langle A_{+}A_{-}\rangle_{\rm LR}}
\newcommand {\ApAmUD}	{\langle A_{+,{\rm UD}}A_{-,{\rm UD}}\rangle}
\newcommand {\ApAmLR}	{\langle A_{+,{\rm LR}}A_{-,{\rm LR}}\rangle}
\newcommand {\AaAbUD}	{\langle A_{\alpha}A_{\beta}\rangle_{\rm UD}}
\newcommand {\AaAbLR}	{\langle A_{\alpha}A_{\beta}\rangle_{\rm LR}}

\newcommand {\ApsqLRst}	{\langle A^{2}_{+,{\rm LR,stat}}\rangle}
\newcommand {\AmsqLRst}	{\langle A^{2}_{-,{\rm LR,stat}}\rangle}
\newcommand {\ApmsqSt}{\langle A^{2}_{\pm,{\rm stat}}\rangle}
\newcommand {\ApmsqUDst}{\langle A^{2}_{\pm,{\rm UD,stat}}\rangle}
\newcommand {\ApmsqLRst}{\langle A^{2}_{\pm,{\rm LR,stat}}\rangle}

\newcommand {\ApsqRn}	{\langle A^{2}_{+,{\rm stat+det}}\rangle}
\newcommand {\AmsqRn}	{\langle A^{2}_{-,{\rm stat+det}}\rangle}
\newcommand {\ApsqUDrn}	{\langle A^{2}_{+,{\rm UD,stat+det}}\rangle}
\newcommand {\AmsqUDrn}	{\langle A^{2}_{-,{\rm UD,stat+det}}\rangle}
\newcommand {\ApsqLRrn}	{\langle A^{2}_{+,{\rm LR,stat+det}}\rangle}
\newcommand {\AmsqLRrn}	{\langle A^{2}_{-,{\rm LR,stat+det}}\rangle}

\newcommand {\ApmsqRn}  {\langle A^{2}_{\pm,{\rm stat+det}}\rangle}
\newcommand {\ApmsqUDrn}{\langle A^{2}_{\pm,{\rm UD,stat+det}}\rangle}
\newcommand {\ApmsqLRrn}{\langle A^{2}_{\pm,{\rm LR,stat+det}}\rangle}
\newcommand {\AAUDrn}   {\langle A_{+}A_{-}\rangle_{\rm UD,stat+det}}
\newcommand {\AALRrn}   {\langle A_{+}A_{-}\rangle_{\rm LR,stat+det}}

\newcommand {\AsqWD}	{\langle A^{2}_{\phiw\pm\dphiw}\rangle}
\newcommand {\ApsqWD}	{\langle A^{2}_{+,\phiw\pm\dphiw}\rangle}
\newcommand {\AmsqWD}	{\langle A^{2}_{-,\phiw\pm\dphiw}\rangle}
\newcommand {\ApmsqWD}	{\langle A^{2}_{\pm,\phiw\pm\dphiw}\rangle}

\newcommand {\ApmsqWDrn}{\langle A^{2}_{\pm,\phiw\pm\dphiw,{\rm stat+det}}\rangle}
\newcommand {\AAWD}	{\langle A_{+}A_{-}\rangle_{\phiw\pm\dphiw}}
\newcommand {\AAWDrn}   {\langle A_{+}A_{-}\rangle_{\phiw\pm\dphiw,{\rm stat+det}}}

\newcommand {\APAMUD}	{\langle A_{+}A_{-}\rangle_{90^{\circ}\pm\dphiw}}
\newcommand {\APAMLR}	{\langle A_{+}A_{-}\rangle_{0^{\circ}\pm\dphiw}}

\newcommand {\ASqUD}	{\langle A^{2}_{90^{\circ}\pm\dphiw}\rangle}
\newcommand {\ASqLR}	{\langle A^{2}_{0^{\circ}\pm\dphiw}\rangle}

\newcommand {\ASq}	{\langle A^{2}_{\dphiw}\rangle}
\newcommand {\APAM}	{\langle A_{+}A_{-}\rangle_{\dphiw}}
\newcommand {\ASqFix}	{\langle A^{2}_{\phiw\pm15^{\circ}}\rangle}
\newcommand {\APAMFix}	{\langle A_{+}A_{-}\rangle_{\phiw\pm15^{\circ}}}
\newcommand {\DELTA}	{\Delta(\dphiw)}

\newcommand {\daa}	{\delta}

\begin{document}


\title{Measurement of Charge Multiplicity Asymmetry Correlations in High Energy Nucleus-Nucleus Collisions at $\snn=200$~GeV}

\affiliation{AGH University of Science and Technology, Cracow, Poland}
\affiliation{Argonne National Laboratory, Argonne, Illinois 60439, USA}
\affiliation{University of Birmingham, Birmingham, United Kingdom}
\affiliation{Brookhaven National Laboratory, Upton, New York 11973, USA}
\affiliation{University of California, Berkeley, California 94720, USA}
\affiliation{University of California, Davis, California 95616, USA}
\affiliation{University of California, Los Angeles, California 90095, USA}
\affiliation{Universidade Estadual de Campinas, Sao Paulo, Brazil}
\affiliation{Central China Normal University (HZNU), Wuhan 430079, China}
\affiliation{University of Illinois at Chicago, Chicago, Illinois 60607, USA}
\affiliation{Cracow University of Technology, Cracow, Poland}
\affiliation{Creighton University, Omaha, Nebraska 68178, USA}
\affiliation{Czech Technical University in Prague, FNSPE, Prague, 115 19, Czech Republic}
\affiliation{Nuclear Physics Institute AS CR, 250 68 \v{R}e\v{z}/Prague, Czech Republic}
\affiliation{University of Frankfurt, Frankfurt, Germany}
\affiliation{Institute of Physics, Bhubaneswar 751005, India}
\affiliation{Indian Institute of Technology, Mumbai, India}
\affiliation{Indiana University, Bloomington, Indiana 47408, USA}
\affiliation{Alikhanov Institute for Theoretical and Experimental Physics, Moscow, Russia}
\affiliation{University of Jammu, Jammu 180001, India}
\affiliation{Joint Institute for Nuclear Research, Dubna, 141 980, Russia}
\affiliation{Kent State University, Kent, Ohio 44242, USA}
\affiliation{University of Kentucky, Lexington, Kentucky, 40506-0055, USA}
\affiliation{Institute of Modern Physics, Lanzhou, China}
\affiliation{Lawrence Berkeley National Laboratory, Berkeley, California 94720, USA}
\affiliation{Massachusetts Institute of Technology, Cambridge, MA 02139-4307, USA}
\affiliation{Max-Planck-Institut f\"ur Physik, Munich, Germany}
\affiliation{Michigan State University, East Lansing, Michigan 48824, USA}
\affiliation{Moscow Engineering Physics Institute, Moscow Russia}
\affiliation{National Institute of Science and Education and Research, Bhubaneswar 751005, India}
\affiliation{Ohio State University, Columbus, Ohio 43210, USA}
\affiliation{Old Dominion University, Norfolk, VA, 23529, USA}
\affiliation{Institute of Nuclear Physics PAN, Cracow, Poland}
\affiliation{Panjab University, Chandigarh 160014, India}
\affiliation{Pennsylvania State University, University Park, Pennsylvania 16802, USA}
\affiliation{Institute of High Energy Physics, Protvino, Russia}
\affiliation{Purdue University, West Lafayette, Indiana 47907, USA}
\affiliation{Pusan National University, Pusan, Republic of Korea}
\affiliation{University of Rajasthan, Jaipur 302004, India}
\affiliation{Rice University, Houston, Texas 77251, USA}
\affiliation{Universidade de Sao Paulo, Sao Paulo, Brazil}
\affiliation{University of Science \& Technology of China, Hefei 230026, China}
\affiliation{Shandong University, Jinan, Shandong 250100, China}
\affiliation{Shanghai Institute of Applied Physics, Shanghai 201800, China}
\affiliation{SUBATECH, Nantes, France}
\affiliation{Temple University, Philadelphia, Pennsylvania, 19122}
\affiliation{Texas A\&M University, College Station, Texas 77843, USA}
\affiliation{University of Texas, Austin, Texas 78712, USA}
\affiliation{University of Houston, Houston, TX, 77204, USA}
\affiliation{Tsinghua University, Beijing 100084, China}
\affiliation{United States Naval Academy, Annapolis, MD 21402, USA}
\affiliation{Valparaiso University, Valparaiso, Indiana 46383, USA}
\affiliation{Variable Energy Cyclotron Centre, Kolkata 700064, India}
\affiliation{Warsaw University of Technology, Warsaw, Poland}
\affiliation{University of Washington, Seattle, Washington 98195, USA}
\affiliation{Wayne State University, Detroit, Michigan 48201, USA}
\affiliation{Yale University, New Haven, Connecticut 06520, USA}
\affiliation{University of Zagreb, Zagreb, HR-10002, Croatia}

\author{L.~Adamczyk}\affiliation{AGH University of Science and Technology, Cracow, Poland}
\author{J.~K.~Adkins}\affiliation{University of Kentucky, Lexington, Kentucky, 40506-0055, USA}
\author{G.~Agakishiev}\affiliation{Joint Institute for Nuclear Research, Dubna, 141 980, Russia}
\author{M.~M.~Aggarwal}\affiliation{Panjab University, Chandigarh 160014, India}
\author{Z.~Ahammed}\affiliation{Variable Energy Cyclotron Centre, Kolkata 700064, India}
\author{A.~V.~Alakhverdyants}\affiliation{Joint Institute for Nuclear Research, Dubna, 141 980, Russia}
\author{I.~Alekseev}\affiliation{Alikhanov Institute for Theoretical and Experimental Physics, Moscow, Russia}
\author{J.~Alford}\affiliation{Kent State University, Kent, Ohio 44242, USA}
\author{C.~D.~Anson}\affiliation{Ohio State University, Columbus, Ohio 43210, USA}
\author{D.~Arkhipkin}\affiliation{Brookhaven National Laboratory, Upton, New York 11973, USA}
% ...

\collaboration{STAR Collaboration}\noaffiliation


\date{\today}

\begin{abstract}

A study is reported of the same- and opposite-sign charge-dependent azimuthal correlations with respect to the event plane in Au$+$Au collisions at $\snn=200$~GeV. The charge multiplicity asymmetries between the up/down and left/right hemispheres relative to the event plane are utilized. The contributions from statistical fluctuations and detector effects were subtracted from the (co-)variance of the observed charge multiplicity asymmetries. In the mid- to most-central collisions, the same- (opposite-) sign pairs are preferentially emitted in back-to-back (aligned on the same-side) directions. 
The charge separation across the event plane, measured by the difference, $\Delta$, between the like- and unlike-sign up/down $-$ left/right correlations, is largest near the event plane. The difference is found to be proportional to the event-by-event final-state particle ellipticity (via the observed second-order harmonic $\vlow$), where $\Delta=(1.3\pm1.4({\rm stat})^{+4.0}_{-1.0}({\rm syst}))\times10^{-5}+(3.2\pm0.2({\rm stat})^{+0.4}_{-0.3}({\rm syst}))\times10^{-3}\vlow$ for 20-40\% Au+Au collisions. The implications for the proposed chiral magnetic effect (\cme) are discussed.
\end{abstract}

\pacs{25.75.-q, 25.75.Dw}

\maketitle

\section{Introduction}

Relativistic heavy-ion collisions at RHIC create a hot and dense medium that exhibits the properties of a strongly coupled Quark Gluon Plasma (sQGP)~\cite{wpBRAHMS,wpPHOBOS,wpSTAR,wpPHENIX}. It is possible that chiral symmetry is restored in an sQGP. In addition, it has been suggested that meta-stable domains capable of undergoing topological charge changes can form in the sQGP, and Parity (P) and Charge conjugation and Parity (CP) symmetries may be locally violated~\cite{TDLee1,TDLee2,Morley,PV,PV1Qsep,PV2Qsep,PV3Qsep,PVquench}. 
Kharzeev \etal~proposed that such a Local Parity Violation (\lpv) can lead to the separation of positively and negatively charged particles with respect to the reaction plane. This charge separation would be with respect to the direction defined by the axis of the classical magnetic field that is created by the passing ions. This P and CP violating process together with the magnetic field has been called the Chiral Magnetic Effect (\cme).

One consequence of the expected charge separation into pairs of back-to-back opposite-sign particles would be a positive two-particle azimuthal correlator, $\mcosijk$, of opposite-sign particle pairs and a negative correlator of same-sign particle pairs, where $\phi_\alpha$ and $\phi_\beta$ are the azimuthal angles of the two particles and $\psiRP$ is the reaction plane angle~\cite{Voloshin}. 
Since the reaction plane angle is not known, this correlator is estimated from the three-particle correlator, $\mcosabc$, with $c$ denotes the third particle, assuming that three-particle correlations unrelated to the reaction plane can be neglected. Previously, the STAR collaboration at the Relativistic Heavy-Ion Collider (RHIC) measured a negative correlator for same-sign pairs and a small, near zero, correlator for opposite-sign pairs~\cite{CorrelatorPRL,CorrelatorPRC}. The same-sign result was qualitatively consistent with the expectation from the \cme~\cite{PV1Qsep,PV2Qsep,PV3Qsep,PVquench}. The opposite-sign result, on the other hand, was inconsistent with the expectation where the opposite- and same-sign pair correlations should be equal in magnitude and opposite in sign~\cite{PV1Qsep,PV2Qsep,PV3Qsep}. However, the near-zero opposite-sign result may be consistent with the \cme\ with an additional contribution from in-medium interactions~\cite{PVquench}. More recently, the ALICE experiment at the Large Hadron Collider (LHC) has measured qualitatively similar correlation signals~\cite{ALICE}.

It is assumed that the charge separation along the orbital angular momentum axis, which is the direction of the magnetic field, due to the \cme\ will induce an asymmetry of positively or negatively charged particle multiplicities between the two (up and down) hemispheres which are separated by the reaction plane. No asymmetries due to the \cme\ are expected in the left and right hemispheres separated by the plane normal to the reaction plane.
In this paper, a new correlation technique is introduced that may be sensitive to the charge separation that was previously investigated in STAR. Consistency is found between the previously published data~\cite{CorrelatorPRL,CorrelatorPRC} and the results from the present analysis when using the same charge correlator observable (see Appendix~\ref{app:comp}). In this paper, a new approach is used to explore the charge asymmetries on an event-by-event basis, which extends the previous STAR measurements. The results obtained from $d$+Au and Au+Au collisions at $\snn=200$~GeV at RHIC measured by the STAR experiment are reported~\cite{QuanWang}. 


This paper is organized as follows. The analysis method is described in Sec.~\ref{sec:method}. The data analysis techniques are described in Sec.~\ref{sec:ana}. The systematic uncertainties of the results are described in Sec.~\ref{sec:syst}. The charge asymmetry correlation results are presented in Sec.~\ref{sec:results}. The implications of these results with respect to \lpv/\cme\ are discussed in Sec.~\ref{sec:disc}. Finally, the summary is presented in Sec.~\ref{sec:summary}. The mathematical correspondence between the present charge asymmetry correlation approach and the previously published three-particle correlators~\cite{CorrelatorPRL,CorrelatorPRC} are presented in Appendix~\ref{app:comp}. Additional details about the data analysis are presented in Appendix~\ref{app:ana}. 

\section{Analysis Method\label{sec:method}}

\begin{figure}[hbt]
\begin{center}
\includegraphics[width=0.3\textwidth]{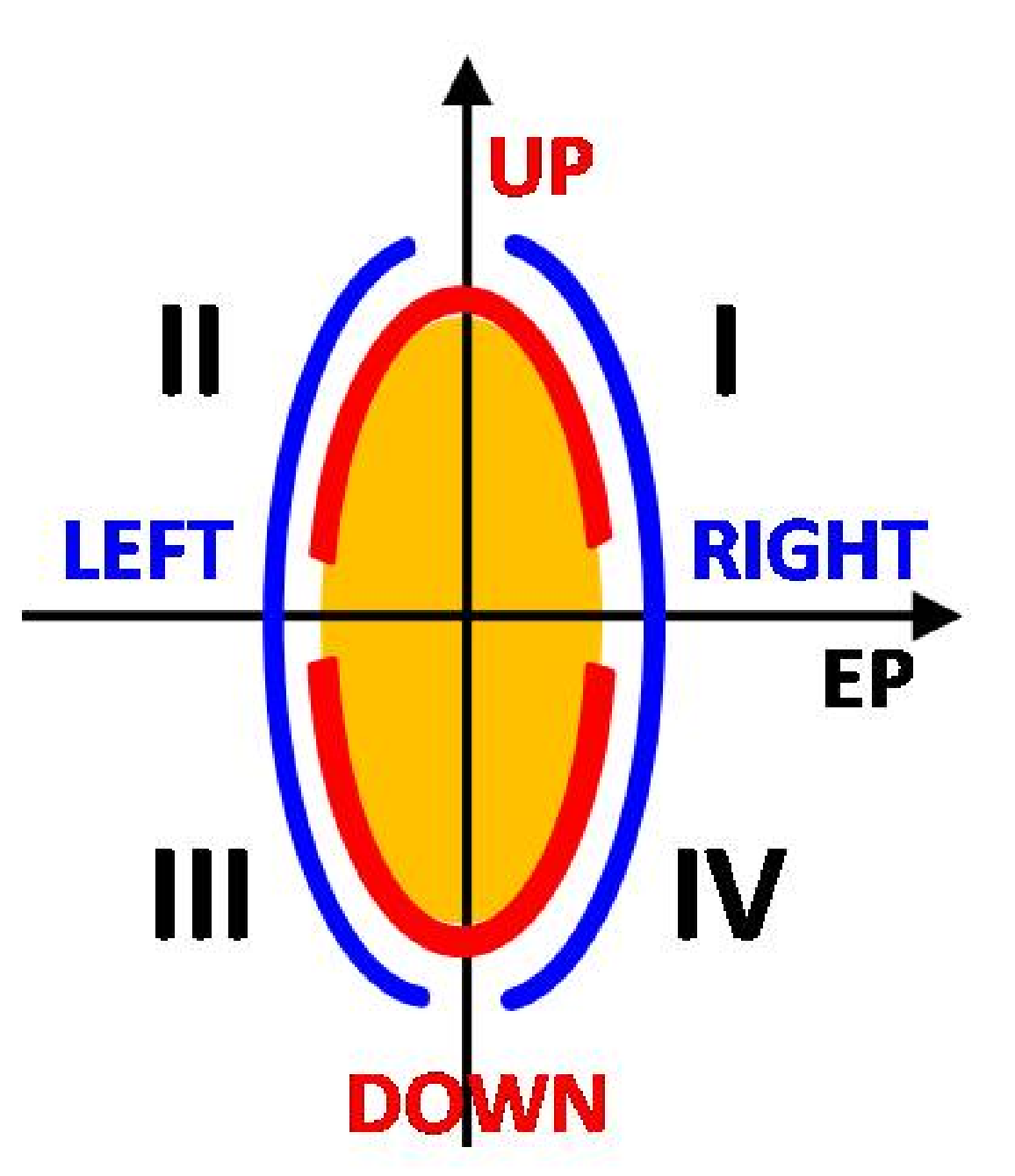}
\end{center}
\caption{(Color online) Schematic depiction of the transverse overlap region of a heavy-ion collision. The event plane (\ep) direction is reconstructed from final state particle momentum space. Four quadrants are defined and are labeled: Up = I+II, Down = III+IV, Left = II+III, Right = I+IV.}
\label{fig:diag}
\end{figure}

Figure~\ref{fig:diag} schematically depicts the transverse overlap region of a heavy-ion collision. The event plane, denoted by `\ep', is reconstructed from the measured charged particle azimuthal distributions. The event plane, `\ep', is not identical to the true reaction plane due to the measurement resolution. This is discussed in more detail in Appendix~\ref{app:EPres}. The particle multiplicity asymmetries are defined on event-by-event basis via,
\bea
\ApUD &=& (\NpU-\NpD) / (\NpU+\NpD)\,,\nonumber\\
\AmUD &=& (\NmU-\NmD) / (\NmU+\NmD)\,,\nonumber\\
\ApLR &=& (\NpL-\NpR) / (\NpL+\NpR)\,,\, {\rm and}\nonumber\\
\AmLR &=& (\NmL-\NmR) / (\NmL+\NmR)\,.
\eea
Here, $\NpU$, $\NpD$, $\NpL$, and $\NpR$ represent the positively charged particle multiplicities in the up (quadrants I and II), down (III and IV), left (II and III), and right (I and IV) hemispheres as depicted in Fig.~\ref{fig:diag}, respectively. The same multiplicities of the negatively charged particles are represented by $\NmU$, $\NmD$, $\NmL$, and $\NmR$.

The topological charge-signs are expected to be random from one metastable domain to another in a single event and in different events~\cite{PV,PV1Qsep,PV2Qsep,PV3Qsep,PVquench}. Also, the reaction plane obtained from the second harmonic of the particle azimuthal distributions cannot distinguish up from down. Thus, while the magnitude of the up-down (\ud) multiplicity asymmetry becomes larger due to the \lpv/\cme, its sign is random. As a result, the average asymmetries remain zero, but the distributions of $\ApmUD$, where $\ApmUD$ is used to denote $\ApUD$ and $\AmUD$, would be wider than those of $\ApmLR$, ({\it i.e.} $\ApLR$ and $\AmLR$). In other words, the variances $\ApmsqUD$ should be larger than the variances $\ApmsqLR$. Therefore, the variances of the charge multiplicity asymmetries is of interest here.

The covariance of the charge multiplicity asymmetries, $\AAUD\equiv\ApAmUD$ and $\AALR\equiv\ApAmLR$, are also studied. The covariance measure is analogous to the traditional parity violation measures as follows. 
The multiplicity asymmetry of one charge sign, {\em e.g.}~$\ApUD$, either positive or negative, can be viewed as to define preferentially the ``parity-axis'' direction, the combined effect of the orbital angular momentum direction and the topological charge sign. The covariance, $\AAUD$, is then a measure of $\AmUD$ with respect to this ``parity-axis." 

The charge asymmetry correlations are, however, parity-even and subject to physics backgrounds similar to those in the charge correlator measurement. These physics backgrounds can be assessed by the left-right (\lr) asymmetry correlations, $\AsqLR$ and $\AALR$, to which the \lpv/\cme\ does not contribute. The \lr\ charge asymmetry correlations could thus serve as the null-result reference. However, as will be discussed in Sec.~\ref{sec:disc:v2}, the physics backgrounds to the \ud\ and \lr\ measurements may be different.

In the present analysis, the charge multiplicity asymmetries, $\ApmUD$ and $\ApmLR$, are computed event-by-event. The variances of these quantities, $\ApmsqUD$ and $\ApmsqLR$, are reported. In order to extract dynamical fluctuations, effects of statistical fluctuations, which are finite in variances, need to be subtracted. In addition, detector effects can introduce ``dynamical fluctuations;'' For example, a deficient detector segment will always produce multiplicity asymmetries and hence their correlations. These effects are largely removed by efficiency corrections (see Appendix~\ref{app:eff}), and the remaining detector effects are small. The details on the contributions from the statistical and detector effects are presented in Appendix~\ref{app:stat}. In the results reported here, the statistical fluctuations and detector effects are subtracted, via 
\bea
\delta\ApmsqUD&=&\ApmsqUD-\ApmsqUDrn\,,\nonumber\\
\delta\ApmsqLR&=&\ApmsqLR-\ApmsqLRrn\,.
\eea
A comparison of $\delta\Apsq$ and $\delta\Amsq$ is described in Appendix~\ref{app:check}, and the two quantities are consistent. Thus, average dynamical variances, 
\bea
\delta\AsqUD&=&(\delta\ApsqUD+\delta\AmsqUD)/2\,,\nonumber\\
\delta\AsqLR&=&(\delta\ApsqLR+\delta\AmsqLR)/2\,,
\eea
are presented. The covariances, $\AAUD$ and $\AALR$, are also presented. The statistical fluctuations do not contribute to the covariances. Detector effects on covariances, after efficiency corrections, are consistent with zero. The statistical fluctuations and detector effects are analyzed together in Appendix~\ref{app:stat}, and are nevertheless removed from the covariances:
\bea
\delta\AAUD&=&\AAUD-\AAUDrn\,,\nonumber\\
\delta\AALR&=&\AALR-\AALRrn\,.
\eea

The differences between the \ud\ and \lr\ measurements which may be directly sensitive to the \cme\ will be reported. Namely,
\bea
\Delta\Asq&\equiv&\delta\AsqUD-\delta\AsqLR\,,\nonumber\\
\Delta\ApAm&\equiv&\delta\AAUD-\delta\AALR\,.
\eea
The $\Delta\Asq$ and $\Delta\ApAm$ are the same as $\AsqUD-\AsqLR$ and $\AAUD-\AALR$, resepectively, because the statistical fluctuation and detector effects cancel in the differences. The difference \udlr\ correlations between same- and opposite-sign charges,
\be
\Delta\equiv\Delta\Asq-\Delta\ApAm,
\ee
will also be reported, which may quantify the charge separation effects.

\begin{figure}[hbt]
\begin{center}
\includegraphics[width=0.45\textwidth]{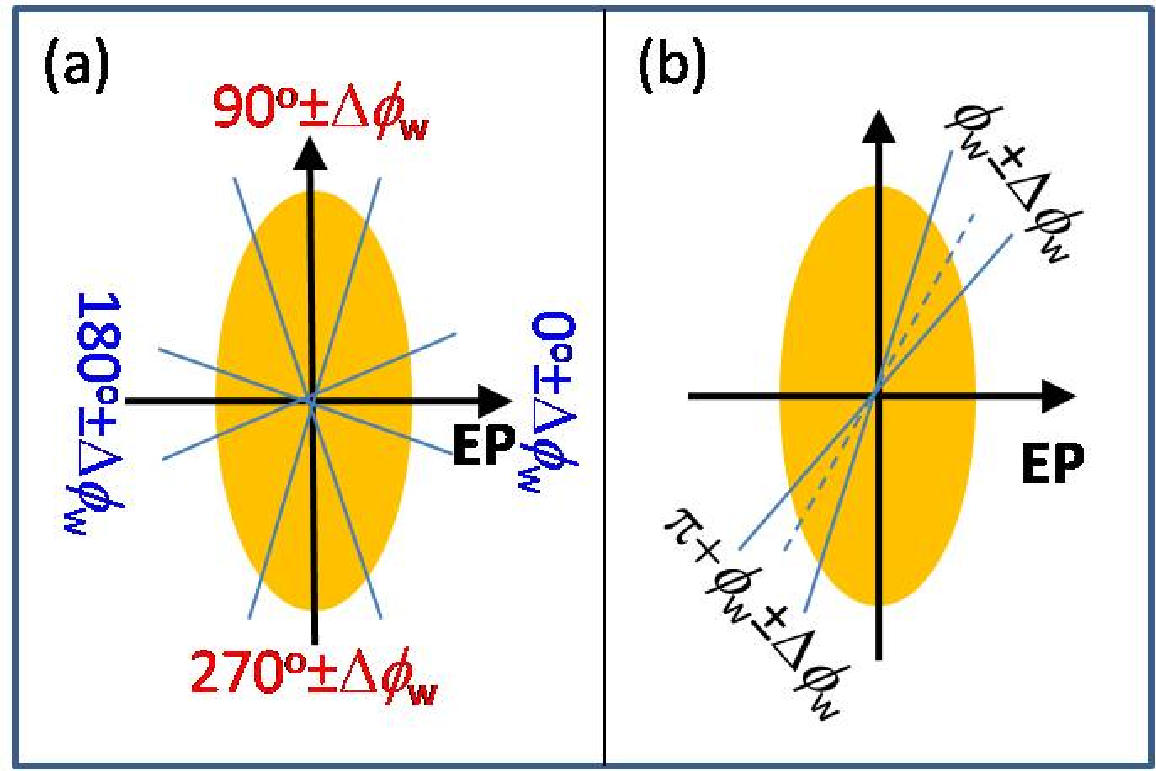}
\end{center}
\caption{(Color online) Schematic view of the transverse overlap region and the back-to-back wedges (azimuthal regions) where the charge asymmetries are computed. The event plane (\ep) direction is reconstructed from final state particle momentum space. (a) Configuration for the study of the wedge size dependence. (b) Configuration for the wedge location dependence.}
\label{fig:diag2}
\end{figure}

In addition to the asymmetries between hemispheres, the asymmetries between azimuthal wedges of smaller sizes will be studied as depicted in Figure~\ref{fig:diag2}(a). The out-of-plane asymmetries between a wedge of size $2\dphiw$ at $\phiw=90^{\circ}$ and the same-size wedge at $\phiw=270^{\circ}$ will be explored. This involves the counting of the particle multiplicity within an azimuth range relative to the \ep\ between $90^{\circ}-\dphiw$ and $90^{\circ}+\dphiw$, and the multiplicity within $270^{\circ}-\dphiw$ and $270^{\circ}+\dphiw$. Similarly for in-plane asymmetries, the particle multiplicities within $0^{\circ}-\dphiw$ and $0^{\circ}+\dphiw$, and within $180^{\circ}-\dphiw$ and $180^{\circ}+\dphiw$ are counted. 

The charge asymmetry correlations will also be calculated within a back-to-back pair of wedges at specific azimuthal locations with respect to the \ep. Figure~\ref{fig:diag2}(b) shows the schematic configuration of the back-to-back wedges with size $2\dphiw$ at location $\phiw$. For these studies, the notations used for the asymmetry correlation variables are refined as follows. The variable $\ApmWD$ stands for the positive/negative particle multiplicity asymmetry between the wedge $\phiw\pm\dphiw$ and its opposite-side partner wedge $(\pi+\phiw)\pm\dphiw$. In this notation, the variables $\ApmUD$ and $\ApmLR$ are equivalent to $\APM{90}{90}$ and $\APM{0}{90}$, respectively.

A disadvantage of using smaller, non-hemispheric, ranges is the fact that the statistical fluctuation and detector effects no longer cancel between variances from out-of-plane wedges and those from in-plane wedges. The statistical fluctuation contributions to $\ApmsqWDrn$ and $\AAWDrn$ are obtained in the same way as described in Appendix~\ref{app:stat}. These contributions are subtracted to obtain the dynamical correlation: 
\bea
\delta\ApmsqWD&=&\ApmsqWD-\nonumber\\
&&\ApmsqWDrn\,,\nonumber\\
\delta\AAWD&=&\AAWD-\nonumber\\
&&\AAWDrn\,.
\eea
Similarly, the average 
\be
\delta\AsqWD=(\delta\ApsqWD+\delta\AmsqWD)/2\,,
\ee
and the differences between the \ud\ and \lr\ correlations will be reported. 

The charge multiplicity asymmetry correlation quantities in the present study can be related to the three-particle charge-dependent azimuthal correlators~\cite{CorrelatorPRL,CorrelatorPRC} as shown in Appendix~\ref{app:comp}. The present analysis differs from that in Refs.~\cite{CorrelatorPRL,CorrelatorPRC} in that the reported correlations represent the contributions from the entire correlation structure (all multipoles), whereas those in the previous studies focused on the lowest-order azimuthal multipole only.

\section{Data Analysis\label{sec:ana}}

The data used in this analysis were taken by the STAR experiment~\cite{STAR} at RHIC at the nucleon-nucleon center of mass energy of $\snn=200$~GeV. The minimum-bias and central triggered Au+Au data were from RHIC Run IV in the year 2004. The reference minimum-bias $d$+Au data used for comparison were from RHIC Run III in 2003. The Au+Au results will also be compared to data taken from Run VII (2007). The Run VII data were used for the study of the asymmetry correlations with respect to the first harmonic event plane.

The details of the STAR experiment can be found in Ref.~\cite{STAR}. The minimum-bias triggers for Au+Au and $d$+Au collisions were provided by the Central Trigger Barrel~\cite{CTB} and the Zero Degree Calorimeters (ZDC)~\cite{ZDC}. A total of 8.8 million Au+Au events with centrality ranging from 0-80\% (Run IV) and a total of 8.9 million $d$+Au events (Run III) were used in this analysis. The Run VII data used for comparison and for the ZDC first harmonic event plane study consists of 70 million minimum-bias Au+Au events. 

The Au+Au collision centrality is defined according to the measured charged particle multiplicity in the main Time Projection Chamber (TPC) within the pseudo-rapidity range $|\eta|<0.5$~\cite{spec200}. 
Results will be presented as a function of centrality in terms of the number of participant nucleons, $\Npart$, which was obtained from a Monte Carlo Glauber calculation \cite{Levente}. The corresponding impact parameters and the uncorrected and corrected charged hadron multiplicities can be found in Ref.~\cite{Levente}.


The main detector used for this analysis was the Time Projection Chamber (TPC)~\cite{TPC1,TPC2}. The TPC is surrounded by a solenoidal magnet providing a uniform magnetic field of 0.5 tesla along the beam direction. Particle tracks were reconstructed by the TPC. The primary collision vertex was reconstructed using TPC tracks passing various quality cuts. Events with a primary vertex location within $\pm 30$~cm of the geometric center of the TPC along the beam axis were used in the analysis. 

In the present asymmetry calculations and TPC event-plane construction, only those ``primary" tracks extrapolated to within 2~cm of the primary vertex were used. The tracks were required to have at least 20 (out of a maximum of 45) hits in the TPC used in track reconstruction. The ratio of the number of hits used in the track reconstruction to the maximum possible number of hits for a given track was required to be greater than 0.51 to eliminate multiple track segments reconstructed from a single particle trajectory. These cuts were varied to assess the systematic uncertainties in the present results which are discussed in Sec.~\ref{sec:syst}.


The second Fourier harmonic in azimuth was used to determine the event-plane angle $\psiEP$~\cite{flowMethod} from the TPC-reconstructed tracks. The event plane was reconstructed for both Au+Au and $d$+Au collisions, and does not necessarily correspond to a specific plane in configuration geometry. The transverse momentum, $\pt$, range of the particles used to determine the event plane was $0.15<\pt<2$~GeV/$c$. The low-$\pt$ cut-off was imposed by the magnetic field strength and the TPC inner radius. The $\pt$-weight method~\cite{flowMethod} was used for the event plane reconstruction as it gives a better event plane resolution than no $\pt$-weight in the presence of the stronger anisotropy at larger values of $\pt$. The event plane reconstruction was done in two different $\eta$ ranges: $-1<\eta<0$ and $0<\eta<1$ (additional details below). The slight non-uniformities of the efficiency and acceptance in azimuth due to the TPC sector boundaries was corrected for in the event-plane construction by using $\phi$-dependent efficiencies (see Appendix~\ref{app:eff}). 
%
%
The azimuthal angle of the \ep\ constructed by the second harmonic ranges from 0 to $\pi$. In half of the events chosen randomly, $\pi$ was added to the reconstructed \ep\ azimuthal angle so that the resulting \ep\ azimuthal angle ranges from 0 to $2\pi$. The first-order harmonic event plane was also measured independently using the ZDC Shower Maximum Detector (SMD). The ZDC-SMD event-plane analysis is described in Ref.~\cite{WangG,ChenJY,QuanWang}.


The particle azimuthal angle $\phi$ relative to \ep\ was properly folded with $\psiEP$ so that $\phi-\psiEP$ was also in the range between 0 and $2\pi$. A particle is assigned to the `up' hemisphere if $0<\phi-\psiEP<\pi$, `down' hemisphere if $\pi<\phi-\psiEP<2\pi$, `left' hemisphere if $\pi/2<\phi-\psiEP<3\pi/2$, and `right' hemisphere if $3\pi/2<\phi-\psiEP<2\pi$ or $0<\phi-\psiEP<\pi/2$. The number of particles, weighted by the efficiency correction factor (described in Appendix~\ref{app:effCorr}), in the upper, lower, left, and right hemispheres was counted. The asymmetries from the corrected particle multiplicities were calculated. Separate calculations of the asymmetries (i) using particles within $0<\eta<1$ with the \ep\ constructed from $-1<\eta<0$, and (ii) using particles within $-1<\eta<0$ with \ep\ constructed from $0<\eta<1$, were performed. These two results were consistent (see Appendix~\ref{app:check}), so their average is discussed.

\section{Systematic Uncertainties \label{sec:syst}}

\begin{table*}[hbt]
\caption{Sources and magnitudes of $\pm$ systematic uncertainties. All the numbers have been multiplied by the corresponding number of participants $\Npart$. The upper section is for the 40-30\% centrality ($\Npart=78.3$) and lower section is for the top 5\% centrality ($\Npart=350.6$).}
\label{tab:syst}
\begin{ruledtabular}
\begin{tabular}{c|cccccc}
Source & $\daa\AsqUD$ & $\daa\AsqLR$ & $\daa\AAUD$ & $\daa\AALR$ & $\Delta\Asq$ & $\Delta\ApAm$ \\\hline
Magnetic field polarity (FF vs RFF) & 0.0004 & 0.0001 & 0.0041 & 0.0034 & 0.0005 & 0.0007\\
Primary vertex $Z_{\rm vtx}$ cut (15 cm vs 30 cm) & 0.0006 & 0.0021 & 0.0048 & 0.0009 & 0.0015 & 0.0039\\
\dca\ cut (1 cm vs 3 cm) & 0.0006 & 0.0009 & 0.0013 & 0.0012 & 0.0015 & 0.0016\\
Min.~number of fit points $N_{\rm fit}$ (15 vs 25) & 0.0003 & 0.0017 & 0.0015 & 0.0001 & 0.0020 & 0.0014\\
TPC side (West vs East) & 0.0010 & 0.0028 & 0.0010 & 0.0009 & 0.0035 & 0.0001\\
Total & 0.0014 & 0.0040 & 0.0067 & 0.0038 & 0.0046 & 0.0045\\
\hline
Magnetic field polarity (FF vs RFF) & 0.0034 & 0.0040 & 0.0006 & 0.0015 & 0.0006 & 0.0021\\
Primary vertex $Z_{\rm vtx}$ cut (15 cm vs 30 cm) & 0.0015 & 0.0050 & 0.0045 & 0.0005 & 0.0035 & 0.0040\\
\dca\ cut (1 cm vs 3 cm) & 0.0017 & 0.0032 & 0.0009 & 0.0007 & 0.0015 & 0.0016\\
Min.~number of fit points $N_{\rm fit}$ (15 vs 25) & 0.0031 & 0.0037 & 0.0023 & 0.0021 & 0.0006 & 0.0021\\
TPC side (West vs East) & 0.0073 & 0.0032 & 0.0022 & 0.0027 & 0.0084 & 0.0005\\
Total & 0.0089 & 0.0087 & 0.0056 & 0.0039 & 0.0093 & 0.0053\\

\end{tabular}
\end{ruledtabular}
\end{table*}

The systematic uncertainties of the results were assessed in the following ways.

The present charge asymmetries were analyzed by rotating the reconstructed \ep\ by $45^{\circ}$. It was found that the \ud\ and \lr\ asymmetry correlations are identical, $\AsqUD=\AsqLR$ and $\AAUD=\AALR$, as expected. The present charge asymmetries were also calculated by randomly discarding a fixed fraction of the particles. Essentially the same $\delta\Asq$, $\daa\ApAm$, $\Delta\Asq$ and $\Delta\ApAm$ results were obtained.

To check for possible directed flow effects, the charge asymmetry correlations within $|\eta|<0.5$ were calculated using the event plane constructed by particles in $0.5<|\eta|<1$. It was found that the observed asymmetry correlation results were consistent with those obtained from $-1<\eta<0$ and $0<\eta<1$.


The greater inefficiency in two of the sectors in the east half of the TPC introduces larger detector effects in the measurement of $\Apmsq$. After subtracting the statistical fluctuations and detector effects, the dynamical asymmetry variances $\delta\Apmsq$ were consistent between the $\eta>0$ and $\eta<0$ regions, as well as between $\delta\Apsq$ and $\delta\Amsq$ from each $\eta$ region (see Fig.~\ref{fig:etaRegion}). The average variances between $\delta\Apsq$ and $\delta\Amsq$ from the two $\eta$ regions are thus reported, and the maximum difference of the individual results from the average is considered as part of the systematic uncertainties. The covariances, $\daa\ApAm$, were also consistent between the two $\eta$ regions, so this paper reports the average $\daa\ApAm$ including half of the difference as part of the systematic uncertainties.

The $\phi$-independent track reconstruction efficiency were not corrected for because it does not affect the present asymmetry measurements. Correcting for the track reconstruction efficiency (which is a function of $\eta$, $\pt$, and centrality) does not significantly affect the values of $\Asq$, $\delta\Asq$, $\Delta\Asq$, $\ApAm$, $\daa\ApAm$, and $\Delta\ApAm$.

The present asymmetry correlations were also studied by varying the event and track quality cuts. Specifically, the event primary vertex position, $Z_{\rm vtx}$ was restricted to within $\pm 15$~cm (default $\pm30$~cm). The maximum distance of closest approach, \dca, cut was also varied between 1-3~cm (default 2~cm) and the minimum number of fit points, $N_{\rm fit}$ requirement was varied between 15-25 (default 20). For these different cut sets, the corresponding $\phi$-acceptance corrections were used. The changes in the present results from the different cuts are generally small, but are included in the systematic uncertainties. The data from the full magnetic field setting (FF) and the reversed full magnetic field setting (RFF) were also analyzed separately. The results are generally consistent within the statistical errors. The difference between the FF and RFF results from the combined FF and RFF data set, whichever is larger, is taken as part of the systematic uncertainties.


Table~\ref{tab:syst} shows the sources and magnitudes of the systematic uncertainties on the charge multiplicity asymmetry correlation measurements for two selected centrality bins. The systematic uncertainties from the various sources are added in quadrature to yield the total systematic uncertainties. The systematic uncertainties are taken to be symmetric between the positive and negative sides. The total systematic uncertainties are shown in the shaded areas in Fig.~\ref{fig:corr} and Fig.~\ref{fig:diff}.

The systematic uncertainties on the charge separation parameter $\Delta$ and its dependence on the observed event-by-event second harmonic parameter $\vlow$ are also studied in the same way. The $\Delta$ values for all events and for events with $|\vlow|<0.04$, and the intercepts and slopes of linear fits to $\Delta(\vlow)$ obtained using different cuts and etc.~are compared. The differences are assigned as asymmetric systematic uncertainties. They are listed in Table~\ref{tab:Delta} as a function of centrality.

\section{Results\label{sec:results}}

\subsection{Charge Asymmetry Correlations\label{sec:asym}}

The single asymmetry quantities $\mean{\ApUD}$, $\mean{\AmUD}$, $\mean{\ApLR}$, $\mean{\AmLR}$ are, by definition, zero because the positions of the up (left) and down (right) hemispheres are random from event to event. The data indeed show zero single asymmetries within the statistical errors. 

\begin{figure}[hbt]
\begin{center}
\includegraphics[width=0.4\textwidth]{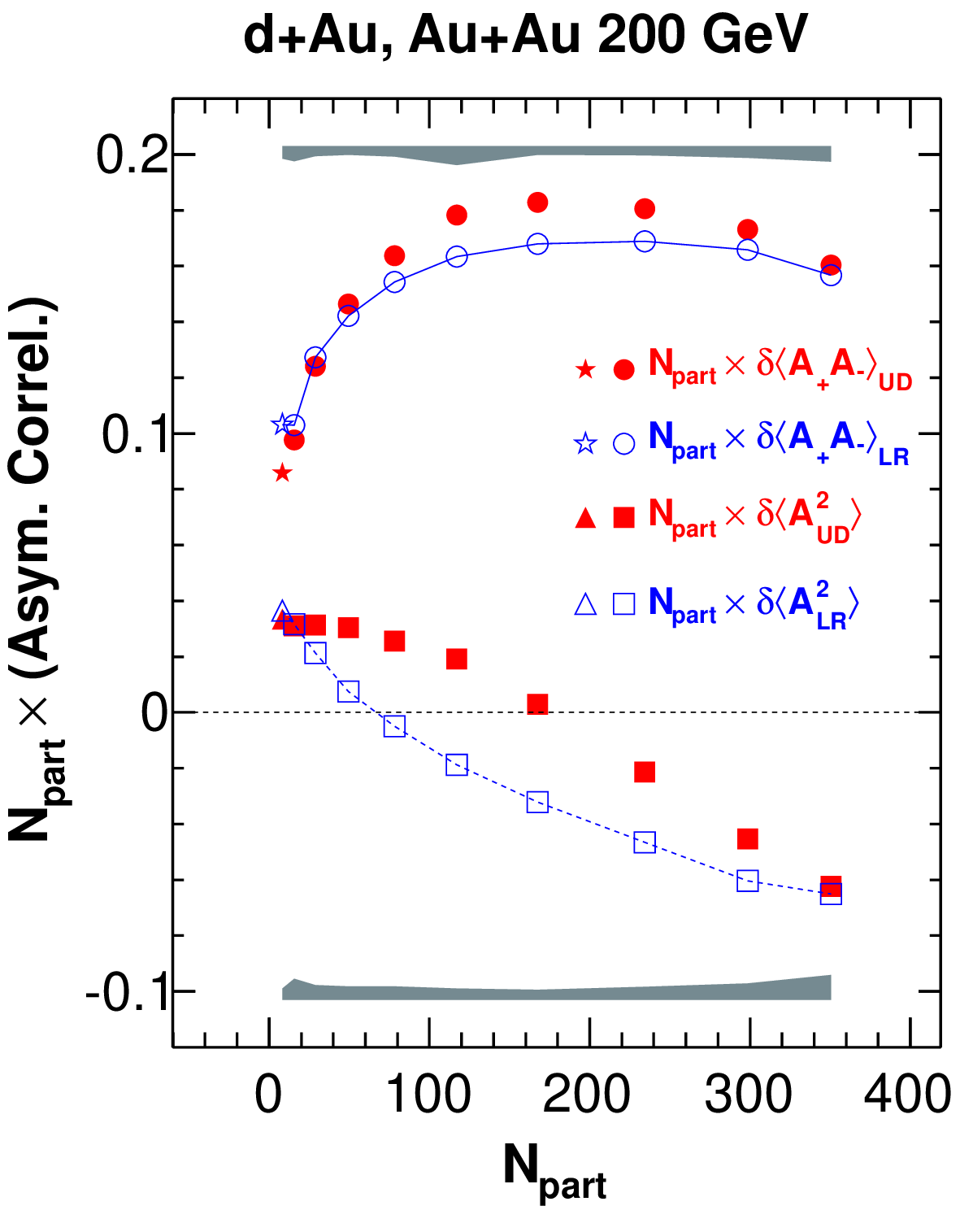}
\end{center}
\caption{(Color online) Centrality dependences of the charge asymmetry dynamical correlations, $\delta\Asq$, and the positive and negative charge asymmetry correlations, $\daa\ApAm$.  The asymmetries are calculated between hemispheres separated by the event plane (\ud) and between those separated by the plane perpendicular to the event plane (\lr). The asymmetry correlations are multiplied by the number of participants $\Npart$. 
The error bars are statistical only. The upper (lower) shaded band shows half of the systematic uncertainty in the $\daa\ApAm$ ($\daa\Asq$); the larger of the \ud\ and \lr\ systematic uncertainties is drawn. The stars and triangles depict the $d$+Au results.}
\label{fig:corr}
\end{figure}

Figure~\ref{fig:corr} shows the dynamical variances, $\delta\AsqUD$ (solid squares) and $\delta\AsqLR$ (hollow squares), and covariances, $\daa\AAUD$ (solid circles) and $\daa\AALR$ (hollow circles), as a function of $\Npart$. Since two-particle correlation measures are typically diluted by a multiplicity factor, the dynamical fluctuation quantities are multiplied by $\Npart$ to reveal the magnitudes over the entire centrality range. The \ud\ and \lr\ quantities are different within all centralities ranges except the most peripheral and most central collisions. If the \ep\ were random and unrelated to the reaction plane, then the \ud\ and \lr\ observables would be the same within the statistical uncertainties. This is not the case, as clearly shown in Fig.~\ref{fig:corr}.

A positive $\delta\Asq$ indicates a broadening of the asymmetry distributions of $\Ap$ and $\Am$ due to dynamical processes, whereas a negative $\delta\Asq$ indicates narrowing of the distributions. In peripheral collisions, both $\delta\AsqUD$ and $\delta\AsqLR$ are positive, suggesting that the same-sign particles within one unit of pseudo-rapidity $|\eta|<1$ are more likely emitted in the same direction. This small-angle correlation is stronger in the up-down hemispheres than in the left-right hemispheres. The small-angle correlation becomes weaker when the collisions are more central. In fact, in more central collisions, the $\delta\AsqUD$ and $\delta\AsqLR$ become negative, {\it i.e.}, the same-sign charge pairs are preferentially emitted back-to-back in those collisions. 

The correlations between $\Ap$ and $\Am$, both \ud\ and \lr, are large and positive implying strong correlations. The correlation is on the order of $\sim10^{-3}$, suggesting that the correlated asymmetry is as large as a few percent. 

Figure~\ref{fig:corr} also shows the asymmetry correlations in $d$+Au collisions. The $d$+Au data lie at the endpoint of the Au+Au curve and are consistent with an extrapolation of that trend.

The data in Fig.~\ref{fig:corr} seems to indicate the following picture. In $d$+Au and peripheral Au+Au collisions, the particles within one unit of pseudo-rapidity are preferentially emitted in the same direction, whether they are the same or opposite charge signs. The magnitude of the small-angle correlation is, however, stronger in the opposite- than in the same-sign pairs, and is stronger in the out-of-plane than in the in-plane direction. In medium-central to central collisions, while the opposite-sign pairs are still preferentially aligned in the same direction and more so than in peripheral collisions, the same-sign pairs are preferentially back-to-back. The small-angle correlation between the opposite-sign pairs is always stronger out-of-plane than in-plane. The tendency of back-to-back emission of same-sign particles is weaker in the out-of-plane than in the in-plane direction.

\begin{figure}[hbt]
\begin{center}
\includegraphics[width=0.4\textwidth]{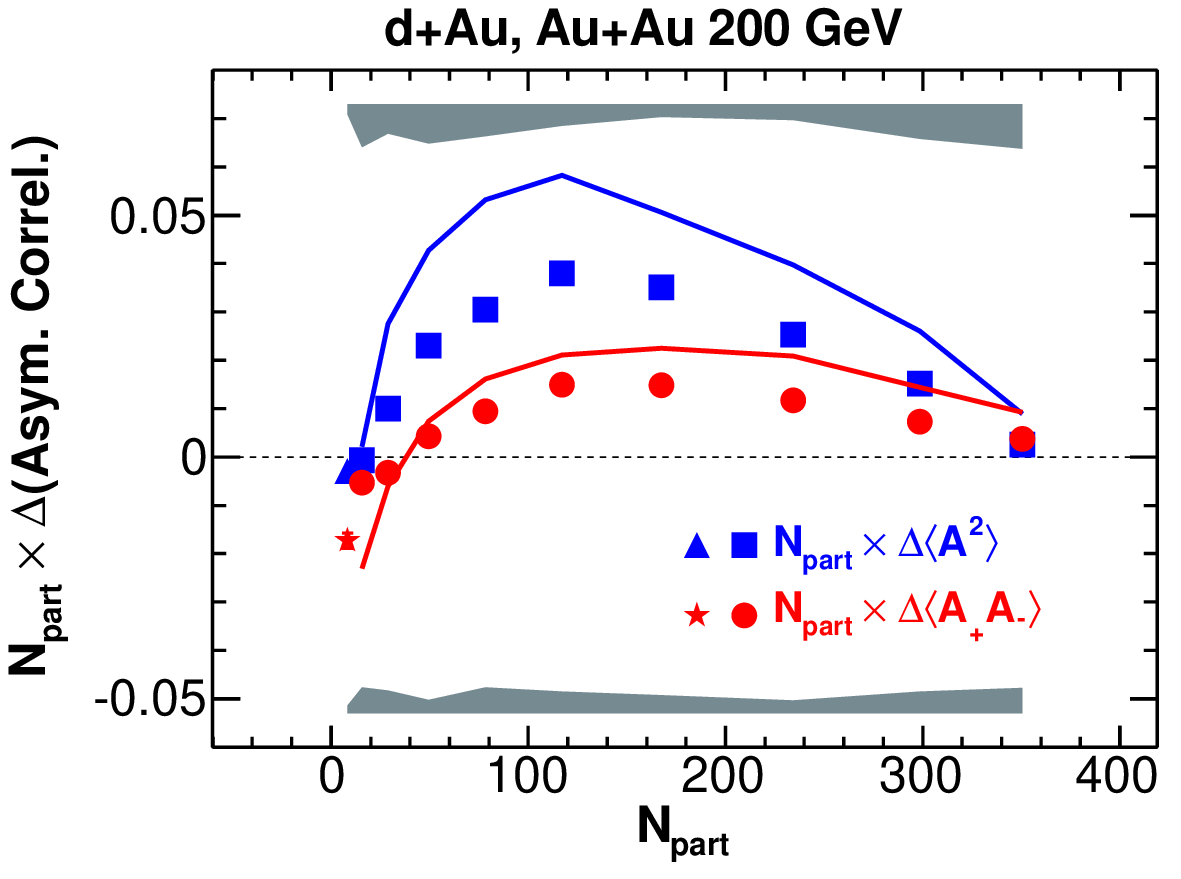}
\end{center}
\caption{(Color online) The correlation differences $\Delta\Asq=\daa\AsqUD-\daa\AsqLR$ and $\Delta\ApAm=\daa\AAUD-\daa\AALR$, scaled by the number of participants $\Npart$, as a function of $\Npart$. 
The error bars are statistical, and the systematic uncertainties are shown in the shaded bands (upper band for $\Delta\ApAm$ and lower band for $\Delta\Asq$). Also shown as the lines are the linear-extrapolated values of $\Delta\Asq$ and $\Delta\ApAm$ corresponding to a perfect event-plane resolution. The star and triangle depict the $d$+Au results.}
\label{fig:diff}
\end{figure}

In order to investigate the possible contributions from the \lpv/\cme, the difference between the \ud\ and \lr\ asymmetry correlations, $\Delta\Asq$ and $\Delta\ApAm$, was studied. The contributions from the detector effects and systematics are largely canceled in these differences (see Sec.~\ref{sec:syst}). Figure~\ref{fig:diff} shows $\Delta\Asq$ and $\Delta\ApAm$ as a function of the centrality. The upper shaded area shows the systematic uncertainty in $\Delta\ApAm$ and the lower shaded area shows that on $\Delta\Asq$. Also shown as the lines are the $\Delta\Asq$ and $\Delta\ApAm$ values which would be expected with a perfect event-plane resolution, which is calculated assuming the linear extrapolations shown in Fig.~\ref{fig:EPres_ext} (right panel). As noted in Appendix~\ref{app:comp}, this linear extrapolation would be correct if high-order harmonic terms in Eq.~(\ref{eq:expansion}) are negligible.

Figure~\ref{fig:diff} shows that $\Delta\Asq$ is larger than zero, {\it i.e.} $\delta\AsqUD>\delta\AsqLR$, in collisions at all centralities. The \ud\ asymmetry distribution is always broader than the \lr\ one. This implies that there are more small-angle same-sign pairs in the out-of-plane direction than in the in-plane direction. Equivalently, there are more back-to-back pairs in-plane than out-of-plane. The \udlr\ difference in $\daa\ApAm$ is small relative to the correlations themselves. This indicates that the majority of the strong correlations between opposite-sign pairs is unrelated to the reaction plane. On the other hand, $\Delta\ApAm$ is also larger than zero, {\it i.e.} $\daa\AAUD>\daa\AALR$ in all centralities except peripheral collisions. The opposite-sign particle pairs are more strongly emitted in the same direction out-of-plane compared to in-plane. 

The general trend of $\Delta\Asq$ with centrality is as follows. It increases with centrality, reaching a maximum in mid-central collisions, and then decreases with increasing centrality. The trend for $\Delta\ApAm$ is similar, except that it is slighlty negative in peripheral collisions. The negative values of $\Delta\ApAm$ in peripheral collisions (and also in $d$+Au collisions) are likely due to ``non-flow" ({\it e.g.} di-jets). The decrease for more central collisions may be due to the experimental event-plane resolution because the difference should disappear in zero impact parameter collisions where the reaction plane is undefined. 

\begin{figure}[hbt]
\begin{center}
\includegraphics[width=0.4\textwidth]{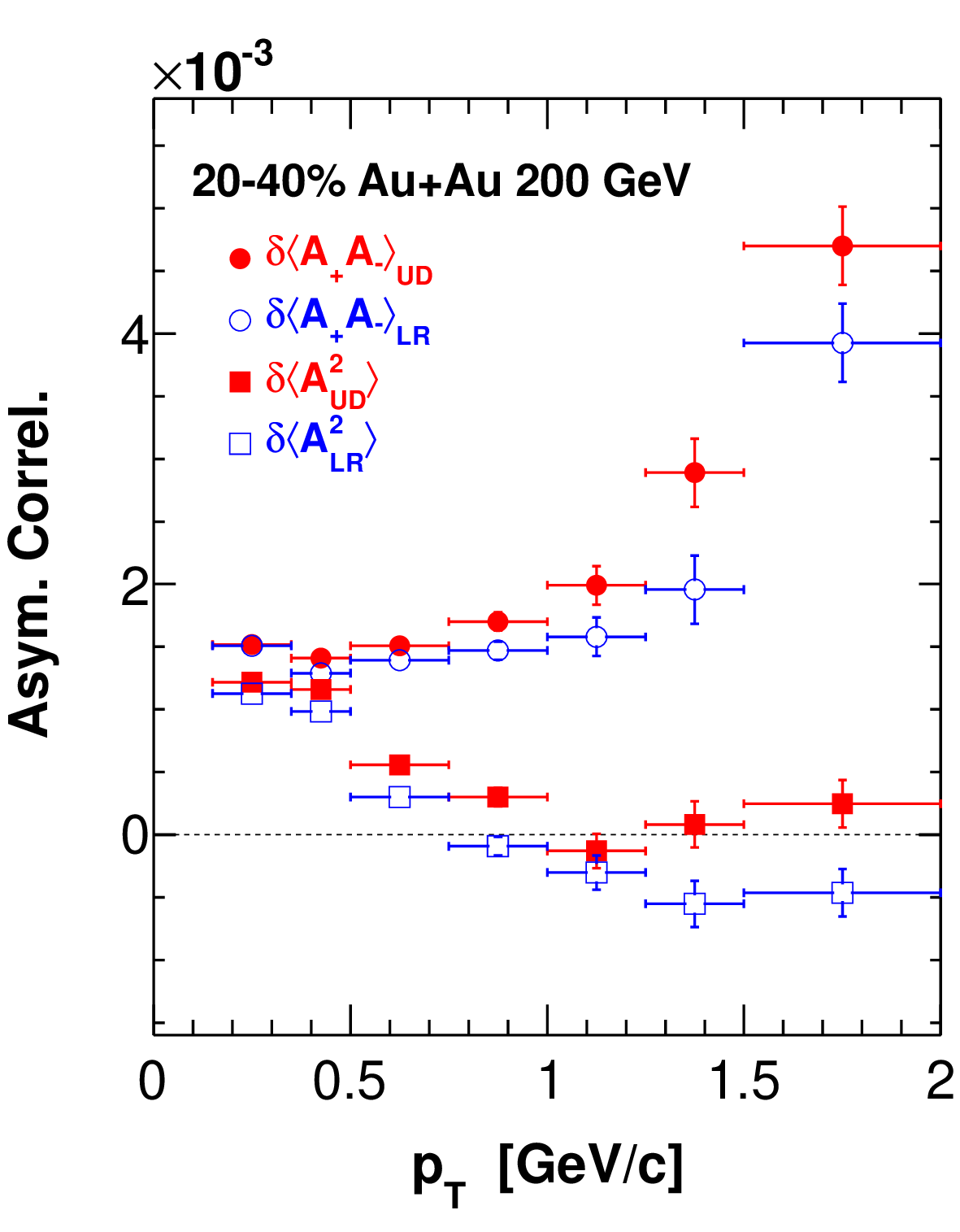}
\end{center}
\caption{(Color online) The $\pt$ dependence of the charge asymmetry dynamical correlations, $\delta\Asq$, and the positive and negative charge asymmetry correlations, $\daa\ApAm$. The data are from 20-40\% central Au+Au collisions. The asymmetries are calculated between hemispheres separated by the event plane (UD) and between those separated by the plane perpendicular to the event plane (LR). 
The error bars are statistical only.}
\label{fig:pt}
\end{figure}

Figure~\ref{fig:pt} shows the $\pt$ dependence of the charge asymmetry correlations in 20-40\% and 0-20\% central Au+Au collisions. 
The values of $\delta\Asq$ are positive at low $\pt$, and decrease sharply with increasing $\pt$ up to about 1~GeV/$c$. In this centrality bin, the $\delta\AsqLR$ values become negative and $\delta\AsqUD$ is approximately zero for $\pt>1$~GeV/$c$. This indicates that low-$\pt$ pairs are emitted preferentially in the same direction. The back-to-back emission of same-sign pairs increases with increasing $\pt$. On the other hand, $\daa\ApAm$ remains relatively constant at low $\pt$ up to roughly 1~GeV/$c$, and then increases sharply with increasing $\pt$. The aligned emission of opposite-sign pairs in the same direction increases strongly with $\pt$ above 1~GeV/$c$. The observed features in $\pt$ are qualitatively similar for other centrality bins. It is worthwhile to note that the charge asymmetry correlations in each $\pt$ bin, shown in Fig.~\ref{fig:pt}, are calculated solely from the particles within that $\pt$ bin. On the other hand, the charge asymmetry correlations as a function of centrality, shown in Fig.~\ref{fig:corr}, are calculated from all particle pairs with $\pt<2$~GeV/$c$.
As such, the results shown in Fig.~\ref{fig:corr} for a given centrality bin cannot be trivially obtained from those in Fig.~\ref{fig:pt} for the same centrality bin.

\begin{figure}[hbt]
\begin{center}
\includegraphics[width=0.4\textwidth]{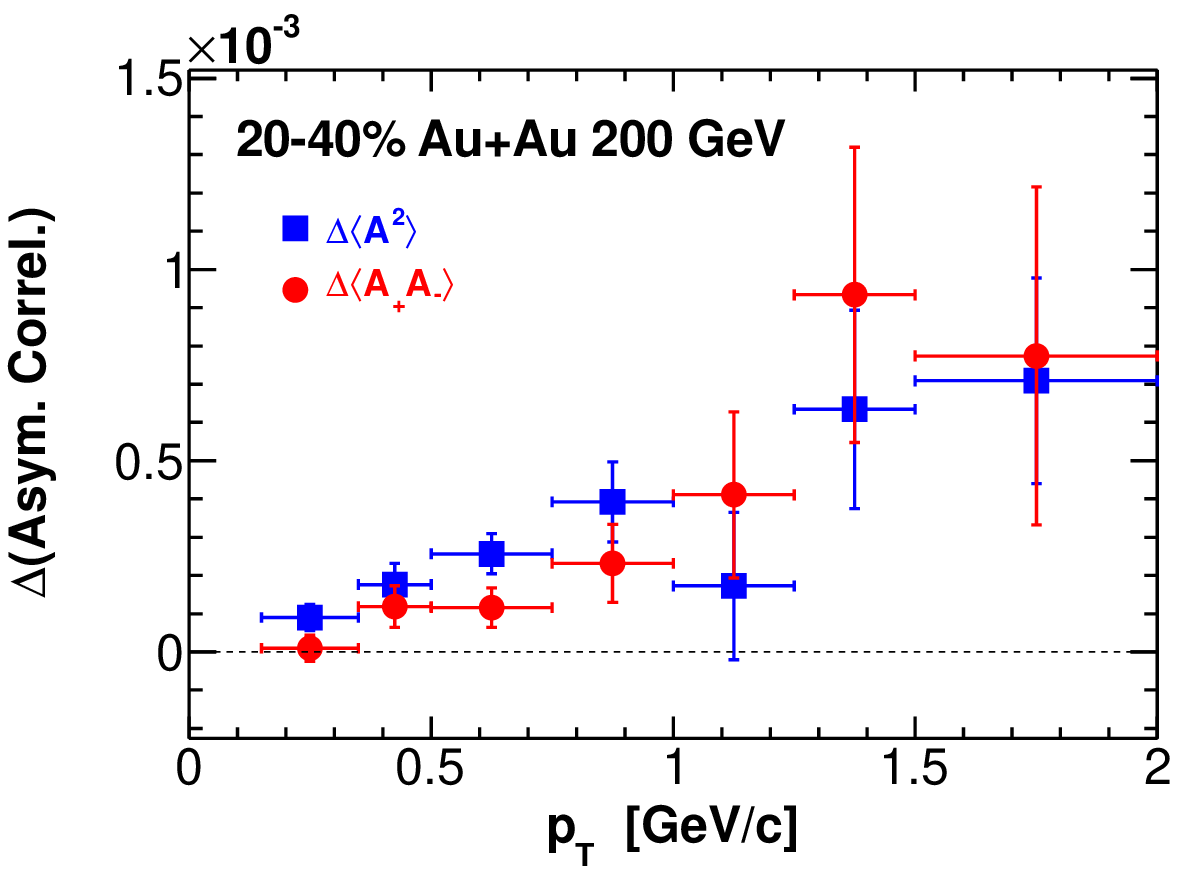}
\includegraphics[width=0.4\textwidth]{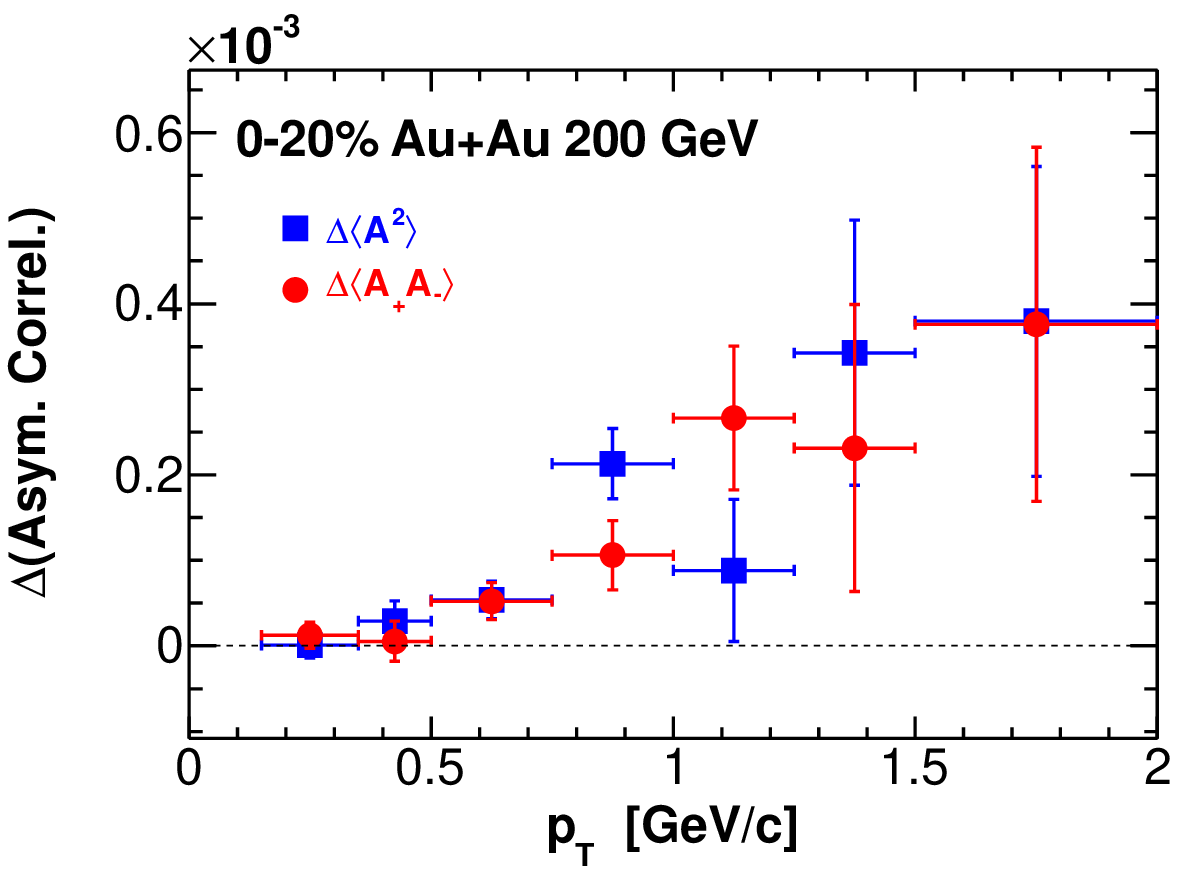}
\end{center}
\caption{(Color online) The correlation differences $\Delta\Asq=\delta\AsqUD-\delta\AsqLR$ and $\Delta\ApAm=\daa\AAUD-\daa\AALR$ as a function of $\pt$. The data are from 20-40\% central (upper) and 0-20\% central (lower) Au+Au collisions. 
The error bars are statistical only.}
\label{fig:diff_pt}
\end{figure}

Figure~\ref{fig:diff_pt} shows the \udlr\ correlations as a function of $\pt$ for 20-40\% central Au+Au collisions. The values of $\Delta\Asq$ and $\Delta\ApAm$ increase with $\pt$. There is qualitatively no difference in the $\pt$ dependence between the same- and opposite-sign correlations. On the other hand, it is generally expected that the \udlr\ difference should be most prominent at low $\pt$ if the \cme\ is responsible. Such a low-$\pt$ feature is not observed in these data. This was also qualitatively observed in the three-particle correlators~\cite{CorrelatorPRL,CorrelatorPRC}.

\subsection{Dependence on Event-by-Event Anisotropies\label{sec:v2}}

\begin{figure*}[hbt]
\begin{center}
\includegraphics[width=0.329\textwidth]{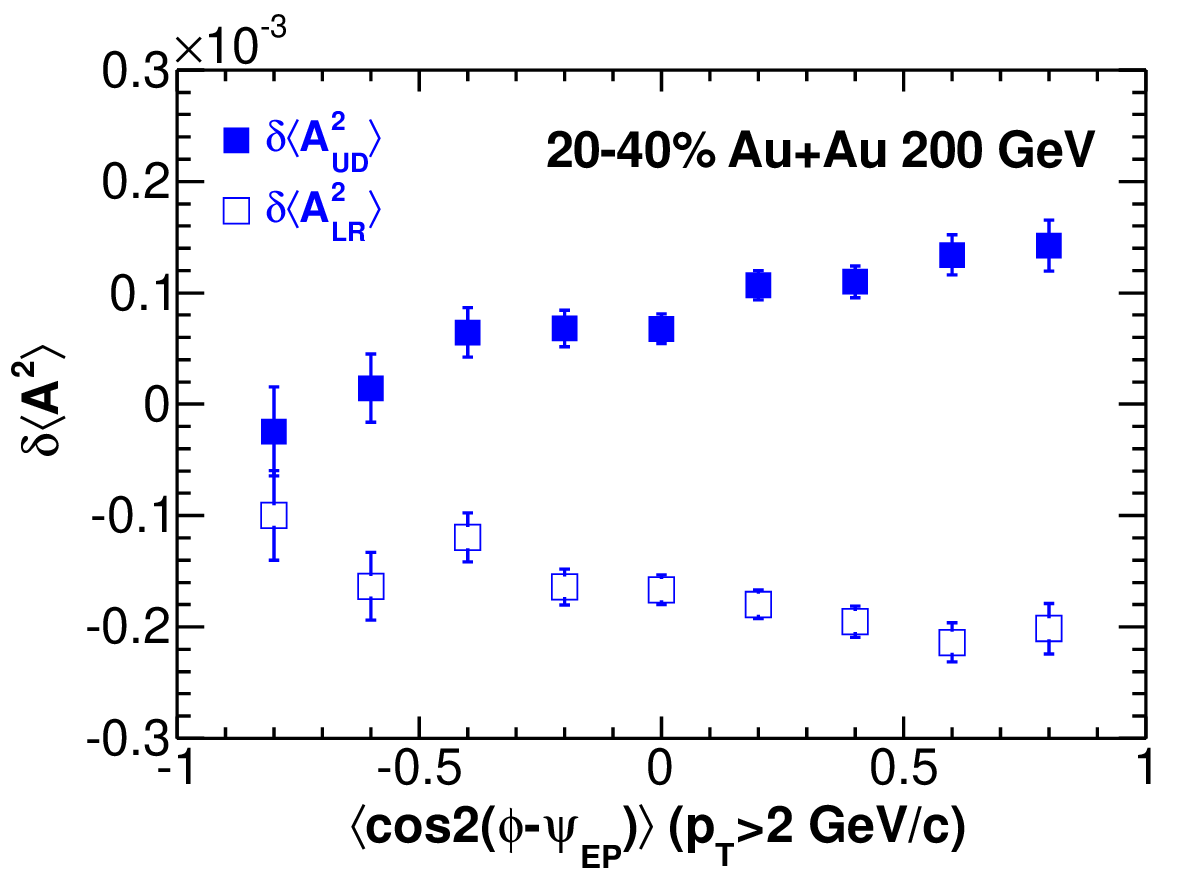}
\includegraphics[width=0.329\textwidth]{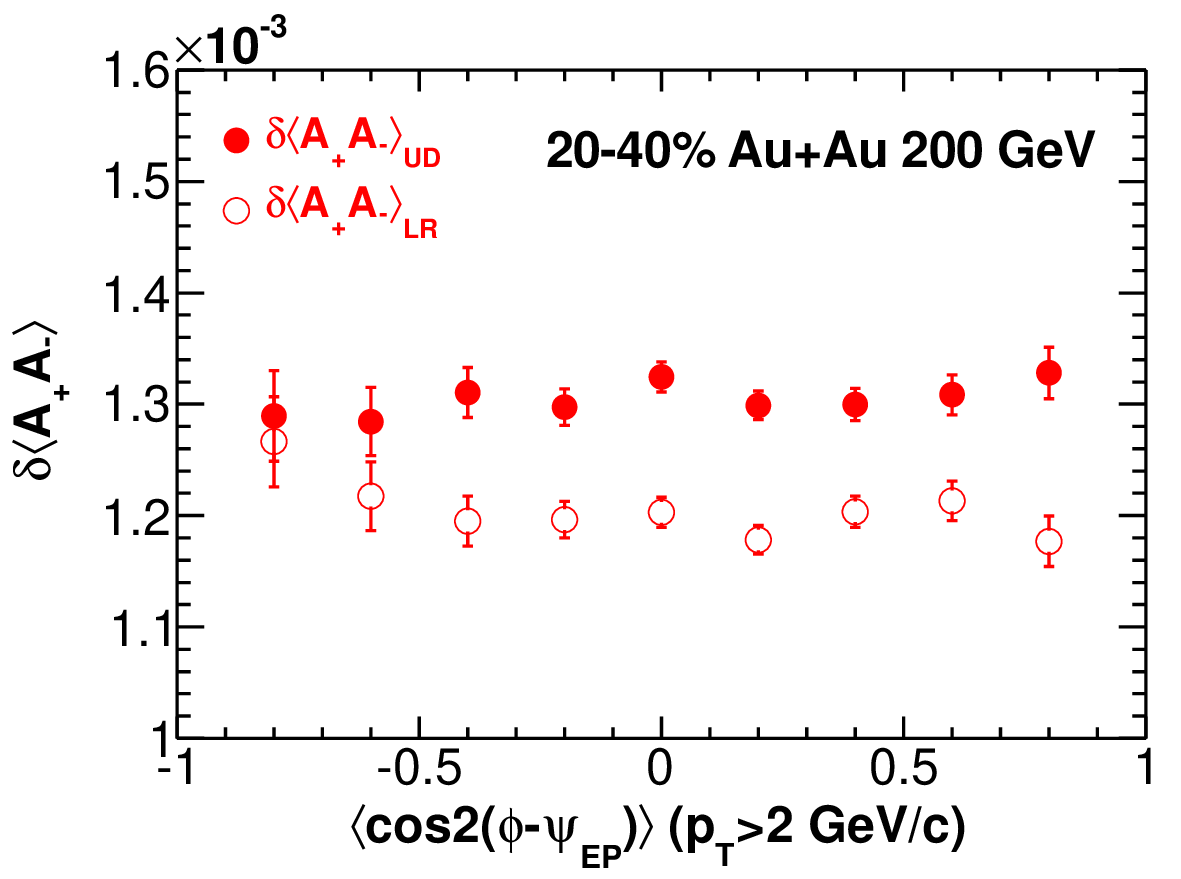}
\includegraphics[width=0.329\textwidth]{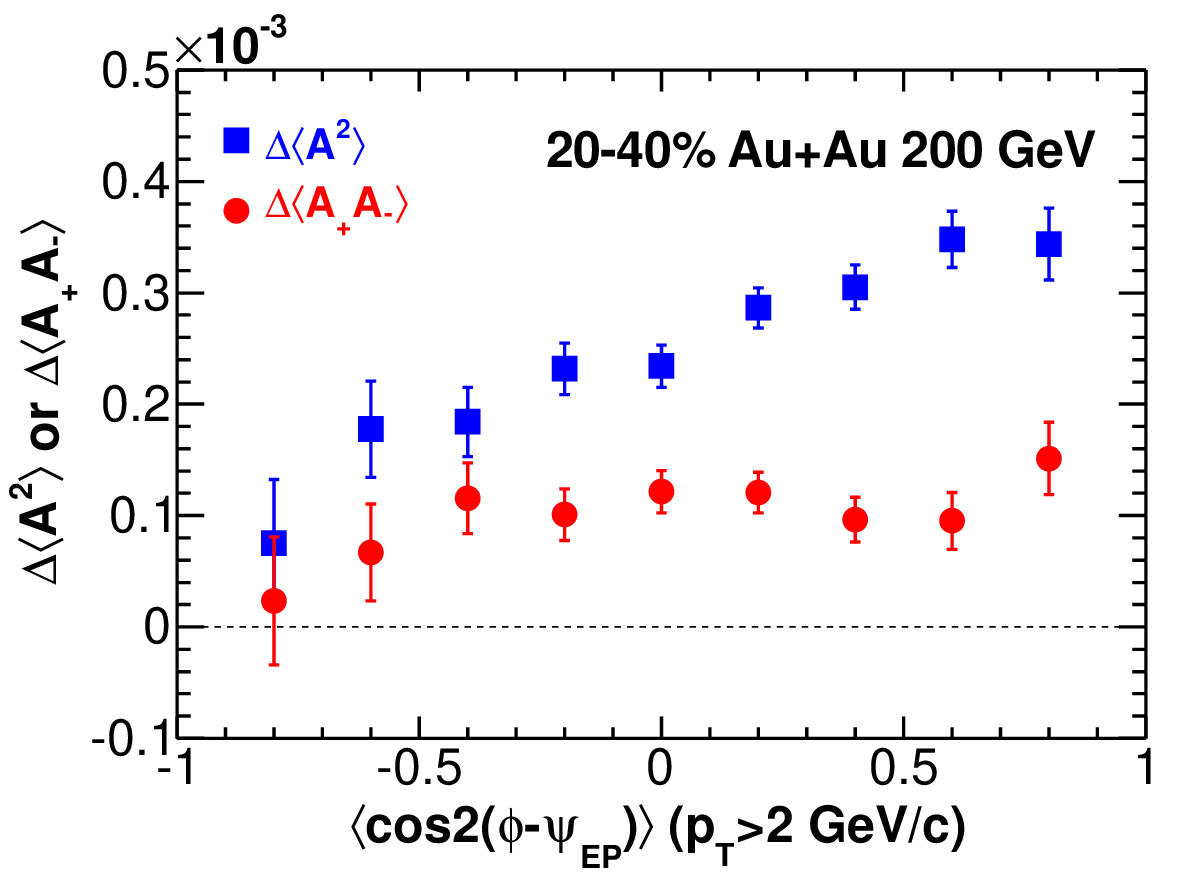}
\includegraphics[width=0.329\textwidth]{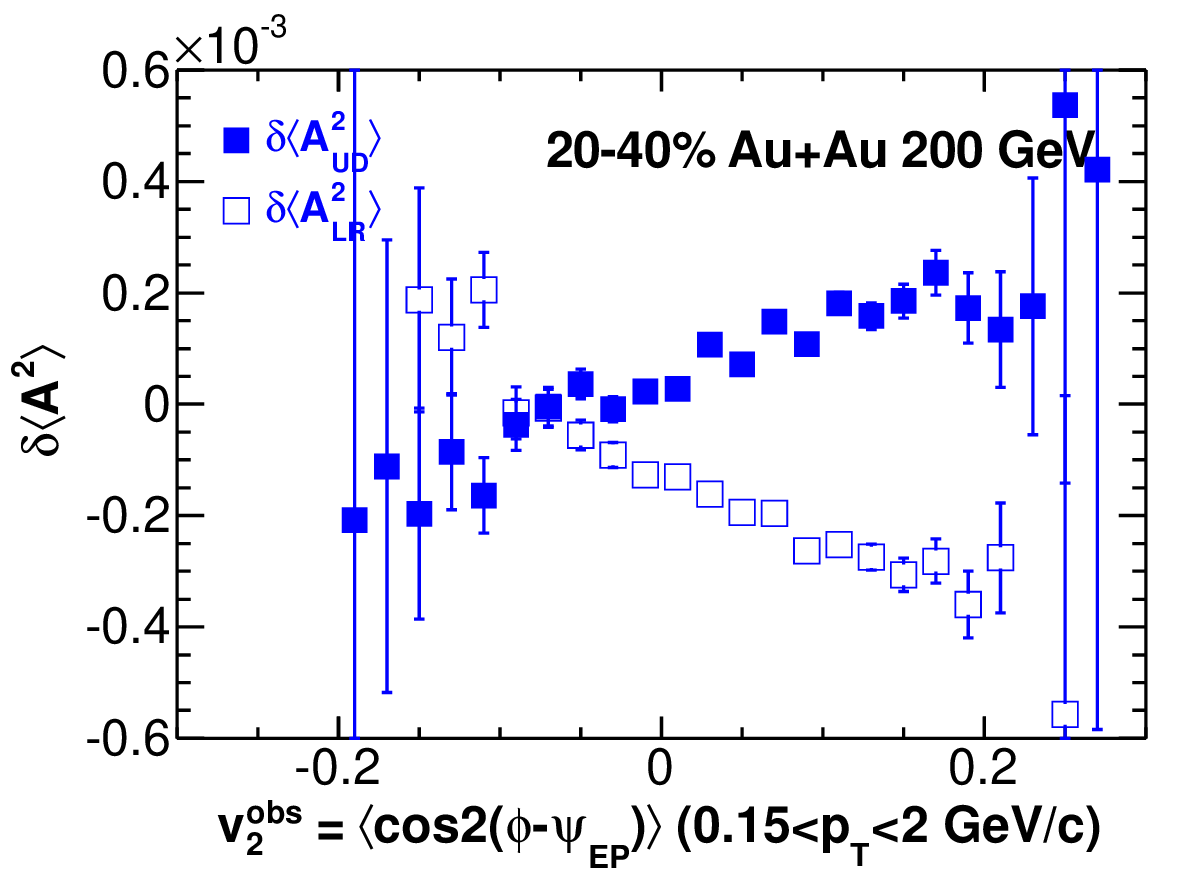}
\includegraphics[width=0.329\textwidth]{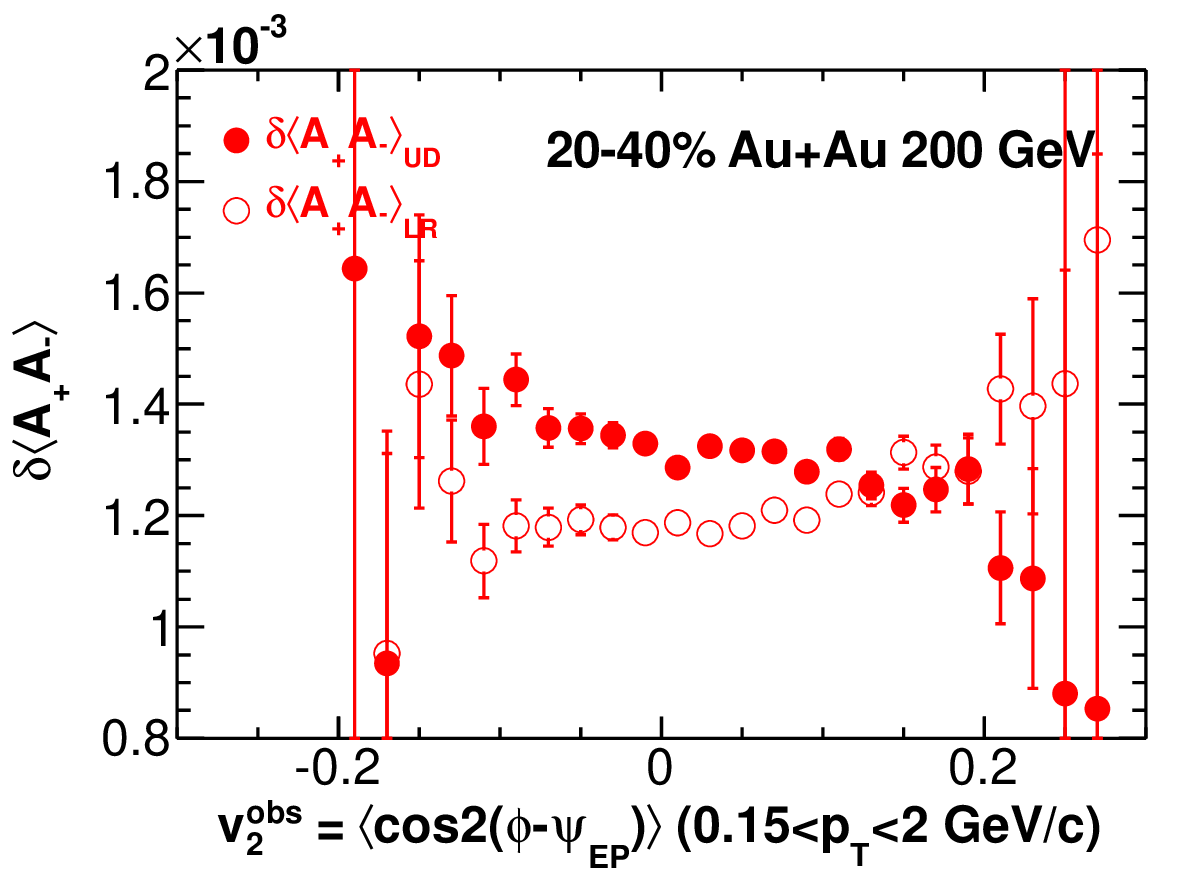}
\includegraphics[width=0.329\textwidth]{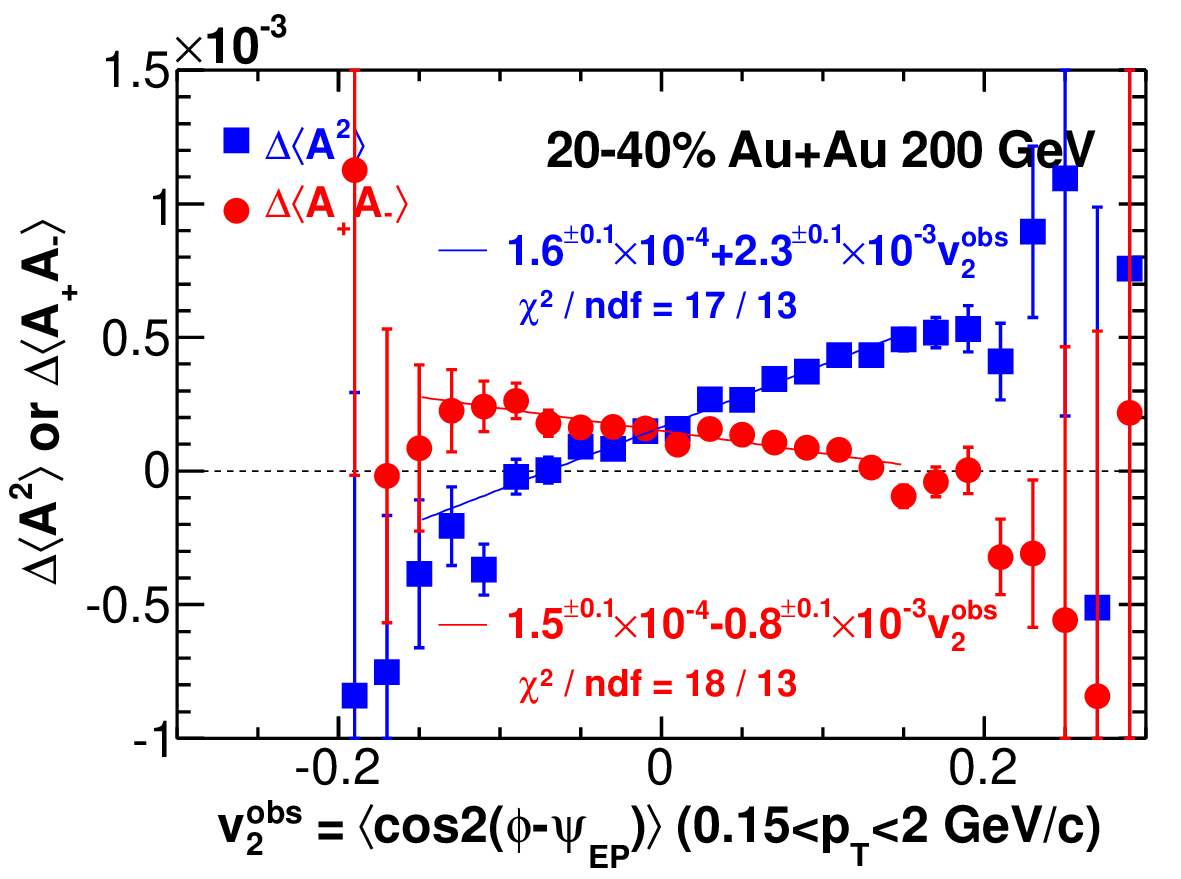}
\end{center}
\caption{(Color online) The charge asymmetry correlations $\delta\Asq$ (left panels) and $\delta\ApAm$ (center panels), and correlation differences $\Delta\Asq=\daa\AsqUD-\daa\AsqLR$ and $\Delta\ApAm=\daa\AAUD-\daa\AALR$ (right panels), as a function of the azimuthal elliptic anisotropy of high-$\pt$ ($\pt>2$~GeV/$c$) particles (upper panels) and low-$\pt$ ($\pt<2$~GeV/$c$) particles (lower panels). Data are from 20-40\% Au+Au collisions. The particle $\pt$ range of $0.15<\pt<2$~GeV/$c$ is used for both \ep\ construction and asymmetry calculation. Error bars are statistical.}
\label{fig:AvsV2}
\end{figure*}

There could be many physics mechanisms contributing to the event-plane dependent charge correlations. Voloshin estimated that resonance decays with anisotropies were insignificant to charge dependent correlations relative to the reaction plane~\cite{Voloshin}. Wang suggested that general cluster particle correlations with anisotropies could generate a sizable difference between the in-plane and out-of-plane particle correlations~\cite{Wang}. Pratt {\it et al.} argued that momentum conservation and local charge conservation together with elliptic flow could yield event-plane dependent correlations that differ between same- and opposite-sign pairs~\cite{Pratt}. A path-length dependent jet-quenching effect~\cite{jetspec,Horner,3part,ridge,Pawan,Aoqi} could also contribute ~\cite{Petersen}. 
%
To test these ideas experimentally, the values of $\delta\Asq$, $\daa\ApAm$, $\Delta\Asq$, and $\Delta\ApAm$ were studied as a function of the azimuthal elliptic anisotropy of high-$\pt$ and low-$\pt$ particles. The high-$\pt$ anisotropy may be most sensitive to the jet-quenching effect, while the low-$\pt$ anisotropy characterizes the bulk event shape. 
The event elliptic anisotropy is computed via $\langle\cos2(\phi-\psiEP)\rangle$ at low $\pt$ and high $\pt$. For low $\pt$, only the particles within one half of the TPC (also used in the asymmetry measurements) were used to compute the elliptic anisotropy, while the angle $\psiEP$ was reconstructed using the particles in the other half of the TPC. The variable $\vlow$ is used to stand for the low-$\pt$ anisotropy and is defined via $\vlow=\langle\cos2(\phi(\pt)-\psiEP)\rangle|_{0.15<\pt<2~{\rm GeV/}c}$. For high $\pt$, particles with $\pt>2$~GeV/$c$ from the entire TPC ($|\eta|<1$) were used to increase the statistics. The variable $\vhigh$ is used to stand for the high-$\pt$ anisotropy and is defined via $\vhigh=\langle\cos2(\phi(\pt)-\psiEP)\rangle|_{\pt>2~{\rm GeV/}c}$.

Figure~\ref{fig:AvsV2} (upper panels) shows the asymmetry correlation results in 20-40\% central Au+Au collisions as a function of $\vhigh$. Over the large range in $\vhigh$, relatively small variations are observed. This may indicate that path-length dependent jet-quenching does not have a significant impact on the charge asymmetry correlations. This is consistent with the theoretical study described in Ref.~\cite{Petersen}.

Figure~\ref{fig:AvsV2} (lower panels) shows the asymmetry correlation results as a function of $\vlow$. Significant changes are observed in the variances. The values of $\delta\AsqLR$ decrease with increasing $\vlow$ while the values of $\delta\AsqUD$ increase. This results in a strong increase in the difference $\Delta\Asq$ with increasing $\vlow$. 
The variations in the covariances are significantly weaker. This results in a weaker dependence of $\Delta\ApAm$ on $\vlow$. However, the change appears to be in the opposite direction, decreasing with increasing $\vlow$. The superimposed linear fits in Fig.~\ref{fig:AvsV2} lower right panel will be discussed in Sec.~\ref{sec:disc:v2}.

It is interesting to note that the $\Delta\Asq$ and $\Delta\ApAm$ cross at $\vlow\approx0$ and the crossing point is at a positive value. For $\vlow=0$, no difference is apparent between the same- and opposite-sign pair correlations. Of course, the average $\vlow$ in these data is nonzero, but positive. As a result, the asymmetry variance of all events is larger than the covariance. As suggested in Appendix~\ref{app:EPres}, it is possible that, in events with negative $\vlow$, the reconstructed \ep\ does not reflect the true reaction plane, perhaps being orthogonal to rather than aligned with the reaction plane. This would mean that the \ud\ and \lr\ are flipped for those events with a significantly negative $\vlow$.
%
%
The magnitudes of the asymmetry correlations depend on centrality, but their qualitative features versus $\vhigh$ and $\vlow$ are similar for different centralities. 

The measurements shown in the lower panels of Fig.~\ref{fig:AvsV2} were also performed with a random event plane. 
The results with the random event plane are similar to those shown in Fig.~\ref{fig:AvsV2}. They have the same dependence on $\vlow$, but the same- and opposite-charge results in the right lower panel cross at zero intercept. 

\subsection{Dependence on Wedge Size\label{sec:size}}

Above, the charge multiplicity asymmetry correlations between hemispheres have been described. One advantage of these asymmetries is that they are calculated from the same set of particles, and are only divided either \ud\ or \lr. The statistical fluctuations and detector effects cancel in the difference between \ud\ and \lr, so $\Delta\Asq\equiv\delta\AsqUD-\delta\AsqLR=\AsqUD-\AsqLR$. However, measurements of multiplicity fluctuations within hemispheres are not sensitive to possibly smaller scale angular structures of the charge separation. For example, the correlated charged particle pairs from the \cme\ that were initially aligned with the total angular momentum direction may or may not remain aligned (or preferentially aligned) in the same direction~\cite{PVquench}. In order to investigate the angular structure of the charge separation, the charge multiplicity asymmetry measurements were restricted to azimuthal ranges (``wedges"), $\dphiw$, that are smaller than $\pi$ (hemispheres), allowing the study of the charge separation as a function of the wedge size.

\begin{figure}[hbt]
\begin{center}
\includegraphics[width=0.4\textwidth]{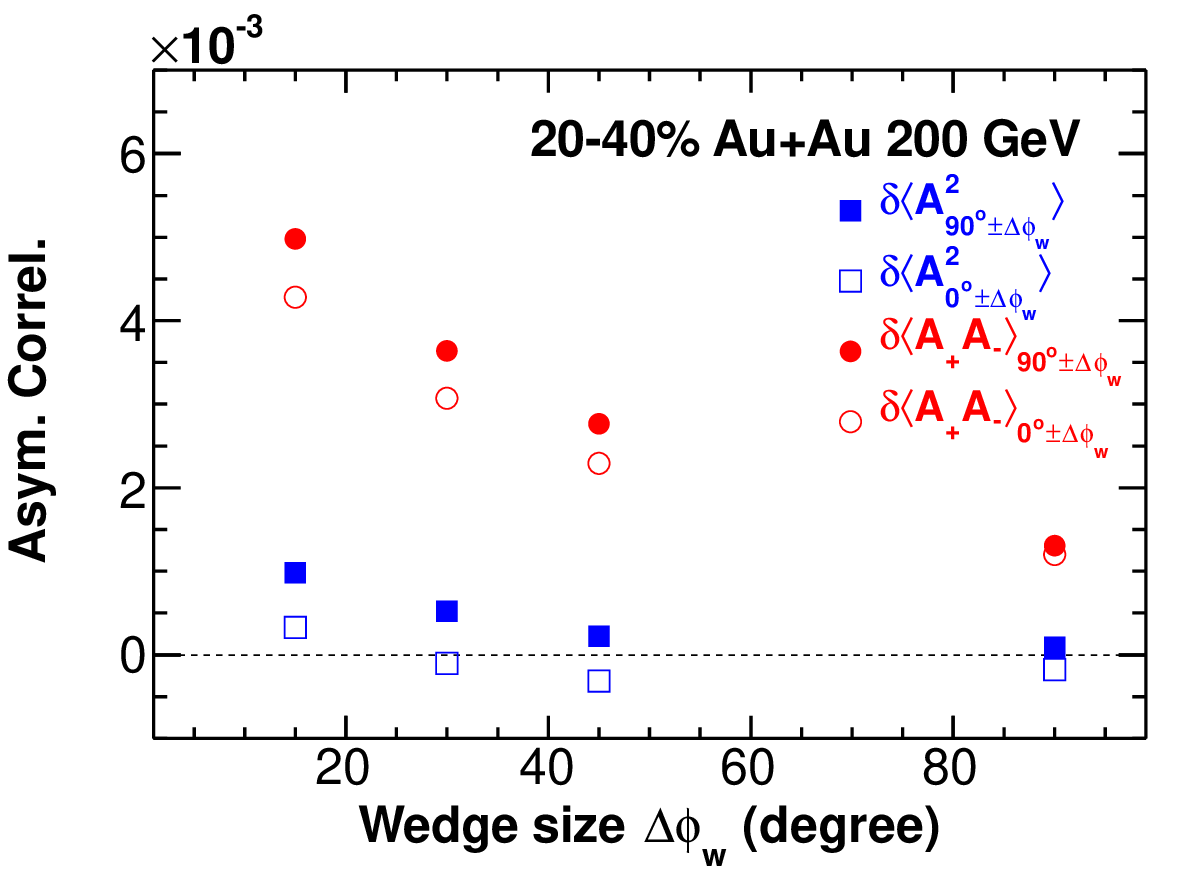}
\includegraphics[width=0.4\textwidth]{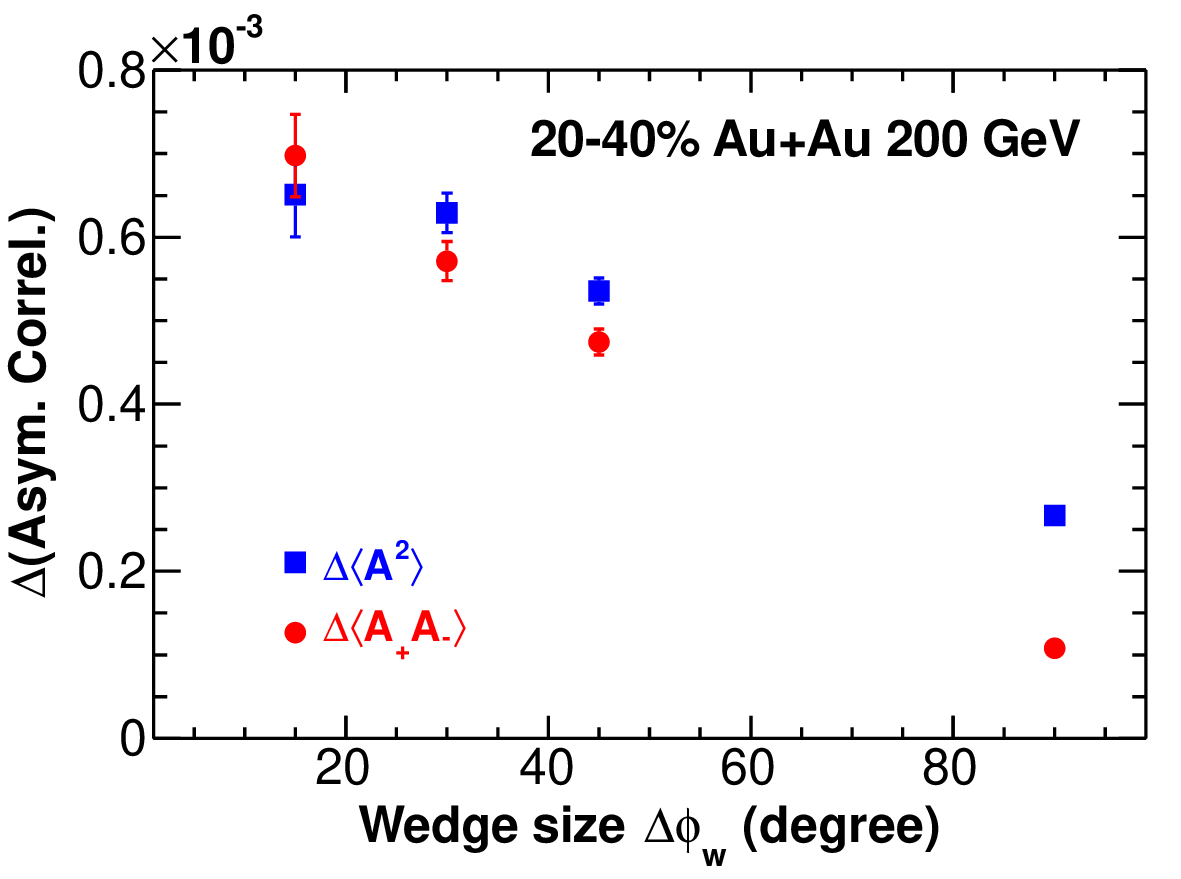}
\end{center}
\caption{(Color online) The wedge size dependences of charge multiplicity asymmetry correlations (upper panel), and their differences between out-of-plane and in-plane, $\Delta\ASq$ and $\Delta\APAM$ (lower panel) for 20-40\% central Au+Au collisions. 
The error bars are statistical only.}
\label{fig:size}
\end{figure}

Figure~\ref{fig:size} (upper panel) shows the asymmetry correlations, $\delta\ASqUD$, $\delta\ASqLR$, $\daa\APAMUD$, and $\daa\APAMLR$, versus the wedge azimuthal size (see Fig.~\ref{fig:diag2}(a)). Both the covariances, $\daa\APAMUD$ and $\daa\APAMLR$, increase with decreasing wedge size, $\dphiw$. This suggests that the major contribution to the opposite-sign charge correlation is local. The variances, $\delta\ASqUD$ and $\delta\ASqLR$, increase with decreasing $\dphiw$.

Figure~\ref{fig:size} (lower panel) shows the difference in the asymmetry correlations, $\Delta\ASq=\delta\ASqUD-\delta\ASqLR$ and $\Delta\APAM=\daa\APAMUD-\daa\APAMLR$, between the out-of-plane and in-plane directions. Both $\Delta\ASq$ and $\Delta\APAM$ increase with decreasing wedge size and have qualitatively similar trends. The difference between the two seems to diminish with decreasing wedge size.



\begin{figure}[hbt]
\begin{center}
\includegraphics[width=0.4\textwidth]{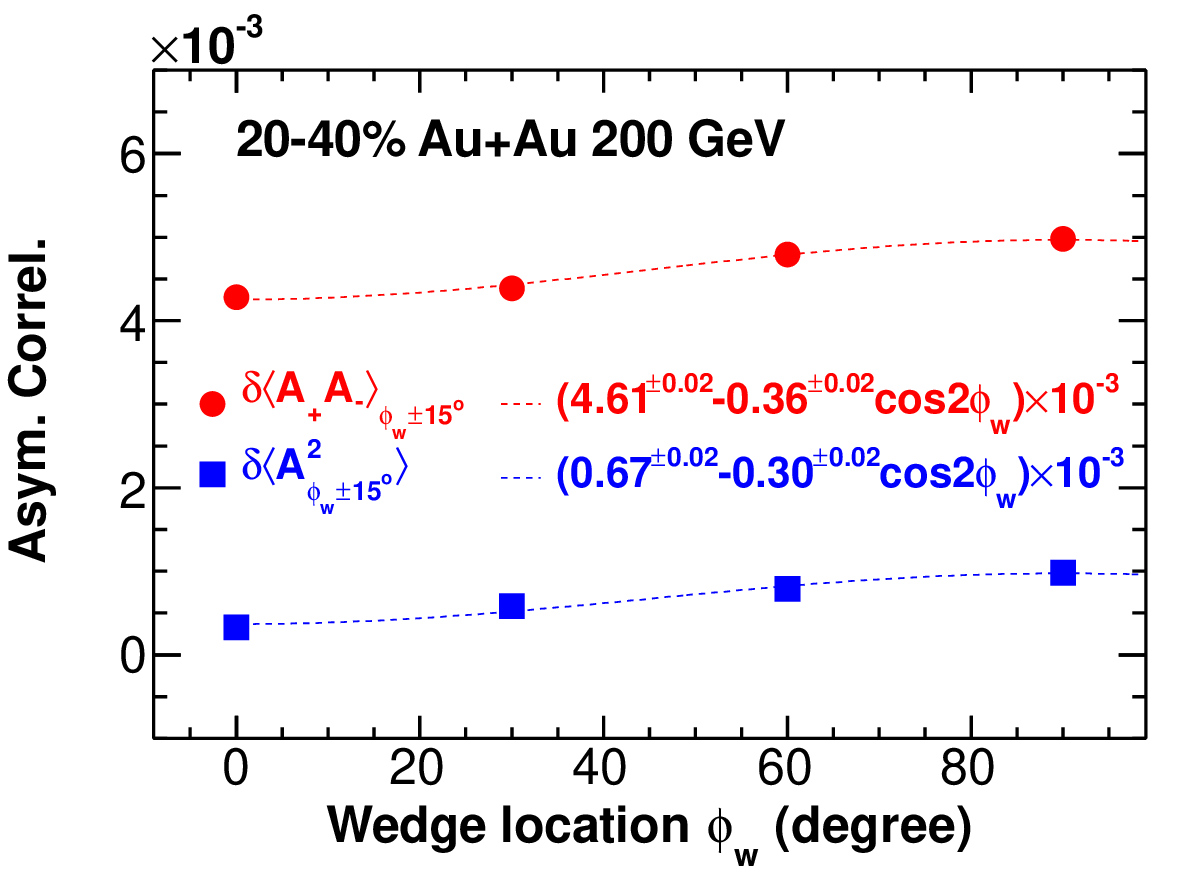}
\end{center}
\caption{(Color online) Charge multiplicity asymmetry correlations as a function of the wedge location, $\phiw$, in 20-40\% central Au+Au collisions. The wedge size is $30^{\circ}$. The curves are the characteristic $\cos(2\phiw$) to guide the eye. 
The error bars are statistical only.}
\label{fig:angle}
\end{figure}

In the above, the focus has been on the difference between the in-plane and out-of-plane wedges. In the following, the charge asymmetry correlations in fixed-size back-to-back wedges, as a function of the wedge azimuth relative to the event plane, are discussed ({\it cf.} Fig.~\ref{fig:diag2}(b)). Figure~\ref{fig:angle} shows the asymmetry correlations between $30^{\circ}$-wide back-to-back wedges, $\delta\ASqFix$ and $\daa\APAMFix$, versus the wedge azimuthal location $\phiw$ relative to the event plane. The data are from 20-40\% central Au+Au collisions. The asymmetry correlations increase from in-plane to out-of-plane, as expected, for both same-sign and opposite-sign charges. The \ep-independent part of the correlations is stronger in opposite-sign charges. 
The dependence appears to follow the characteristic behavior of $\cos(2\phiw)$.

\section{Discussions\label{sec:disc}}

These measurements were motivated by the \lpv/\cme. The \lpv/\cme\ produces quark charge separations along the system's magnetic field axis~\cite{PV,PV1Qsep,PV2Qsep}. This may result in a charge separation reflected by the final state hadrons. A negative correlator for same-sign pairs was observed~\cite{CorrelatorPRL,CorrelatorPRC}. This was qualitatively consistent with the expectation from the local strong parity violation~\cite{PV1Qsep,PV2Qsep,PV3Qsep,PVquench}. However, a small, close to zero, correlator for opposite-sign pairs was also observed~\cite{CorrelatorPRL,CorrelatorPRC}. The result appeared to be inconsistent with the naive \cme\ expectation above. To explain the preliminary version of the results in Ref.~\cite{CorrelatorPRL,CorrelatorPRC}, Kharzeev \etal~\cite{PVquench} suggested that the initial correlation among the quarks from \cme\ would not survive the subsequent dynamic evolution to the final state hadrons, except those emitted from the surface of the collision zone. The back-to-back correlations of opposite-sign pairs from \cme\ would be lost because at least one quark of a pair would be affected by interactions with the evolving medium. Such a medium interaction scenario could qualitatively explain the observed opposite-sign correlator~\cite{PVquench,CorrelatorPRL,CorrelatorPRC}. More recent measurements by STAR~\cite{Dhevan} may indicate that a parity conserving background is more likely to explain the suppression rather than the medium induced back-to-back suppression previously supposed~\cite{CorrelatorPRL,CorrelatorPRC}. 

The \lpv/\cme\ should produce more same-sign charge pairs in the up-down hemispheres, yielding wider (larger variance) asymmetry distributions of both positive and negative particle multiplicities. Therefore, the up-down dynamical variance $\delta\AsqUD$ should be larger than the left-right dynamical variance $\delta\AsqLR$. The \lpv/\cme\ should also produce an anticorrelation between multiplicity asymmetries of positive particles and negative particles in the up-down rather than in the left-right hemisphere. This should result in $\daa\AAUD$ being smaller than $\daa\AALR$. 

The present results show $\delta\AsqUD>\delta\AsqLR$ in Au+Au collisions at all centralities at 200~GeV (see {\it e.g.} Fig.~\ref{fig:diff}). The \ud\ asymmetry distribution is wider than the \lr\ asymmetry distribution. This is qualitatively consistent with the \cme, although the same-sign particles are preferentially back-to-back in medium-central to central collisions regardless of whether they are oriented in-plane or out-of-plane. However, $\daa\AAUD>\daa\AALR$ in all centralities except peripheral collisions. The opposite-sign particle pairs are more strongly emitted in the same hemisphere in the \ud\ than \lr\ direction. This seems contrary to the expectations from the \cme\ alone. 
In $d$+Au and very peripheral Au+Au collisions, $\daa\AAUD<\daa\AALR$. In these small multiplicity collisions, there might exist autocorrelations between particles in the $\eta<0$ and $\eta>0$ regions. One such constraint may be due to ``non-flow," {\it e.g.} di-jets, causing the reconstructed event plane to align preferentially with the plane containing the two jets. In this case, the left and right hemispheres are the two jet-hemispheres, which have large fluctuations in multiplicity and result in large \lr\ asymmetries. The up and down hemispheres, on the other hand, are a roughly symmetric division of the event with respect to the di-jet axis, resulting in smaller asymmetries.

There is little theoretical guidance regarding the quantitative magnitude of the charge asymmetry expected from the \cme. An order of magnitude estimate suggests that the charge asymmetry could be on the order of a few percent~\cite{PV1Qsep,PV2Qsep,PV3Qsep}, charge asymmetry correlations on the order $10^{-4}$-$10^{-3}$. The medium attenuation effects likely reduce the asymmetry correlations by an order of magnitude~\cite{PVquench}. Recently, it has been argued that charge asymmetries can arise from strong magnetic fields and well-known QCD processes without invoking \lpv~\cite{Mueller}. However, the estimated magnitude of the charge asymmetry correlations from those processes is orders of magnitude smaller~\cite{Mueller}.

It is worthwhile to point out that the present charge multiplicity correlation observables are connected to the three-particle correlators in Refs.~\cite{CorrelatorPRL,CorrelatorPRC}, as described in detail in Appendix~\ref{app:comp}. The correlation observables reported in this paper include the entire correlation structure while the three-particle correlators focussed on the lowest-order azimuth multipole only. The comparison of the two observables suggests that the higher order multipoles may be significant in the opposite-sign charge correlations, but they are insignificant in the same-sign correlations. When analyzed using the same three-particle correlator observable, the present data are consistent with those in Refs.~\cite{CorrelatorPRL,CorrelatorPRC}. 

It is also worthwhile to point out that the ``modulated sign correlations'' recently reported in Ref.~\cite{Dhevan} are more closely related to the present charge multiplicity correlation observables than the three-particle correlators in Refs.~\cite{CorrelatorPRL,CorrelatorPRC}. In fact, the opposite-sign modulated sign correlation in Ref.~\cite{Dhevan} is identical to the present opposite-sign multiplicity correlation observable $\Delta\ApAm$ except for a constant multiplicative factor. The opposite-sign modulated sign correlation from the Run VII data in Ref.~\cite{Dhevan} is consistent within errors with the Run IV $\Delta\ApAm$ data reported here.

\subsection{Is the Charge Separation In- or Out-of-Plane?\label{sec:disc:size}}

Although both the same- and opposite-sign \udlr\ correlation results are positive in non-peripheral collisions (see Fig.~\ref{fig:diff}), the same-sign result, $\Delta\Asq$, is larger than the opposite-sign result, $\Delta\ApAm$. It is possible that the various backgrounds may produce correlations that fall in between -- being equal for the different sign pairs ({\it i.e.} the physics backgrounds give zero charge separation). Then, the different results for same- and opposite-sign charges, now having different signs once subtracting the common background, would be consistent with \cme. No medium effects such as those described in Ref.~\cite{PVquench} would be needed. One may simply refer to the larger \udlr\ of same-sign than opposite-sign correlation, $\Delta=\Delta\Asq-\Delta\ApAm>0$, as ``charge separation'' across the event plane.

If the \cme\ results in same-sign pairs in the final state that are still preferentially directed along the magnetic field axis~\cite{PVquench}, or are of the form $a_1\sinijk$~\cite{Voloshin,CorrelatorPRL,CorrelatorPRC}, then the difference should depend on the wedge size. The smaller the wedge size, the larger the \cme\ effect on the difference $\DELTA=\Delta\ASq-\Delta\APAM$. However, it is also possible that, due to final state interactions, the charge separation effect across the reaction plane is no longer preferentially directed in the orbital angular momentum direction, so that $\DELTA$ may not increase with decreasing $\dphiw$. Nevertheless, it is interesting to examine $\DELTA$ as a function of $\dphiw$ which may reveal the angular distribution of the charge separation effect.

\begin{figure}[hbt]
\begin{center}
\includegraphics[width=0.4\textwidth]{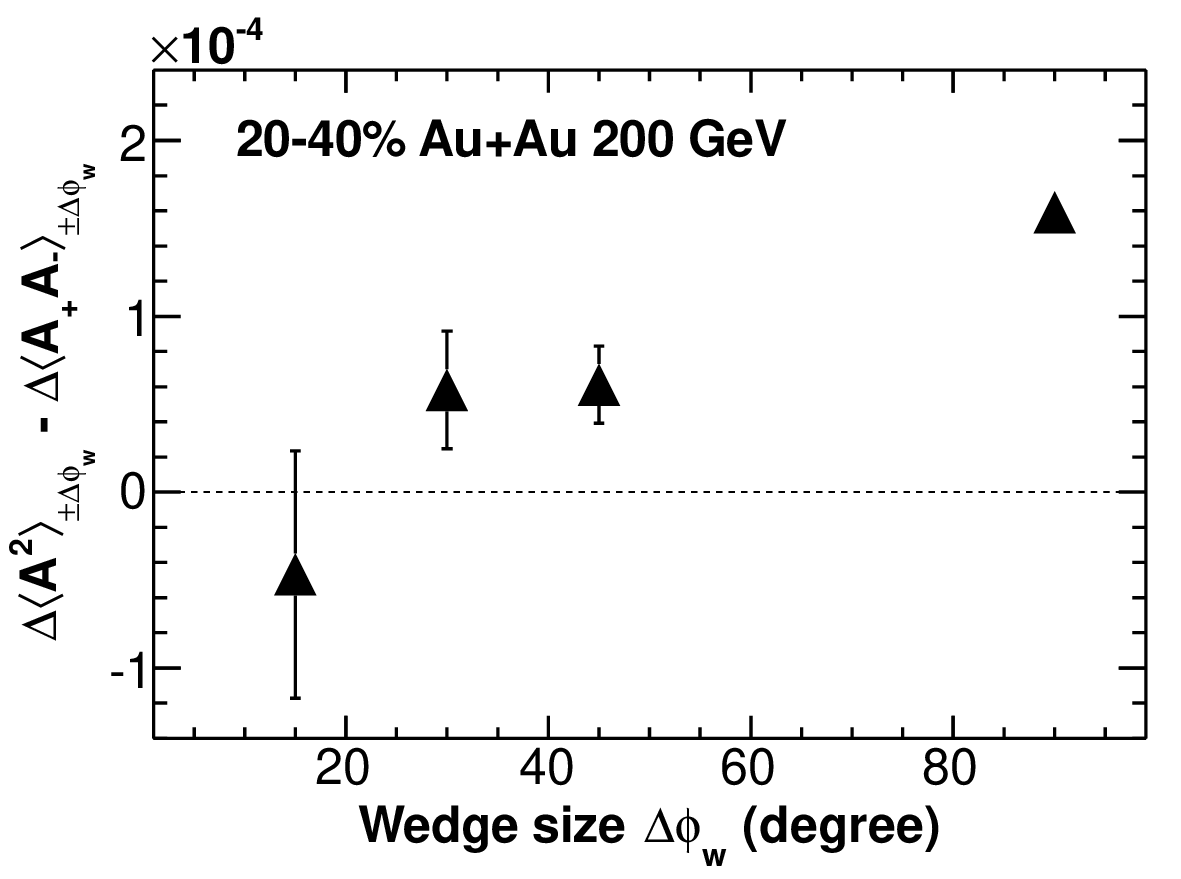}
\end{center}
\caption{The wedge size dependence of the difference between the same-sign and opposite-sign, $\Delta\ASq-\Delta\APAM$, shown in Fig.~\ref{fig:size}. The error bars are statistical only.}
\label{fig:size2}
\end{figure}

Figure~\ref{fig:size2} shows the values of $\DELTA$ versus $\dphiw$. The charge separation effect, $\DELTA$, decreases with decreasing wedge size, $\dphiw$. It appears that the difference between same- and opposite-sign correlation is diminished for small wedge sizes. 
The largest difference between the same- and opposite-sign pairs is obtained when whole hemispheres are used. This suggests that the effect of charge separation across the event plane happens in the vicinity of the in-plane direction rather than out-of-plane.

\begin{figure}[hbt]
\begin{center}
\includegraphics[width=0.4\textwidth]{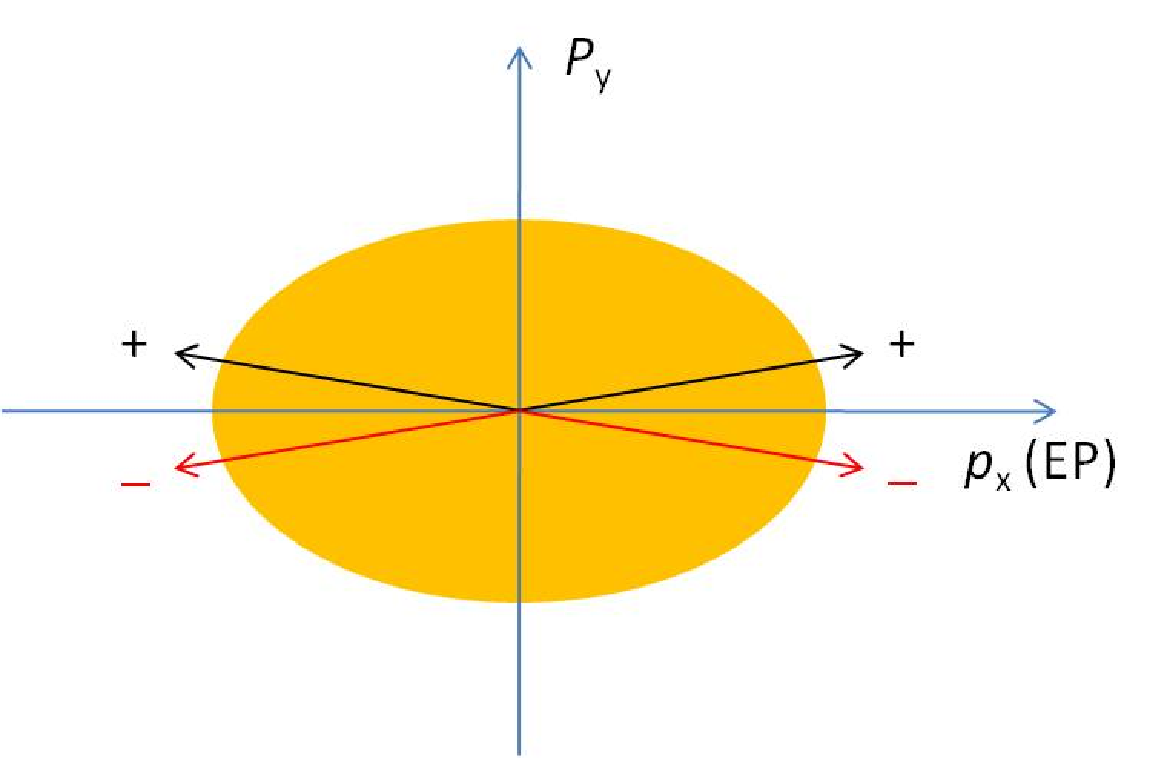}
\end{center}
\caption{(Color online) Schematic depiction of the charge pairs responsible for the observed wedge-size dependence of the difference between the same- and opposite-sign \udlr\ correlations. Note that the same- and opposite-sign pairs that would yield the assumed common physics background falling between the same- and opposite-sign \udlr\ measurements are excluded from this cartoon.}
\label{fig:cartoon3}
\end{figure}

The cartoon in Fig.~\ref{fig:cartoon3} depicts a map of the signal charges consistent with the present results. The supposed common background used in this discussion, which falls in-between the same- and opposite-sign measurements, is excluded from the cartoon. There are preferentially more back-to-back same-sign pairs along the in-plane direction. The positive pairs preferentially occupy one hemisphere either above or below the reaction plane and the negative pairs preferentially occupy the opposite hemisphere. If those hadron pairs are the result of the \cme\ from the initial quarks, then the data would suggest that those quarks initially moving perpendicular to the reaction plane (along the magnetic field direction)~\cite{PVquench} have been deflected (and hadronize) toward the reaction plane direction. The same-sign correlator measure that was previously reported does not distinguish between deflected pairs and the initial pairs without any deflection~\cite{Koch,Koch2}. The opposite-sign correlator would be close to zero for the configuration depicted in the cartoon.

\subsection{Is the Charge Separation a CME Signal or Background?\label{sec:disc:v2}}

\begin{figure}[hbt]
\begin{center}
\includegraphics[width=0.4\textwidth]{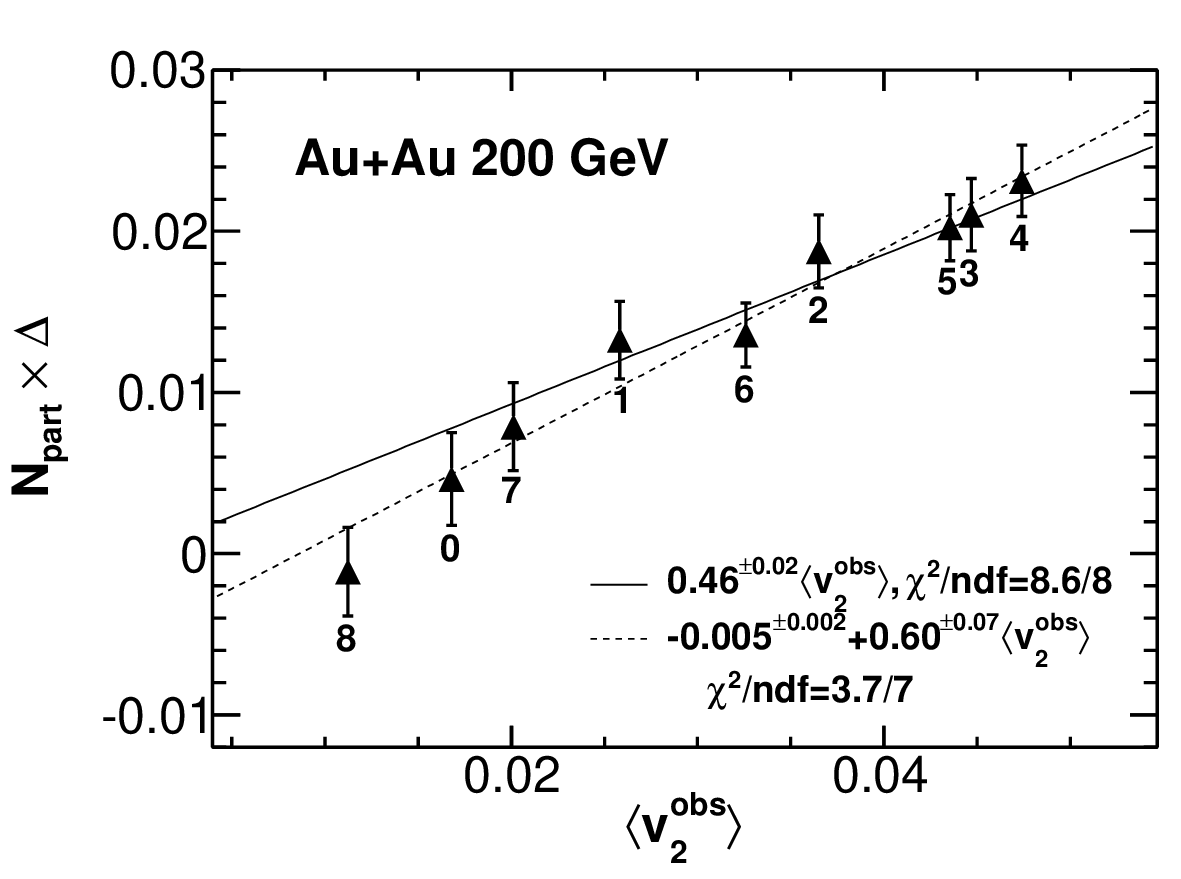}
\end{center}
\caption{The values of $\Delta\Asq-\Delta\ApAm$, scaled by $\Npart$, as a function of the measured average elliptic anisotropy $\mean{\vlow}$ in Au+Au collisions. The centrality bin number is labeled by each data point, 0 for 70-80\% up to 8 for 0-5\%. 
The error bars are statistical only.}
\label{fig:AdiffvsV2ave}
\end{figure}

The charge separation effects described above could be due to the \cme\ if the physics backgrounds for the same- and opposite-sign pair correlations are similar and fall in between the two, although the charge separation seems to happen in the vicinity of the reaction plane. It is, however, possible that the physics backgrounds may be quite different for the same- and opposite-sign pairs, and the same- and opposite-sign difference may be dominated by physics backgrounds. For example, local charge conservation will naturally cause differences between the same- and opposite-sign pairs \cite{Pratt}. In fact, the results shown in Fig.~\ref{fig:diff} indicate that the centrality dependence of the asymmetry correlations is similar to the centrality dependence of the elliptic anisotropy. This is more clearly shown in Fig.~\ref{fig:AdiffvsV2ave} where the difference between the same- and opposite-sign results (scaled by $\Npart$) is plotted as a function of the measured average elliptic anisotropy in each centrality bin. The dependence is roughly linear; the lines in Fig.~\ref{fig:AdiffvsV2ave} show two linear fits, one with the intercept fixed at zero and the other with the intercept as a free parameter. If the charge separation is indeed a correlation background, then the approximate proportionality suggests that the charge-dependent correlation strength is insensitive to centrality. However, the apparent linear relationship does not necessarily mean that the charge separation must be an anisotropy related background. Because the \cme\ and the average anisotropy are both functions of centrality, they can be indirectly related resulting in an apparent relationship between the charge separation and the average anisotropy. 

In order to gain further insights, one wants to fix the centrality hence the possible \cme, and vary the event anisotropy. This can be achieved by the study in Fig.~\ref{fig:AvsV2} of the asymmetry correlations as a function of the event-by-event elliptic anisotropy of the measured particles. 
Figure~\ref{fig:AvsV2} suggests, given a fixed range of centrality, that the bulk event structure may have a significant effect and the backgrounds for same- and opposite-sign pairs may indeed be different. The results in Fig.~\ref{fig:AvsV2} could be interpreted as follows. The values of $\delta\AsqLR$ decrease with increasing $\vlow$, while the values of $\delta\AsqUD$ increase. The trends of $\delta\AsqLR$ could result from a relative abundance of back-to-back same-sign pairs in-plane rather than out-of-plane. The more abundant back-to-back pairs in-plane give a larger $\vlow$ and reduce the \lr\ asymmetry, thereby decreasing $\delta\AsqLR$. Likewise, the $\delta\AsqUD$ trends could result from a reduction in the back-to-back same-sign pairs out-of-plane rather than in-plane, which increases both the $\vlow$ and $\delta\AsqUD$.
The $\vlow$ dependences in $\daa\AAUD$ and $\daa\AALR$ are significantly weaker. The trends seem to be opposite from those in $\delta\AsqUD$ and $\delta\AsqLR$. This may stem from the different nature of the correlations between opposite-sign pairs (small-angle) and same-sign pairs (back-to-back). These behaviors of $\delta\Asq$ and $\daa\ApAm$ with $\vlow$ may be in-line with suggestions that those charge correlations arise from cluster particle correlations overlaid with elliptic anisotropy~\cite{Wang,Pratt}.

\begin{figure*}[hbt]
\begin{center}
\includegraphics[width=0.329\textwidth]{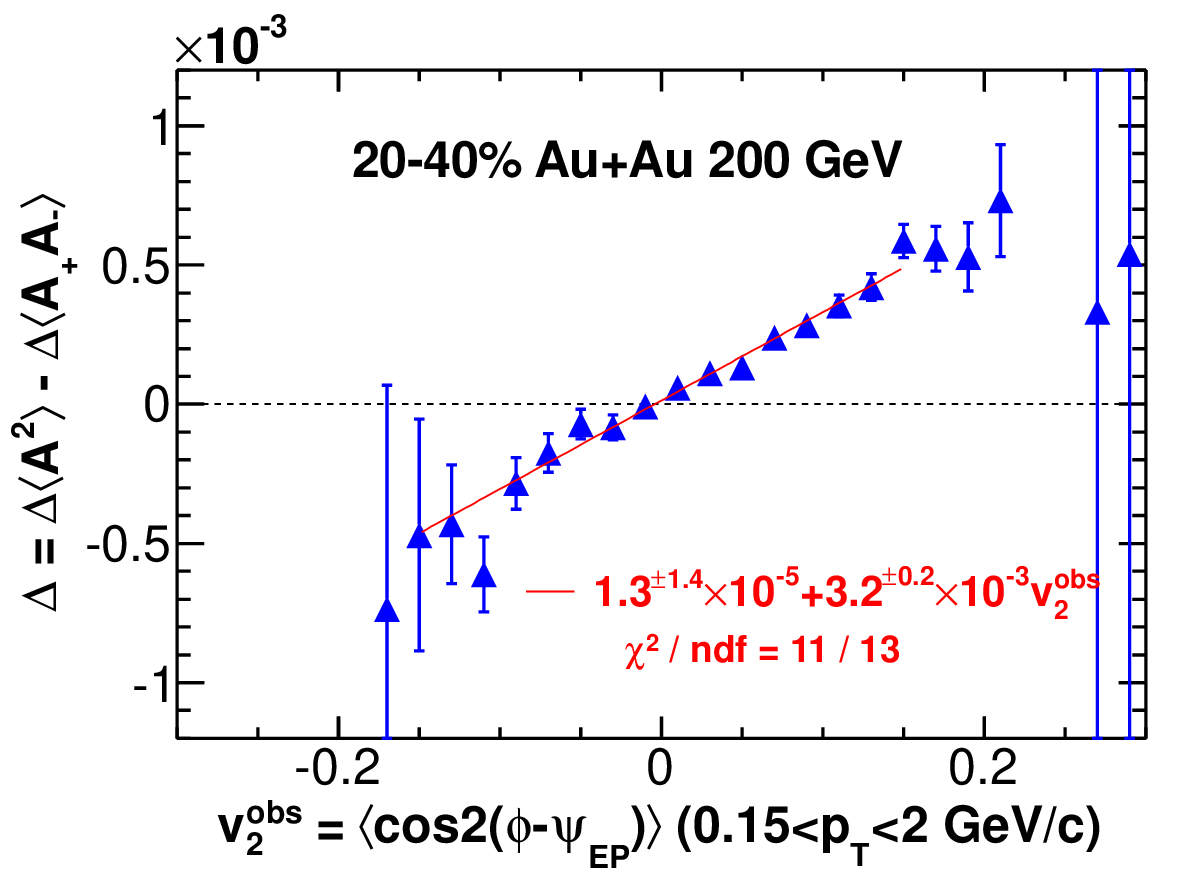}
\includegraphics[width=0.329\textwidth]{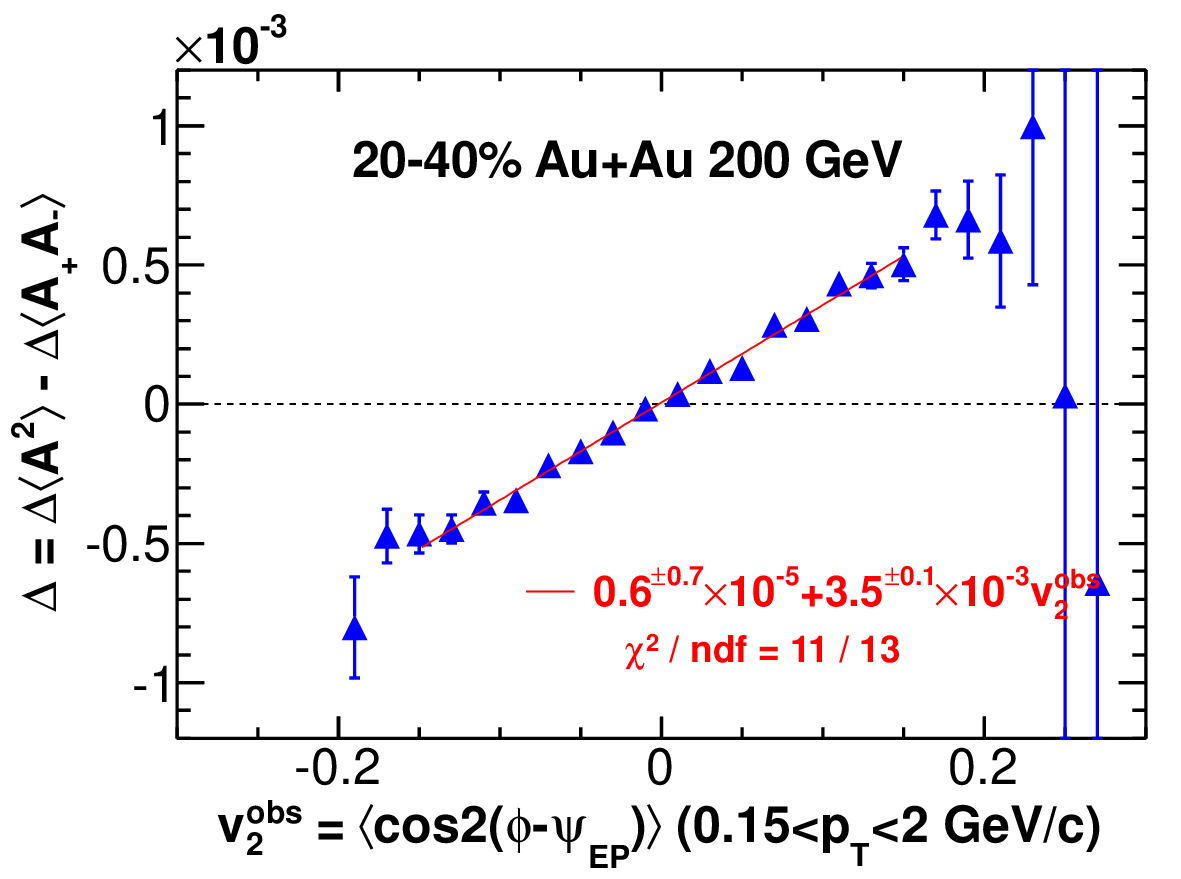}
\includegraphics[width=0.329\textwidth]{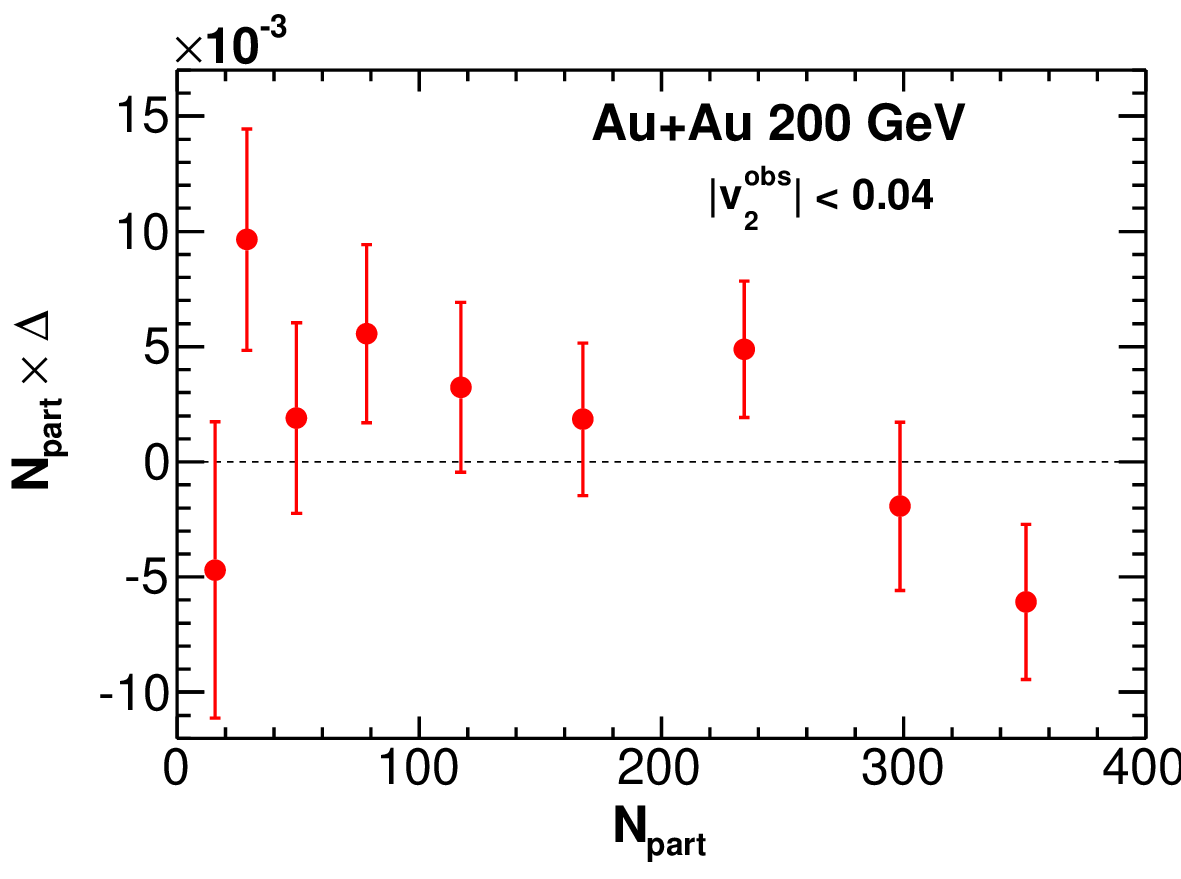}
\end{center}
\caption{(Color online) $\Delta=\Delta\Asq-\Delta\ApAm$ as a function of $\vlow$, the event-by-event elliptical anisotropy of particle distributions relative to the second-harmonic event plane reconstructed from TPC tracks (left panel) and the first harmonic event plane reconstructed from the ZDC-SMD neutron signals (middle panel), in 20-40\% central Au+Au collisions. Right panel: Average $\Delta$ for events with $|\vlow|<0.04$ relative to the TPC event plane as a function of centrality. The error bars are statistical only.}
\label{fig:AdiffvsV2}
\end{figure*}

Figure~\ref{fig:AdiffvsV2} (left panel) shows the difference between same- and opposite-sign correlations, $\Delta=\Delta\Asq-\Delta\ApAm$, as a function of the event-by-event $\vlow$ in 20-40\% central Au+Au collisions. At large positive $\vlow$, $\Delta\Asq>\Delta\ApAm$ is consistent with the \cme. It is possible that at significantly negative $\vlow$, the reconstructed \ep\ may be orthogonal to, rather than aligned with, the real reaction plane so that \ud\ and \lr\ are flipped. As a result, the negative $\Delta$ would really be positive if calculated related to the true reaction plane. This would also be consistent with the \cme. On the other hand, for events with modest negative $\vlow>-0.1$, it is found by the sub-event method that the \ep\ resolution is relatively well defined (see Fig.~\ref{fig:EPresV2} in Appendix~\ref{app:EPres}). However, in the region $-0.1<\vlow\lesssim0$, the values of $\Delta$ are negative. This suggests that the \cme, which should give $\Delta\Asq>\Delta\ApAm$, cannot be entirely responsible for the present observations.

The STAR data from RHIC Run VII have also been analyzed using the first harmonic event plane reconstructed by the ZDC-SMD~\cite{QuanWang,WangG,ChenJY}. The ZDC-SMD event plane resolutions can be found in Appendix~\ref{app:EPres}. The corresponding same- and opposite-sign correlation difference as a function of the event-by-event elliptic anisotropy relative to the ZDC-SMD event plane is shown in Fig.~\ref{fig:AdiffvsV2} (middle panel). This result agrees with that obtained using the second-harmonic TPC event plane.

Several authors~\cite{Wang,Pratt} have suggested that there could be a physics background proportional to the event-by-event $\vlow$ due to the net effect of particle intrinsic correlations and production elliptic anisotropy. These correlations include momentum conservation and local charge conservation~\cite{Pratt}. Specific examples could be decays of flowing resonances or clusters~\cite{Voloshin,Wang}. Thus, the values of $\Delta(\vlow)$ were fit with a straight line versus $\vlow$. The result is 
\begin{widetext}
\be
\Delta(\vlow)=(1.3\pm1.4({\rm stat})^{+4.0}_{-1.0}({\rm syst}))\times10^{-5}+(3.2\pm0.2({\rm stat})^{+0.4}_{-0.3}({\rm syst}))\times10^{-3}\vlow
\ee
for 20-40\% central Au+Au collisions. 
Since momentum conservation effects are the same between same- and opposite-sign charges, the difference between the same- and opposite-sign correlations may be mainly caused by local charge conservation. In this case, the fitted slope parameter would be a measure of the two-particle correlation strength from local charge conservation, scaled by multiplicity $dN_+/d\eta\approx dN_-/d\eta\sim100$~\cite{Levente}. Separate fits to the same- and opposite-charge correlation data in Fig.~\ref{fig:AvsV2}(c) yield 
\bea
\Delta\Asq(\vlow)&=&(16.4\pm1.0({\rm stat})^{+0.9}_{-1.3}({\rm syst}))\times10^{-5}+(2.3\pm0.1({\rm stat})^{+0.4}_{-0.2}({\rm syst}))\times10^{-3}\vlow\,,\\
\Delta\ApAm(\vlow)&=&(15.1\pm1.0({\rm stat})^{+1.0}_{-3.6}({\rm syst}))\times10^{-5}-(0.8\pm0.1({\rm stat})^{+0.4}_{-0.2}({\rm syst}))\times10^{-3}\vlow\,,
\eea
\end{widetext}
respectively.
The same-sign correlation slope parameter may be a measure of the effect of momentum conservation. The weaker dependence of the opposite-sign correlation on event anisotropy may be the net effect of two competing mechanisms of momentum and local charge conservations. 

\begin{table*}[hbt]
\caption{The charge separation parameter $\Delta=\Delta\Asq-\Delta\ApAm$ for all events and for events with the observed second harmonic parameter $|\vlow|<0.04$, and the linear fit intercept and slope to $\Delta(\vlow)$, as a function of centrality ($\Npart$ is the corresponding number of participants). The first error is statistical and the second asymmetric error is systematic.}
\label{tab:Delta}
\begin{ruledtabular}
\begin{tabular}{cc|cccc}
Centrality & $\Npart$ & $\Delta$ & $\Delta(|\vlow|<0.04)$ & intercept & slope \\\hline
80-70\% & 15.7 & $(3.0\pm1.8^{+5.8}_{-2.7})\times10^{-4}$ & $(-3.0\pm4.1^{+9.4}_{-2.4})\times10^{-4}$ & $(0.2\pm2.0^{+7.7}_{-2.1})\times10^{-4}$ & $(1.4\pm0.2^{+0.5}_{-0.5})\times10^{-2}$\\
70-60\% & 28.8 & $(4.6\pm0.8^{+2.1}_{-1.0})\times10^{-4}$ & $(3.3\pm1.7^{+0.4}_{-2.5})\times10^{-4}$ & $(2.4\pm0.9^{+1.1}_{-1.7})\times10^{-4}$ & $(8.6\pm0.9^{+0.7}_{-1.7})\times10^{-3}$\\
60-50\% & 49.3 & $(3.8\pm0.5^{+0.8}_{-1.0})\times10^{-4}$ & $(0.4\pm0.8^{+1.9}_{-2.5})\times10^{-4}$ & $(1.1\pm0.5^{+0.9}_{-1.1})\times10^{-4}$ & $(6.9\pm0.5^{+0.3}_{-1.0})\times10^{-3}$\\
50-40\% & 78.3 & $(2.7\pm0.3^{+0.9}_{-0.7})\times10^{-4}$ & $(0.7\pm0.5^{+1.1}_{-1.6})\times10^{-4}$ & $(0.4\pm0.3^{+1.1}_{-1.0})\times10^{-4}$ & $(5.1\pm0.4^{+0.7}_{-0.6})\times10^{-3}$\\
40-30\% & 117.1 & $(2.0\pm0.2^{+0.6}_{-0.1})\times10^{-4}$ & $(2.8\pm3.1^{+7.2}_{-6.1})\times10^{-5}$ & $(1.5\pm2.2^{+7.2}_{-1.9})\times10^{-5}$ & $(3.8\pm0.3^{+0.1}_{-0.4})\times10^{-3}$\\
30-20\% & 167.6 & $(1.2\pm0.1^{+0.2}_{-0.1})\times10^{-4}$ & $(1.1\pm2.0^{+3.1}_{-3.6})\times10^{-5}$ & $(0.8\pm1.5^{+2.2}_{-1.3})\times10^{-5}$ & $(2.6\pm0.2^{+0.4}_{-0.5})\times10^{-3}$\\
20-10\% & 234.2 & $(5.8\pm0.8^{+0.2}_{-1.2})\times10^{-5}$ & $(2.1\pm1.3^{+1.0}_{-3.2})\times10^{-5}$ & $(1.4\pm1.0^{+0.2}_{-2.4})\times10^{-5}$ & $(1.4\pm0.2^{+0.4}_{-0.2})\times10^{-3}$\\
10-5\% & 298.6 & $(2.6\pm0.9^{+1.6}_{-0.7})\times10^{-5}$ & $(-0.6\pm1.2^{+3.2}_{-0.9})\times10^{-5}$ & $(0.5\pm1.0^{+2.2}_{-0.6})\times10^{-5}$ & $(1.1\pm0.2^{+0.2}_{-0.4})\times10^{-3}$\\
5-0\% & 350.6 & $(-0.3\pm0.8^{+2.2}_{-0.9})\times10^{-5}$ & $(-1.7\pm1.0^{+3.2}_{-0.4})\times10^{-5}$ & $(-1.3\pm0.8^{+2.0}_{-1.2})\times10^{-5}$ & $(8.5\pm2.2^{+2.4}_{-1.0})\times10^{-4}$\\

\end{tabular}
\end{ruledtabular}
\end{table*}


Charge correlations as a function of elliptic anisotropy have also been studied in A Multi-Phase Transport (AMPT) model~\cite{GLMa}, motivated by phenomenological studies~\cite{Wang,Koch,Pratt} and the preliminary version of the data reported here. The AMPT model is a useful tool for this study because it can mostly reproduce elliptic flow data, it includes decays of resonances possessing anisotropic flow, and it should contain correlations caused by momentum and local charge conservations. The AMPT results indeed show a linear dependence between the same-sign charge correlations and the elliptic anisotropy~\cite{GLMa}, qualitatively consistent with data. It is possible that the linear dependence observed in AMPT is due to the net effect of elliptic anisotropy and a difference between the same- and opposite-sign charged particle correlations~\cite{Wang,Pratt}.

If the linear dependence of the data is entirely due to background correlations of the type suggested in Ref.~\cite{Wang,Pratt} and the \cme\ does not contribute appreciably to the final state $\vlow$, then the intercept is the most sensitive to possible CME effects, and the slope is a measure of the background correlation strength.
In this case, multiplicity asymmetries of particles in phase space which are less elliptically distributed (near $\vlow=0$)~\footnote{Note that in events with zero ellipticity there can be other harmonic shapes, such as triangularity. In other words, events with zero ellipticity are not necessarily spherical.}, with respect to an event plane reconstructed elsewhere in phase space, will be more sensitive to possible \cme\ effects. Shown in Fig.~\ref{fig:AdiffvsV2} (right panel) is the charge separation $\Delta=\Asq-\Delta\ApAm$ in events with $|\vlow|<0.04$ as a function of the centrality. The values of $\Delta$ are multiplied by $\Npart$ for clarity. The charge separation seems to be consistent with zero within the present statistical precision, suggesting no substantial charge separation in those events with approximately zero ellipticity of the measured particles. For the mid-central 20-40\% events, it was found that 
\be
\Delta(|\vlow|<0.04)=(1.9\pm1.9({\rm stat})^{+3.6}_{-3.8}({\rm syst}))\times10^{-5}\,.
\ee
For comparison, the charge separation averaged over all events is 
\be
\Delta=(1.6\pm0.1({\rm stat}))\times10^{-4}\,.
\ee

Table~\ref{tab:Delta} lists the Run IV TPC data of the charge separation parameter $\Delta$, the charge separation parameter within the range of $|\vlow|<0.04$, and the linear fit intercept and slope as a function of centrality. The first quoted error is statistical and the second quoted error is systematic. The details of the systematic uncertainty study are given in Sec.~\ref{sec:syst}.

\begin{figure*}[hbt]
\begin{center}
\includegraphics[width=0.329\textwidth]{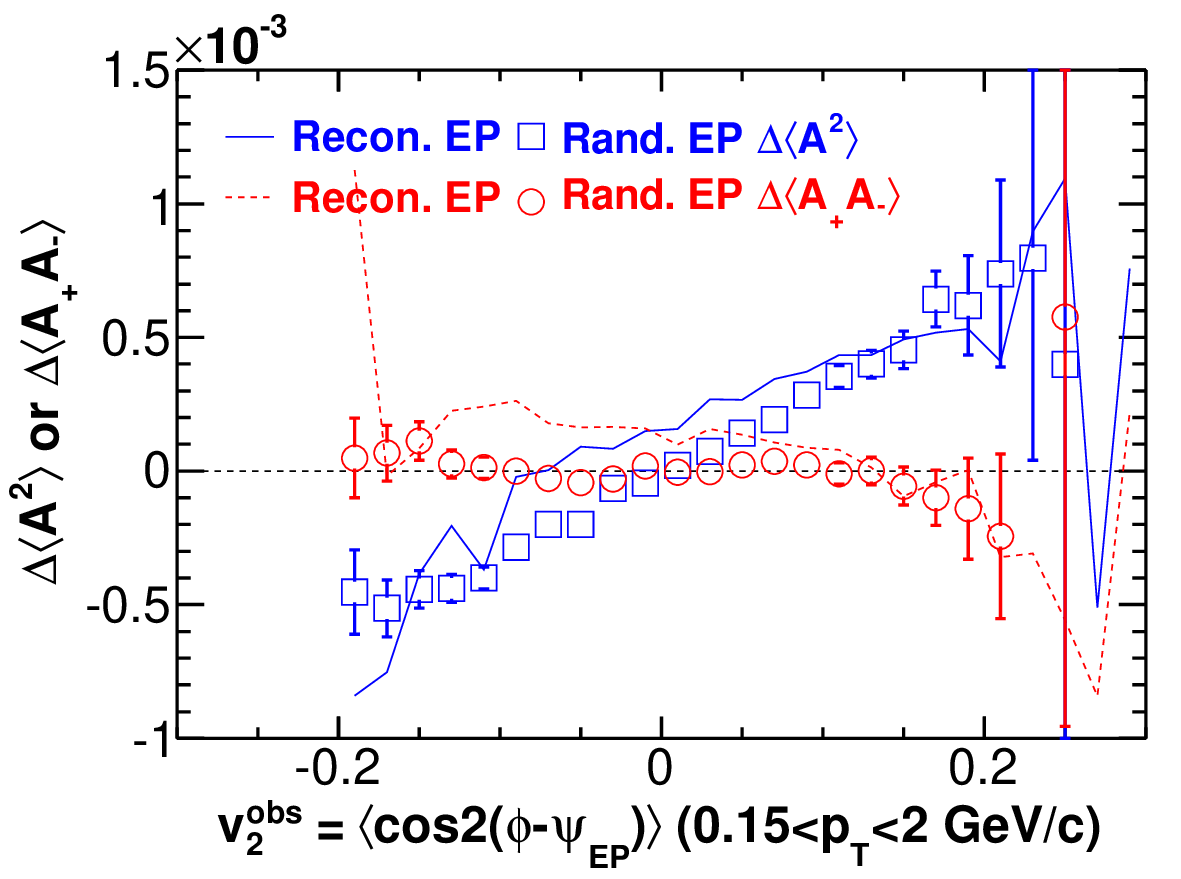}
\includegraphics[width=0.329\textwidth]{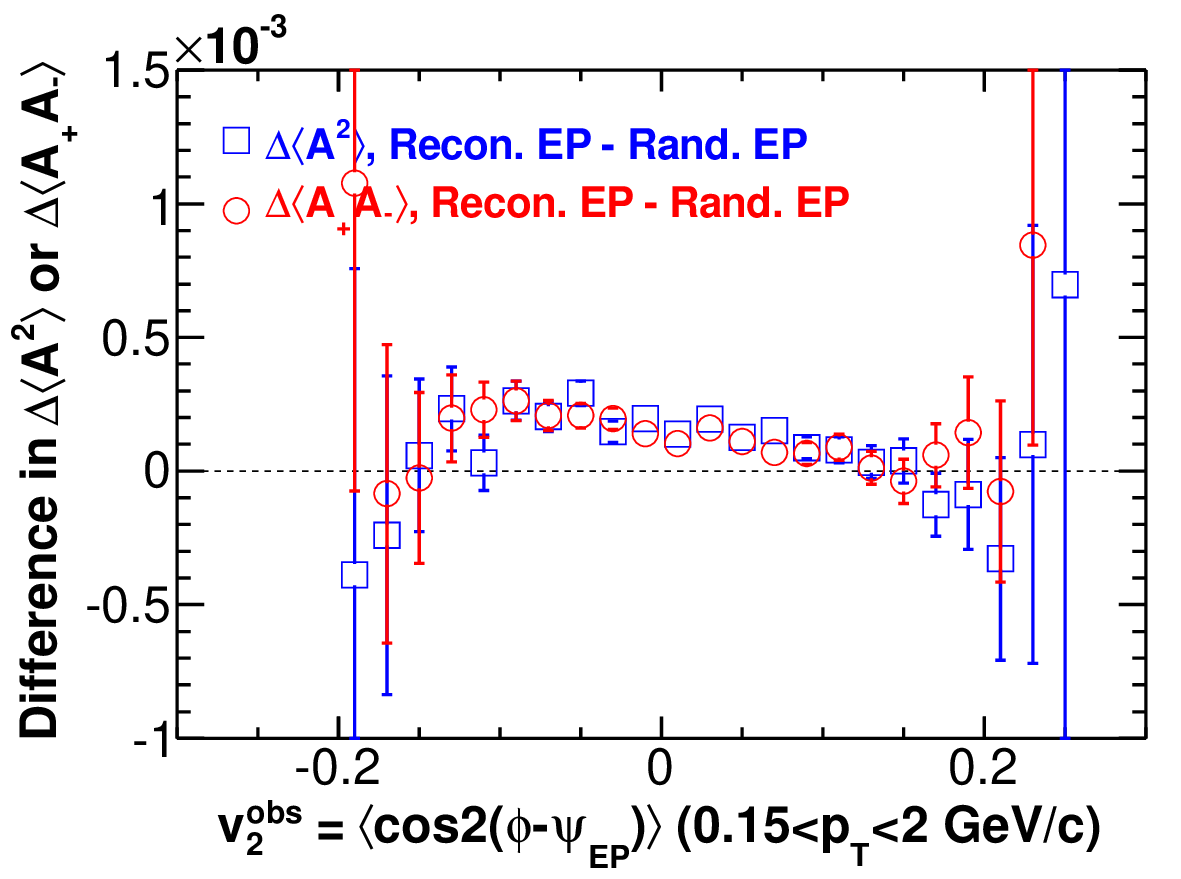}
\includegraphics[width=0.329\textwidth]{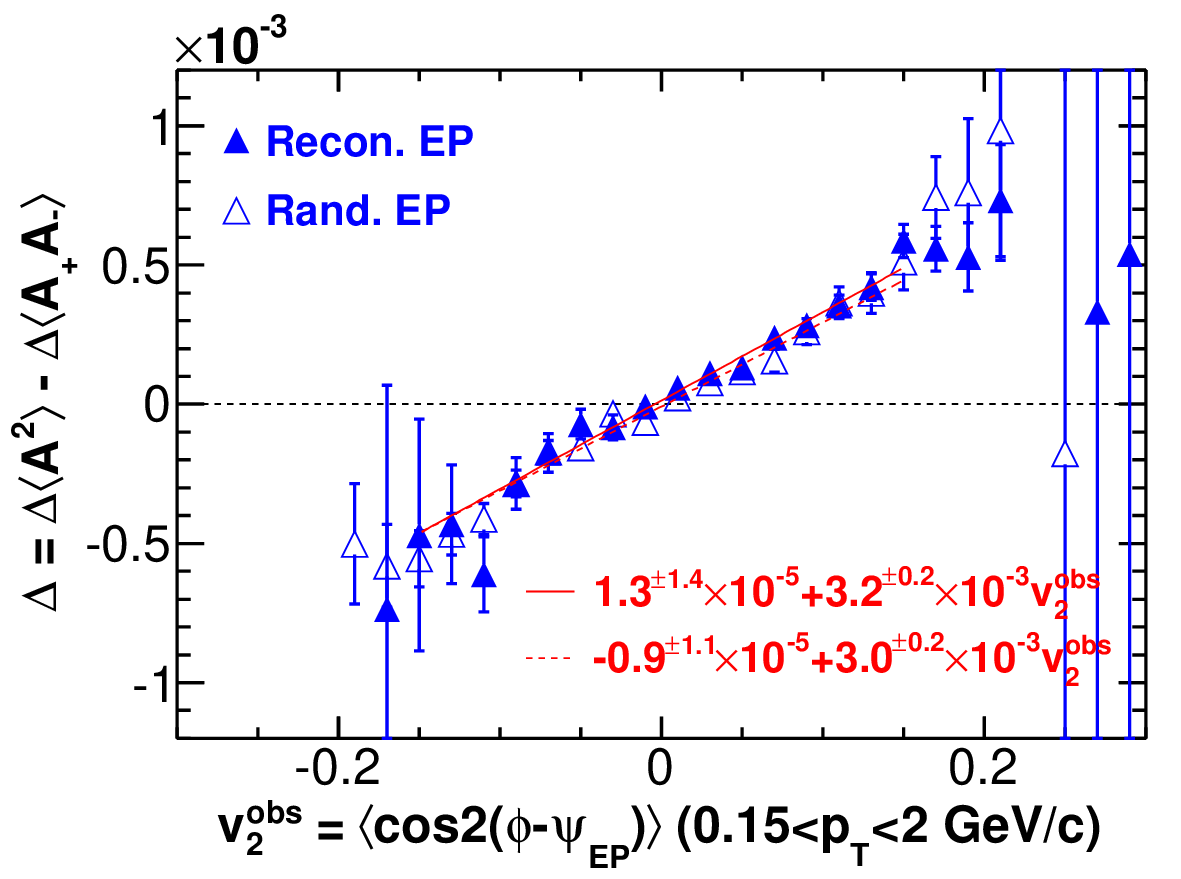}
\end{center}
\caption{(Color online) Left panel: $\Delta\Asq$ and $\Delta\ApAm$ as a function of $\vlow$ with a random \ep\ in 20-40\% Au+Au collisions. The lines depict the results with reconstructed \ep\ from Fig.~\ref{fig:AvsV2} lower right panel for comparison. Middle panel: The differences in $\Delta\Asq$ and $\Delta\ApAm$ between the reconstructed \ep\ and random \ep\ results, respectively. Right panel: $\Delta=\Delta\Asq-\Delta\ApAm$ as a function of $\vlow$ with a random \ep\ (open triangles). The solid triangles are results with reconstructed \ep\ from Fig.~\ref{fig:AdiffvsV2} left panel for comparison. 
Error bars are statistical.}
\label{fig:AvsV2rn}
\end{figure*}

To gain more insight, the charge asymmetry variances and covariances using \udlr\ differences with respect to a {\em random} azimuthal plane as a function of the event elliptic anisotropy relative to the randomly chosen plane were also studied. The dependences of \udlr\ versus $\vlow$ are shown in the left panel of Fig.~\ref{fig:AvsV2rn} and are qualitatively similar to those with respect to the reconstructed \ep. The $\Delta\Asq$ and $\Delta\ApAm$ from the random plane cross at $\vlow\approx0$ with approximately zero intercept, while those from the recontructed \ep\ cross at a positive intercept as aforementioned. The differences between the reconstructed \ep\ and random plane results are shown in the middle panel of Fig.~\ref{fig:AvsV2rn}. The differences are sensitive to \ep-dependent correlations, and appear to be independent of charge signs. The difference between the same- and opposite-sign charge \udlr\ fluctuations, shown by the open triangles in the right panel of Fig.~\ref{fig:AvsV2rn}, are also linearly dependent on the observed $\vlow$ relative to the random plane. The dashed line is a linear fit to the open data points. For comparison the data with the reconstructed \ep\ and the corresponding linear fit from the left panel of Fig.~\ref{fig:AdiffvsV2} are superimposed as the solid points and solid line, respectively. The linear dependences of the charge separation $\Delta(\vlow)$ are equal between the reconstructed \ep\ and the random plane within the statistical uncertainties. This confirms that the observed charge separation is indeed correlated with the observed final-state event shape. 

It is worthwhile to note that, with respect to the random plane, the average $\mean{\vlow}$ is zero (the event probability distribution is symmetric about $\vlow=0$), and the event-integrated charge separation is zero as expected. With respect to the reconstructed \ep, the center-of-gravity of the event probability distribution is no longer at zero but at positive $\mean{\vlow}$ (i.e.~the observed average elliptic anisotropy) and the event-integrated charge separation is finite and positive as reported in the present work.

Further comparisons of oriented and random event plane results will be interesting. Additional studies to disentangle effects from the final-state event shape and the magnetic field have been proposed by using collisions of uranium nuclei~\cite{Voloshin_UU} which have a large intrinsic deformation. These studies will provide complementary experimental data for understanding the origin of the observed charge separation. 

\section{Summary\label{sec:summary}}

Correlations of positive ($\Ap$) and negative ($\Am$) charge multiplicity asymmetries with respect to the event plane and the plane perpendicular to the event plane have been measured. The asymmetries are measured in one half of the STAR TPC (one unit in pseudo-rapidity) while the event plane is reconstructed in the other half of the TPC. The dynamical variance $\delta\Apmsq$ and covariance $\daa\ApAm$ are obtained by subtracting the contributions from the statistical fluctuations and detector effects, and were presented in $d$+Au and Au+Au collisions.

The charge asymmetry dynamical variances, $\delta\AsqUD$ and $\delta\AsqLR$, are positive in $d$+Au and peripheral Au+Au collisions and become negative in medium-central to central collisions ({\it cf.} Fig.~\ref{fig:corr}). The positive dynamical variances in $d$+Au and peripheral Au+Au collisions indicate that the dynamical physics processes broaden the charge asymmetry distributions. The same-sign pairs are preferentially emitted in the same direction. The negative dynamical variances in medium-central to central collisions indicate that the dynamical physics processes narrow the charge asymmetry distributions. The same-sign pairs are now preferentially emitted back-to-back. The charge asymmetry covariances, $\daa\AAUD$ and $\daa\AALR$, are positive and large, indicating that significant correlations are present between positive and negative particles in the events ({\it cf.} Fig.~\ref{fig:corr}). The positive and large covariances indicate that the opposite-sign pairs are preferentially emitted in the same direction, and the aligned emission is largely independent of the reaction plane.


The charge asymmetry dynamical variances, $\delta\AsqUD$ and $\delta\AsqLR$, decrease with increasing $\pt$ ({\it cf.} Fig.~\ref{fig:pt}). They are positive at low $\pt$, indicating that the low-$\pt$ same-sign pairs are preferentially emitted in the same direction. They are preferentially back-to-back at large $\pt$ in central collisions. The charge asymmetry covariances are approximately independent of $\pt$ at low $\pt$, and rapidly increase with $\pt$. The aligned same-side emission of opposite-charge pairs is significantly stronger at large $\pt$.

The dynamical charge asymmetry variance, $\delta\AsqUD$, is larger than $\delta\AsqLR$. The charge asymmetry covariance, $\daa\AAUD$, is larger than $\daa\AALR$ in non-peripheral Au+Au collisions. The differences $\Delta\Asq=\delta\AsqUD-\delta\AsqLR$ and $\Delta\ApAm=\daa\AAUD-\daa\AALR$ increase with centrality in peripheral collisions, reaching a maximum in medium central collisions, and then decreases towards more central collisions ({\it cf.} Fig.~\ref{fig:diff}). 
The differences, $\Delta\Asq$ and $\Delta\ApAm$, in a given centrality bin increase with $\pt$ ({\it cf.} Fig.~\ref{fig:diff_pt}). The centrality and $\pt$ dependences are qualitatively similar between the same- and opposite-sign \udlr\ correlations. 

To gain more insight into the contributing physics mechanisms, two aspects of the charge asymmetry correlations were investigated. First, the charge asymmetry correlations as a function of the event-by-event elliptic anisotropy of the particles in the asymmetry measurements, $\vlow$, was studied ({\it cf.} Fig.~\ref{fig:AvsV2}). The same-sign \udlr\ correlation, $\Delta\Asq$, increases with increasing $\vlow$, and the opposite-sign \udlr\ correlation $\Delta\ApAm$ decreases slightly. They cross at the same positive value at $\vlow\approx0$. Also studied was the charge asymmetry correlations as a function of the size of the azimuthal region for the asymmetry measurements ({\it cf.} Fig.~\ref{fig:size}). It was found that the \udlr\ differences of both the variance and covariance, $\Delta\ASq$ and $\Delta\APAM$, increase with decreasing wedge size. However, the difference between the two diminishes with decreasing wedge size.



The present measurements were motivated by the Chiral Magnetic Effect, which could yield a charge separation of final state hadrons across the reaction plane. Previous STAR measurements of the three-particle correlators were consistent with charge separation by the \cme\ together with medium interactions. However, the possible physics backgrounds for the measured correlator results were not fully explored. Similarly, the \cme\ charge separation would yield a positive same-sign correlation $\Delta\Asq$ and a negative, or zero in the case of medium interactions, opposite-sign correlation $\Delta\ApAm$. The present measurements show $\Delta=\Delta\Asq-\Delta\ApAm>0$, an indication of charge separation, although both $\Delta\Asq$ and $\Delta\ApAm$ are positive. For the mid-central 20-40\% events, it was observed that $\Delta=(1.6\pm0.1({\rm stat}))\times10^{-4}$.

It is possible that the physics backgrounds for same-sign and opposite-sign correlations are equal and fall in-between the present \udlr\ results. In such a case, only the difference $\Delta=\Delta\Asq-\Delta\ApAm$ is sensitive to the \cme\ and their average behavior is due to other physical mechanisms. The measured values of $\Delta$ were largest when whole hemispheres were used in the asymmetry measurements and diminishes when smaller wedge sizes were used ({\it cf.} Fig.~\ref{fig:size2}). This suggests that the charge separation across the event plane is mainly in-plane and same-sign pairs are roughly back-to-back in the upper or lower hemisphere.

It is also possible and more likely that the physics backgrounds are different for the same- and opposite-sign \udlr\ correlations. In fact, the charge separation, $\Delta=\Delta\Asq-\Delta\ApAm$, is roughly proportional to the average elliptic anisotropy in each centrality bin ({\it cf.} Fig.~\ref{fig:AdiffvsV2ave}). It is further found that the values of $\Delta$ are approximately proportional to the event-by-event elliptic anisotropy of the particles measured in the charge multiplicity asymmetries: 
$\Delta(\vlow)=(1.3\pm1.4({\rm stat})^{+4.0}_{-1.0}({\rm syst}))\times10^{-5}+(3.2\pm0.2({\rm stat})^{+0.4}_{-0.3}({\rm syst}))\times10^{-3}\vlow$. 
This proportionality was also observed using the first-order harmonic event plane reconstructed using the ZDC ({\it cf.} Fig.~\ref{fig:AdiffvsV2}). This suggests that the physics backgrounds may be the net effect of particle production elliptic anisotropy and a difference in particle intrinsic correlations between same- and opposite-sign charge pairs. These intrinsic particle correlations include momentum conservation and local charge conservation, and the local charge conservation presents a difference between same- and opposite-sign charge pairs. 
For the particular case of events of nearly zero particle ellipticity ($|\vlow|<0.04$) where such a background may be absent, the charge separation effect was observed to be 
$\Delta(|\vlow|<0.04)=(1.9\pm1.9({\rm stat})^{+3.6}_{-3.8}({\rm syst}))\times10^{-5}$ for the 20-40\% centrality. 
Thus, in an event-by-event analysis, a linearly decreasing amount of charge separation is observed as a function of event ellipticity; and the trend has an intercept consistent with zero. These data serve as an interesting benchmark in the phenomenology of the Chiral Magnetic Effect and will hopefully stimulate further developments in both theory and experiment.

\section*{Acknowledgments}
We thank the RHIC Operations Group and RCF at BNL, the NERSC Center at LBNL and the Open Science Grid consortium for providing resources and support. This work was supported in part by the Offices of NP and HEP within the U.S.~DOE Office of Science, the U.S.~NSF, the Sloan Foundation, CNRS/IN2P3, FAPESP CNPq of Brazil, Ministry of Ed.~and Sci.~of the Russian Federation, NNSFC, CAS, MoST, and MoE of China, GA and MSMT of the Czech Republic, FOM and NWO of the Netherlands, DAE, DST, and CSIR of India, Polish Ministry of Sci.~and Higher Ed., National Research Foundation (NRF-2012004024), Ministry of Sci., Ed.~and Sports of the Rep.~of Croatia, and RosAtom of Russia.

\appendix

\section{Connections to Correlators\label{app:comp}}

The differences $\ApmsqUD-\ApmsqLR$ and $\AAUD-\AALR$ are related to the previously reported three-particle azimuth correlators ~\cite{CorrelatorPRL,CorrelatorPRC}, but with important differences. By using a Fourier series of a step-function in $\phi-\psiEP$ (the particle azimuth relative to the event plane), the present charge asymmetry observables can be expressed as
\bea
\ApmUD&=&\frac{4}{\pi\Npm}\sum_{i=1}^{\Npm}\sum_{n=0}^{\infty}\frac{\sin(2n+1)(\phi_{\pm,i}-\psiEP)}{2n+1}\,,\nonumber\\
\ApmLR&=&\frac{4}{\pi\Npm}\sum_{i=1}^{\Npm}\sum_{n=0}^{\infty}\frac{\cos(2n+1)(\phi_{\pm,i}-\psiEP)}{2n+1}\,.
\eea
The difference in asymmetry correlations is
\begin{widetext}
\be
\AaAbLR-\AaAbUD=\left(\frac{4}{\pi}\right)^2\left\langle\frac{1}{N_{\alpha}N_{\beta}}\sum_{i,j=0}^{N_{\alpha},N_{\beta}}\sum_{n.m=0}^{\infty}\frac{\cos[(2n+1)(\phi_{\alpha,i}-\psiEP)+(2m+1)(\phi_{\beta,j}-\psiEP)]}{(2n+1)(2m+1)}\right\rangle\,,
\label{eq:expansion}
\ee
where $\alpha$ and $\beta$ stand for `$+$' or `$-$' particles. This is similar in form to the azimuth correlator observable $\mcosijk\approx\mcosabcv$ of Ref.~\cite{CorrelatorPRL,CorrelatorPRC}:
\be
\mcosijk=\left\langle\frac{1}{N_{\alpha}(N_{\beta}-\delta_{\alpha\beta})}\sum_{i,j=0,i\neq j}^{N_{\alpha},N_{\beta}}\cosijk\right\rangle\,,
\label{eq:correlator}
\ee
\end{widetext}
where $\psiRP$ is the reaction plane angle and $\delta_{\alpha\beta}$ is the Kronecker delta. The asymmetry correlation differences contain all possible (including mixed) harmonic terms, while the correlator observables contain only one of the infinite number of terms. While the correlators are in terms of azimuthal angle relative to $\psiRP$, the asymmetry correlation observables are in terms of azimuthal angle relative to $\psiEP$, and are therefore affected by \ep\ resolution.

\begin{figure}[hbt]
\begin{center}
\includegraphics[width=0.4\textwidth]{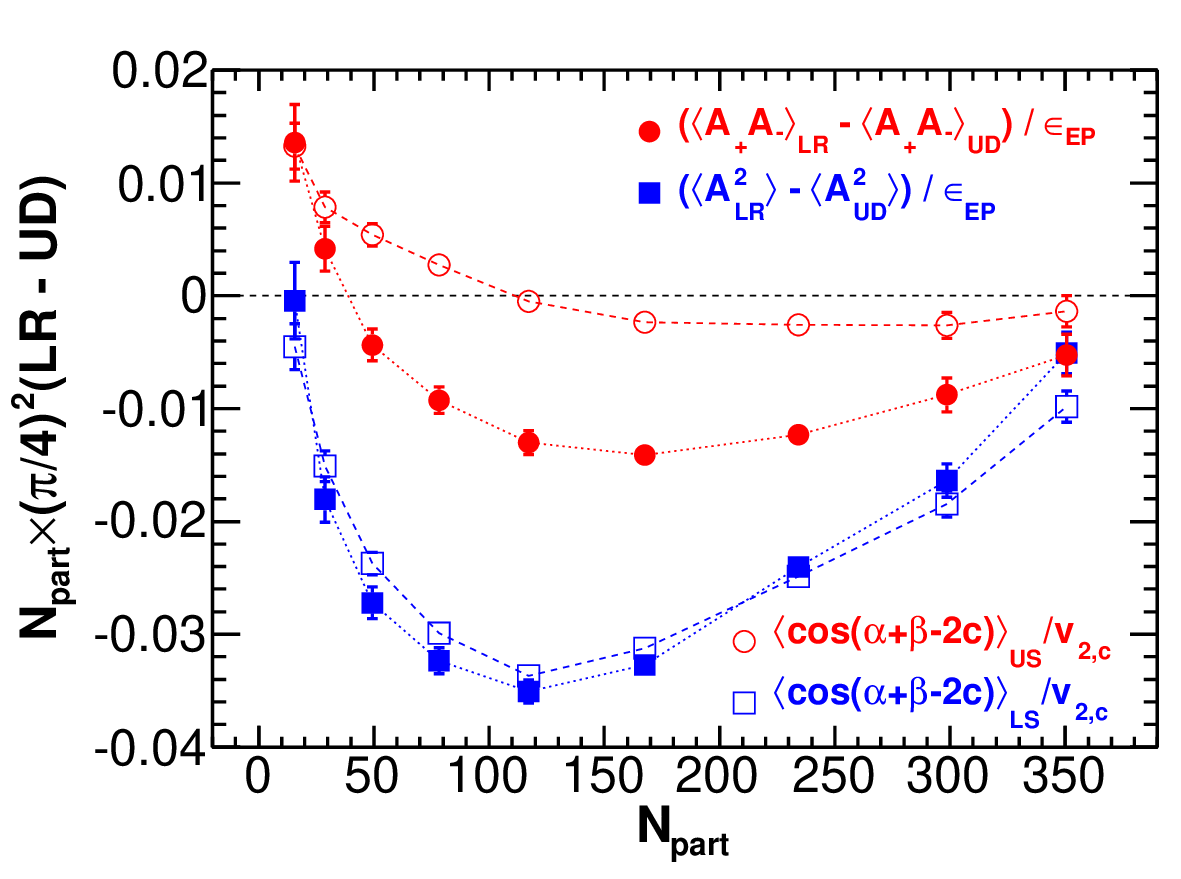}
\includegraphics[width=0.4\textwidth]{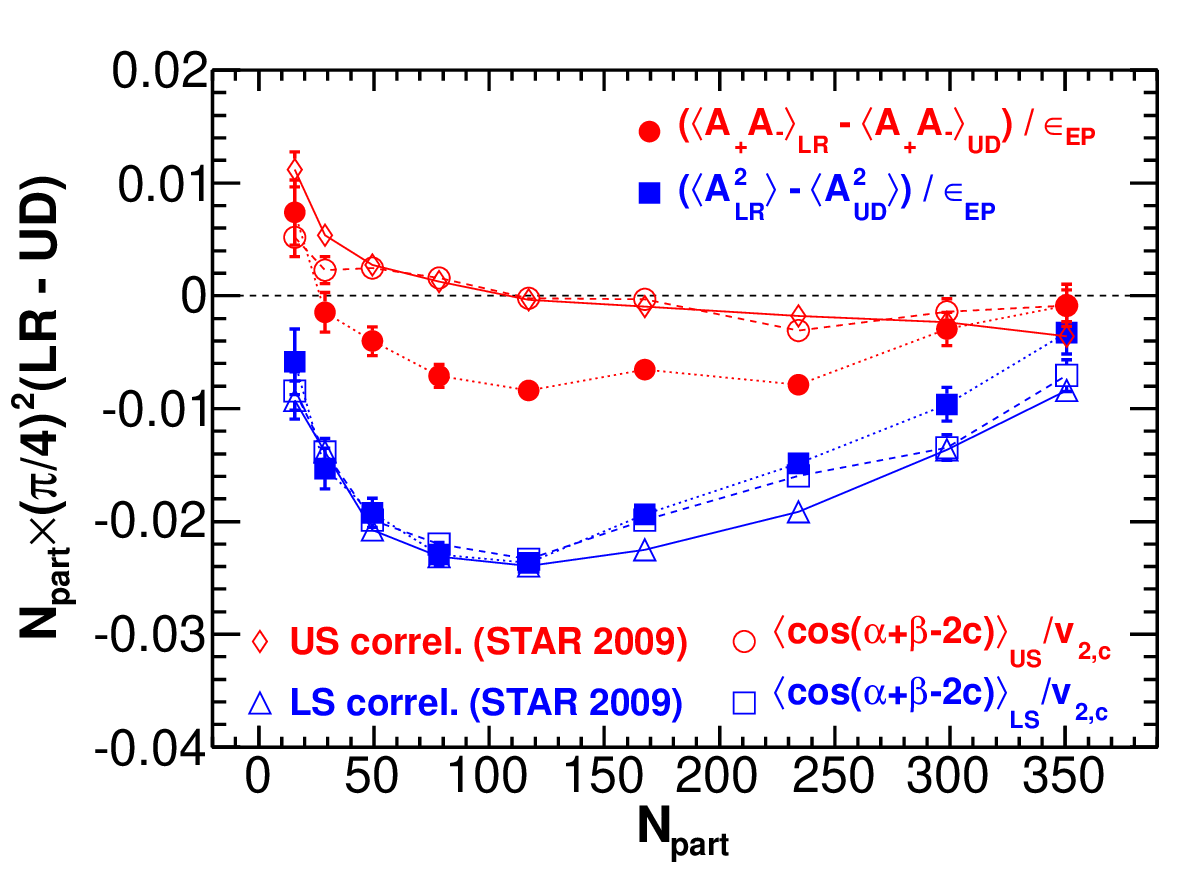}
\end{center}
\caption{(Color online) The correlation differences, $\AsqLR-\AsqUD$ and $\AALR-\AAUD$, scaled by the number of participants $(\pi/4)^2\Npart$. 
Also shown are $\mcosabcv\approx\mcosijk$ of same- and opposite-sign particle pairs ($\alpha$ and $\beta$) calculated from particles used for the charge asymmetry correlations with particle $c$ from those used for \ep\ construction. The upper panel shows the default results where particles are divided according to $\eta$; the lower panel shows results with particles divided randomly into two halves and compared to published correlators (STAR 2009~\cite{CorrelatorPRL,CorrelatorPRC}). The error bars are statistical only.}
\label{fig:comp}
\end{figure}

The present charge asymmetry correlations and the azimuthal correlators are related but differ as shown in Eq.~(\ref{eq:expansion}) and Eq.~(\ref{eq:correlator}). To gain more insight into the relationships and differences, the upper panel of Fig.~\ref{fig:comp} shows the comparison between the asymmetry correlation differences (\lr\ $-$ \ud) and the azimuthal correlator $\mcosijk$. The charge asymmetry correlation differences \lr\ $-$ \ud\ in Fig.~\ref{fig:comp} have already been divided by the event-plane resolution.
Those correlators are calculated using the $\alpha$ and $\beta$ particles from the same $\eta$ region as were used for the charge asymmetry measurement with particle $c$ from the other $\eta$ region used for \ep\ construction. The correlator results shown in the upper panels of Fig~\ref{fig:comp} are not identical to those published previously~\cite{CorrelatorPRL,CorrelatorPRC} where the three particles $\alpha$, $\beta$, and $c$ are from the entire TPC acceptance of $-1<\eta<1$. The present correlator measurements, which are from a smaller range in $\eta$, hence a smaller range in $\Delta\eta$, are larger than those in~\cite{CorrelatorPRL,CorrelatorPRC} because the measured correlators decrease with increasing range of $\Delta\eta$~\cite{CorrelatorPRL,CorrelatorPRC}.

To compare directly to the azimuthal correlators in Refs. \cite{CorrelatorPRL,CorrelatorPRC}, the present charge asymmetries were formed using the entire TPC acceptance. To avoid the self-correlation, the event was divided randomly into two sub-events. One sub-event was used to reconstruct the event plane and the other was used to calculate the charge asymmetries. The results are shown in the lower panel of Fig.~\ref{fig:comp}, and are compared to the azimuthal correlators from~\cite{CorrelatorPRL,CorrelatorPRC}.  The azimuthal correlators have also been calculated using two particles, $a$ and $b$, from the sub-event used for the asymmetry measurements and the third particle $c$ from the other sub-event used for \ep\ reconstruction. The azimuthal correlators (shown as open points in Fig.~\ref{fig:comp}) are consistent with those of Ref.~\cite{CorrelatorPRL,CorrelatorPRC}. 

As shown in Eqs.~(\ref{eq:expansion}) and (\ref{eq:correlator}), the correlators contain an infinite number of harmonic terms in the asymmetry correlation difference observables. The present values of $\AsqLR-\AsqUD$ are comparable to the same-sign correlator $\mcosijk$. This suggests that the high-order harmonic terms in $\Delta\Asq$ may be small. This in turn suggests that the event-plane resolution correction by linear extrapolation may be sufficiently accurate for the measurements of $\AsqLR-\AsqUD$. 
On the other hand, the present values of $\AALR-\AAUD$ is significantly different from the opposite-sign correlator. It is important to note that the difference $\AALR-\AAUD$ is significantly negative, while the correlator magnitude is close to zero. This suggests that the high-order terms in Eq.~(\ref{eq:expansion}) are important. Because of these terms, the event-plane resolution correction via the linear extrapolation applied on $\AALR-\AAUD$ in Fig.~\ref{fig:comp} is likely invalid. However, the imperfect event-plane resolution can only reduce the magnitude of $\AAUD-\AALR$ that is shown in Fig.~\ref{fig:diff}. Thus, the true $\AALR-\AAUD$ with respect to the real reaction plane may be even more negative than that shown.

\section{Analysis Details\label{app:ana}}

\subsection{Detector Efficiency\label{app:eff}}

\begin{figure}[hbt]
\begin{center}
\includegraphics[width=0.4\textwidth]{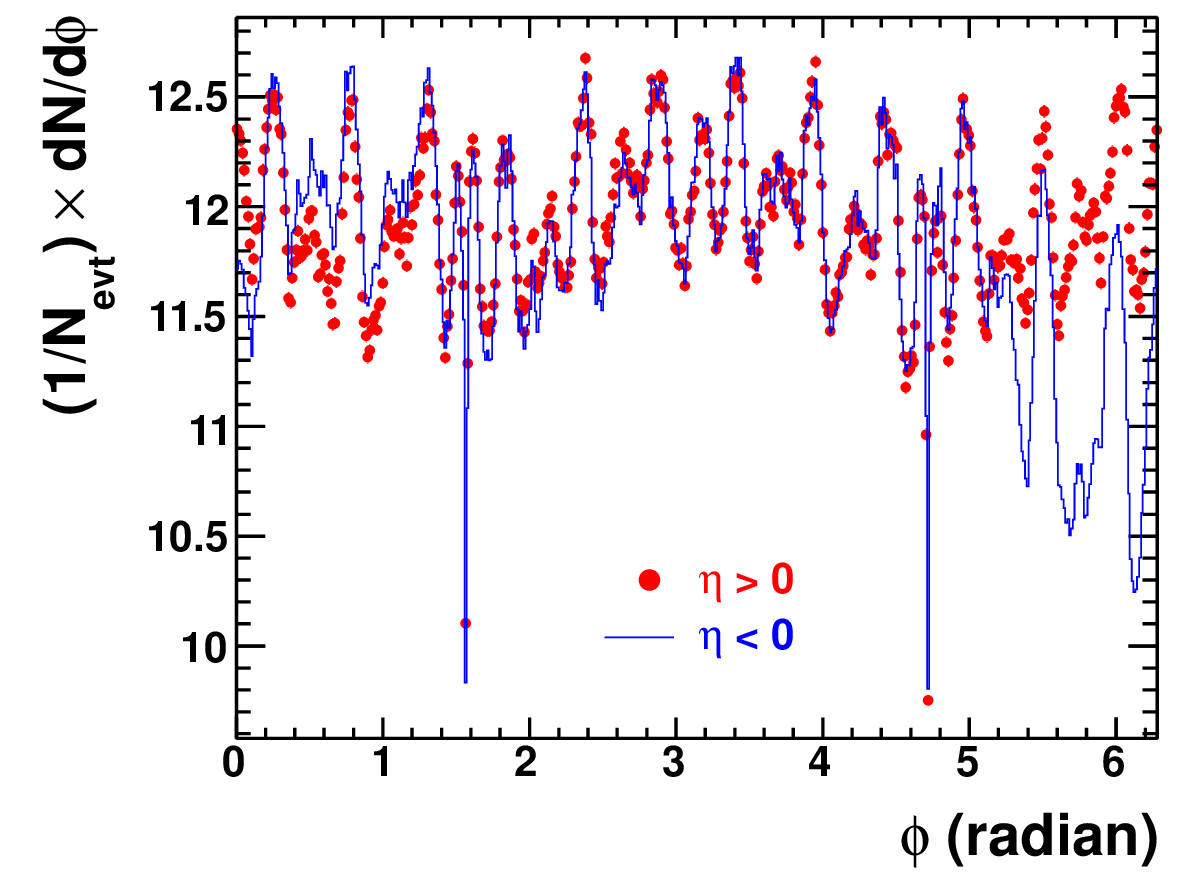}
\includegraphics[width=0.4\textwidth]{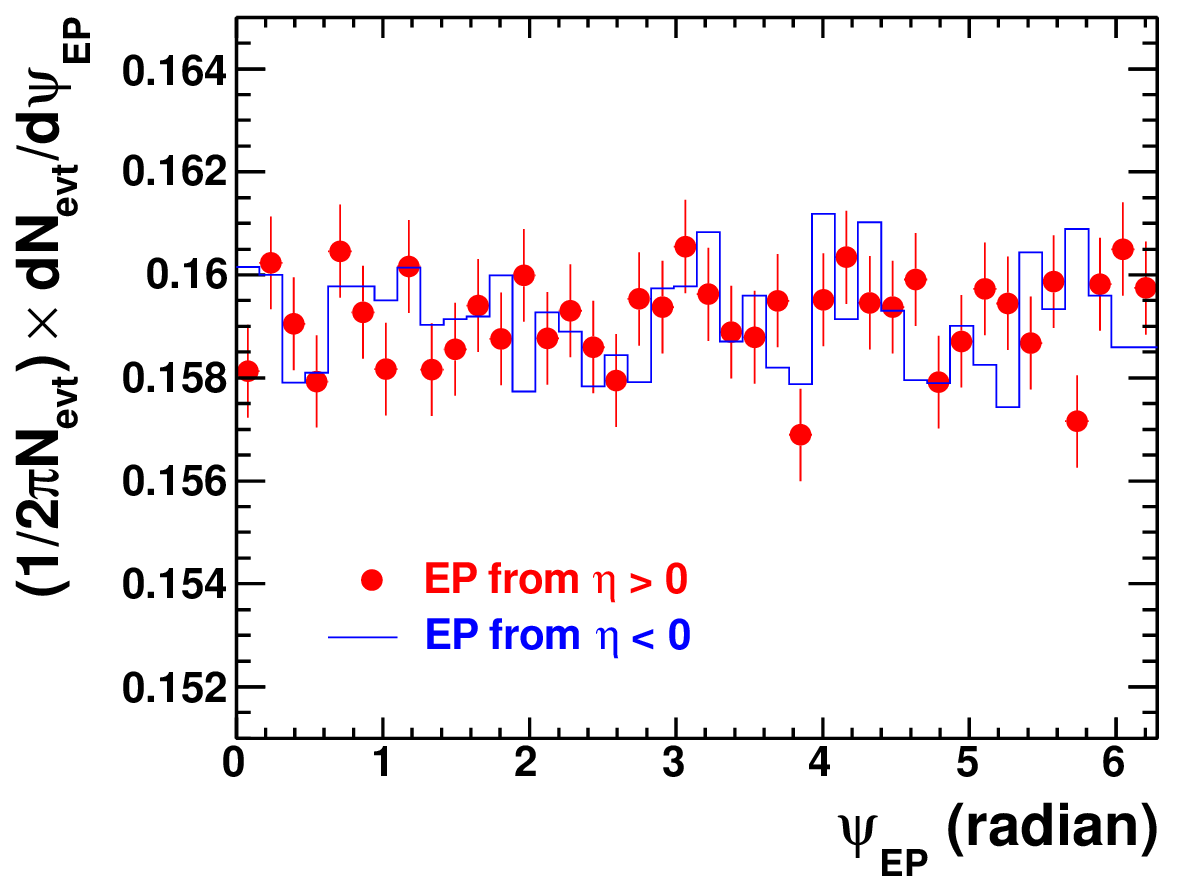}
\end{center}
\caption{(Color online) Upper panel: Positively charged particle multiplicity distributions versus the azimuthal angle, $\phi$, in 30-40\% central Au+Au collisions. The data are shown separately for $\eta>0$ and $\eta<0$. The magnetic polarities of the STAR magnet have been summed and the $\pt$ is integrated over the range $0.15<\pt<2$~GeV/$c$. The inverse of these distributions, properly normalized, were used to correct for the track efficiency versus $\phi$. Lower panel: Reconstructed event-plane azimuthal angle distributions in 30-40\% central Au+Au collisions. Particles within $0.15 < \pt < 2$~GeV/$c$ from $\eta<0$ and $\eta>0$ were used separately to reconstruct the \ep. The error bars are statistical only.}
\label{fig:tech}
\end{figure}

Figure~\ref{fig:tech} (upper panel) shows examples of the $\phi$-distributions of the positively charged particle multiplicity within the $\eta>0$ and $\eta<0$ regions separately for 30-40\% central Au+Au collisions. The negative particle results are similar. The results from other centralities are also similar. In Fig.~\ref{fig:tech}, the magnetic field polarities have been summed and the particles are integrated over the range $0.15<\pt<2$~GeV/$c$. The regular-pattern of the TPC sector boundaries is clearly seen for the both positive and negative $\eta$ particles. The $\eta<0$ particles have an additional inefficiency in the region of $11\pi/6<\phi<2\pi$, due to an inefficiency in the electronics for two of the sectors in the east side of the TPC. This inefficiency persisted over the whole period over which the present data was collected but showed no significant variation over time within this period.

The single particle $\phi$-dependent inefficiencies, primarily due to the TPC sector boundaries, were corrected for in reconstructing the \ep\ and in calculating the multiplicity asymmetries. Figure~\ref{fig:tech} (lower panel) shows the event-plane $\psiEP$ distributions reconstructed separately from the $\eta>0$ and $\eta<0$ particles in 30-40\% central Au+Au collisions. The distributions are uniform.  Fitting the $\eta>0$ and $\eta<0$ event-plane distributions with a constant value resulted in values of $\chi^2/$ndf of 39/39 and 48/39, respectively.

The efficiency corrections were obtained in the following way. The multiplicity distributions in $\phi$, such as the ones in Fig.~\ref{fig:tech} (upper panel), were normalized to average unity. These normalized distributions are referred to as the acceptance $\times$ efficiency, $\epsilon(\phi)$. These $\epsilon(\phi)$ distributions are separated according to the magnetic field polarities, particle charge signs, and centrality bins. They are further separated for positive and negative $\eta$ (approximately corresponding to the two halves of the TPC depending on the primary vertex position), and for the following $\pt$ bins: 0.15-0.5, 0.5-1.0, 1.0-1.5, and 1.5-2.0~GeV/$c$. The detector acceptance $\times$ inefficiency correction factor is taken as $1/\epsilon(\phi)$. 
The correction factor $1/\epsilon(\phi)$ here can be larger or smaller than unity.

The overall single particle tracking efficiency is not corrected for during the event-plane construction. The present results were checked when using an event-plane reconstruction for which the particle multiplicities were corrected by the centrality and $\pt$-dependent tracking efficiency. This did not affect the present results.

\subsection{Self-Correlation\label{app:self}}

\begin{figure}[hbt]
\begin{center}
\includegraphics[width=0.4\textwidth]{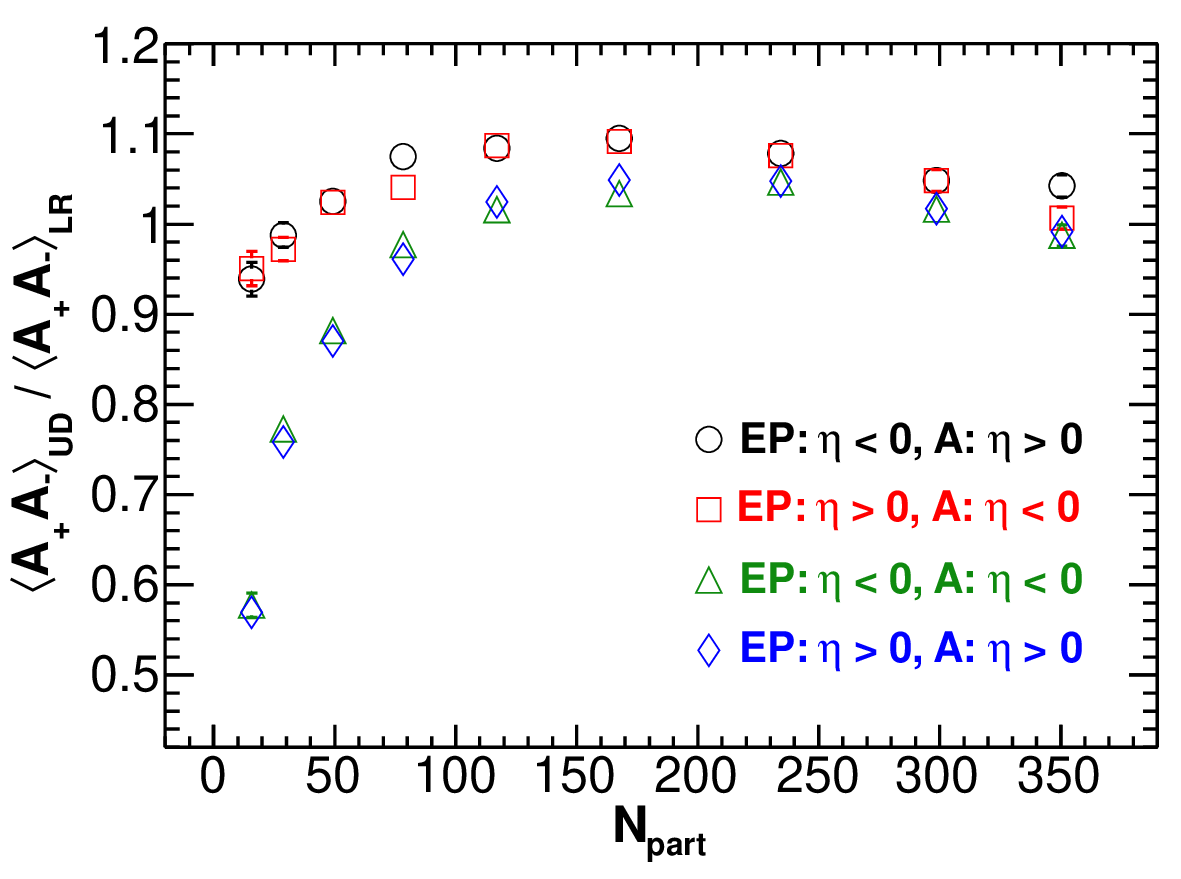}
\end{center}
\caption{(Color online) The relative charge asymmetry correlations, $\AAUD/\AALR$, as a function of the number of participants, $\Npart$, for four combinations of $\eta$ ranges used for \ep\ reconstruction and asymmetry calculation. 
The error bars are statistical only.}
\label{fig:autoCorrel}
\end{figure}

Four asymmetries were calculated separately, (i) using particles within $0<\eta<1$ with the \ep\ reconstructed from $-1<\eta<0$, (ii) using particles within $-1<\eta<0$ with the \ep\ reconstructed from $0<\eta<1$, (iii) using particles within $-1<\eta<0$ with the \ep\ reconstructed from $-1<\eta<0$, and (iv) using particles within $0<\eta<1$ with the \ep\ reconstructed from $0<\eta<1$. The \ud\ charge asymmetry covariance relative to that of \lr, $\AAUD/\AALR$, is shown in Fig.~\ref{fig:autoCorrel} for all four cases. Differences are observed between those using the same set of particles for the \ep\ construction and asymmetry calculation (cases (iii) and (iv)) and those using different sets of particles (cases (i) and (ii)). The differences are more significant in peripheral collisions. The reason for this difference is as follows. The reconstructed \ep\ divides the multiplicity of the event into two roughly evenly populated halves. Therefore, the positive charged particle asymmetry and that for negative particles calculated using the same set of particles used for the \ep\ reconstruction are anticorrelated between \ud. This does not affect those of \lr. This results in a relatively smaller $\AAUD$ than $\AALR$ for cases (iii) and (iv). To avoid this self-correlation, data from cases (i) and (ii) were used, where the particles used for the \ep\ construction and for the asymmetry calculation are different. On the other hand, no apparent self-correlation is observed in the variances $\ApmsqUD$ and $\ApmsqLR$. This is because, in the cases of (iii) and (iv), only half of the particles used in the \ep\ reconstruction are used for asymmetry calculation. However, cases (iii) and (iv) are not used in the analysis of the variances $\ApmsqUD$ and $\ApmsqLR$ to be consistent with that of the covariances $\AAUD$ and $\AALR$.

\subsection{$\phi$-Efficiency Correction\label{app:effCorr}}

The TPC sector boundaries could create ``dynamical'' event-by-event correlations even though the \ep\ angles are uniform. Specifically, the east TPC sector with systematically lower efficiency introduces a dynamical event-by-event correlation (see below). To reduce this sort of fake dynamical correlation, the sector boundary effects were corrected for by the $\phi$-dependent correction $1/\epsilon(\phi)$. This correction is applied on average over the entire dataset. In principle, any time variation in $\epsilon(\phi)$ would introduce dynamical event-by-event correlations if a single set of correction factors is applied. However, it was found that $\epsilon(\phi)$ is stable over the entire run period of the present data. 

\begin{figure*}[hbt]
\begin{center}
\includegraphics[width=0.329\textwidth]{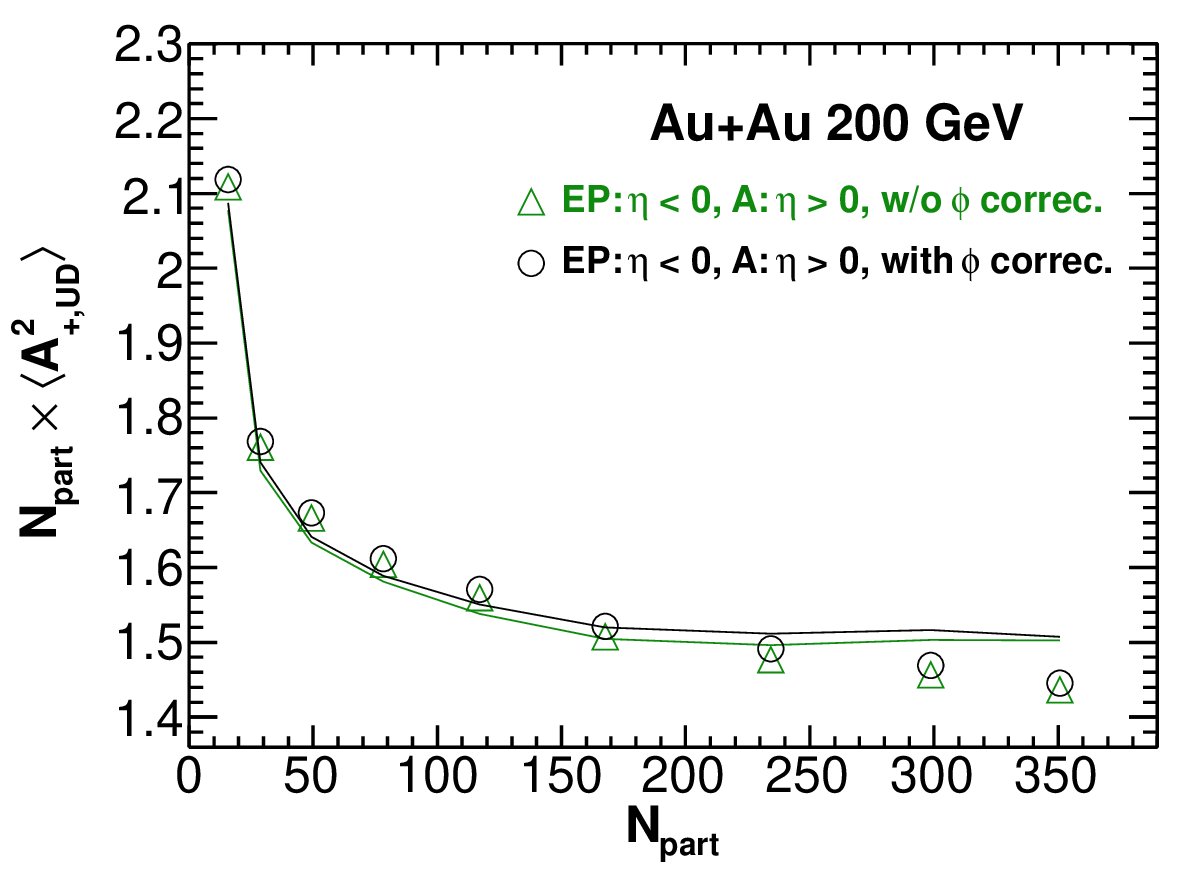}
\includegraphics[width=0.329\textwidth]{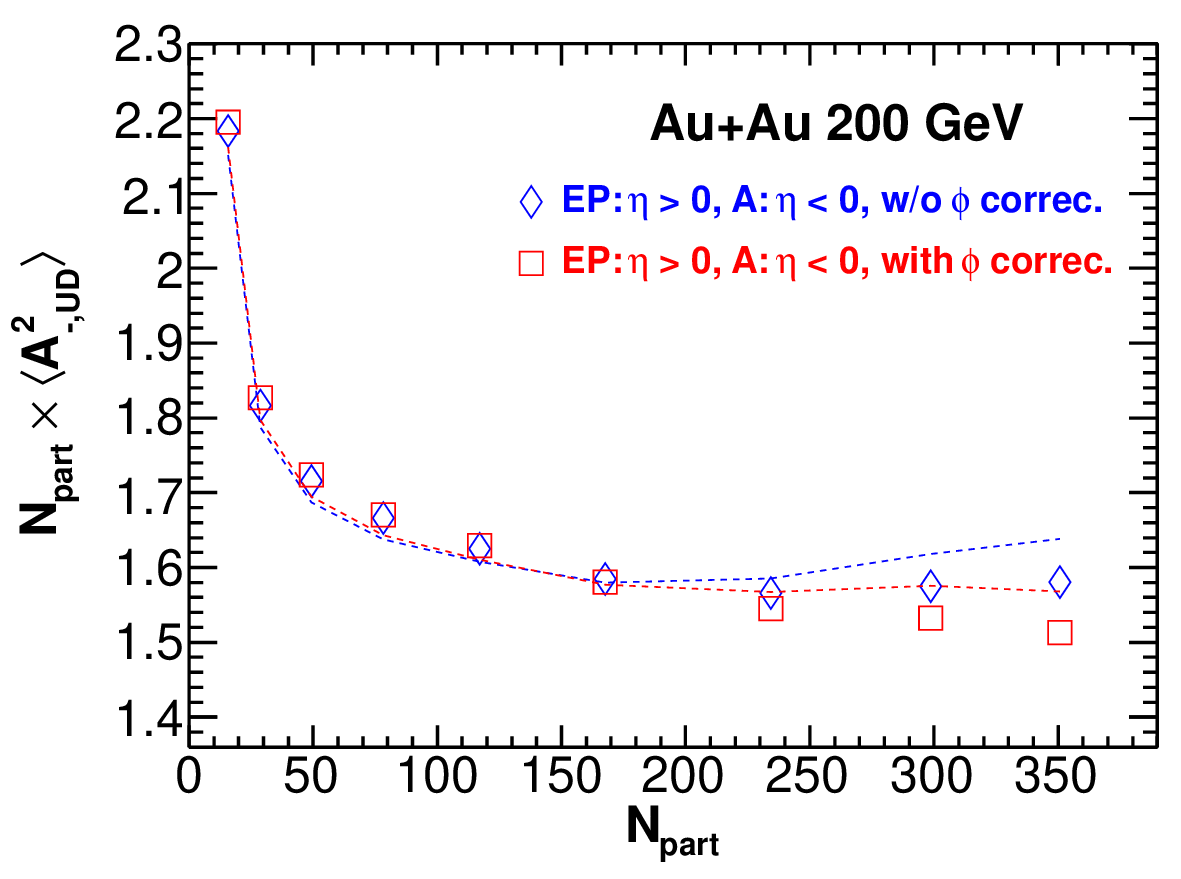}
\includegraphics[width=0.329\textwidth]{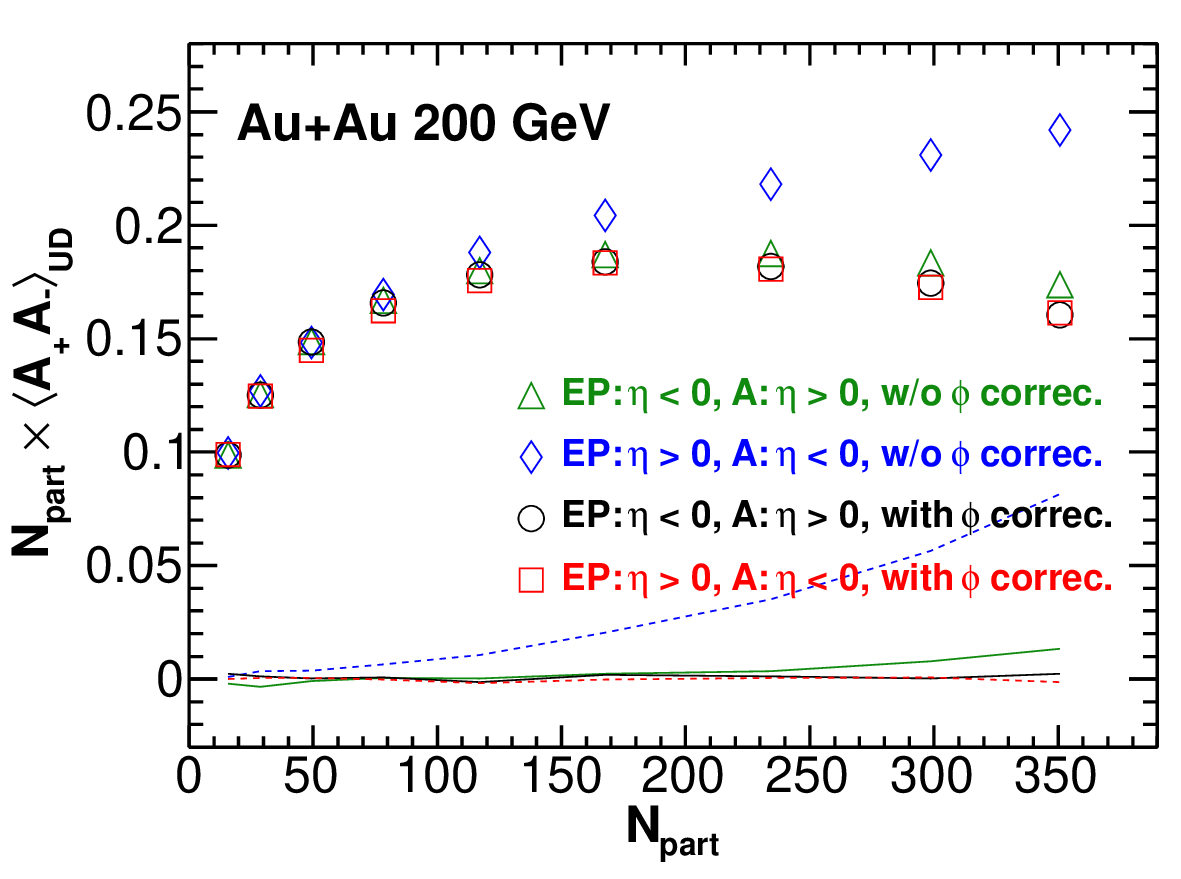}
\end{center}
\caption{(Color online) The asymmetry correlations, $\ApsqUD$ (left panel), $\AmsqUD$ (middle panel), and $\AAUD$ (right panel), multiplied by the number of participants $\Npart$, before and after the corrections for the $\phi$-dependent acceptance $\times$ efficiency. The colored curves are the corresponding statistical fluctuations and detector effects. The results are separated for $\eta>0$ and $\eta<0$. Only $\eta>0$ is shown for $\ApsqUD$ and $\eta<0$ for $\AmsqUD$ for clarity. 
The error bars are statistical only.}
\label{fig:phiCorrec}.
\end{figure*}

Figure~\ref{fig:phiCorrec} shows the charge asymmetry correlations $\ApsqUD$, $\AmsqUD$ and $\AAUD$ before and after the $\phi$-efficiency corrections. The asymmetry correlations are multiplied by the number of participants, $\Npart$, to better show the magnitude. For clarity, only the $\eta>0$ region is shown for $\ApsqUD$ and only the $\eta<0$ region is shown for $\AmsqUD$. The values of $\ApsqUD$ and $\AmsqUD$ are similar in the same $\eta>0$ or $\eta<0$ region. The \lr\ asymmetry correlations are similar to those of \ud. 
The corrections for $\eta<0$ are larger than that for $\eta>0$. This is due to the greater non-uniformity from the TPC electronics inefficiency at $11\pi/6<\phi<2\pi$ in the $\eta<0$ region (see Fig.~\ref{fig:tech}). 

\begin{figure*}[hbt]
\begin{center}
\includegraphics[width=0.329\textwidth]{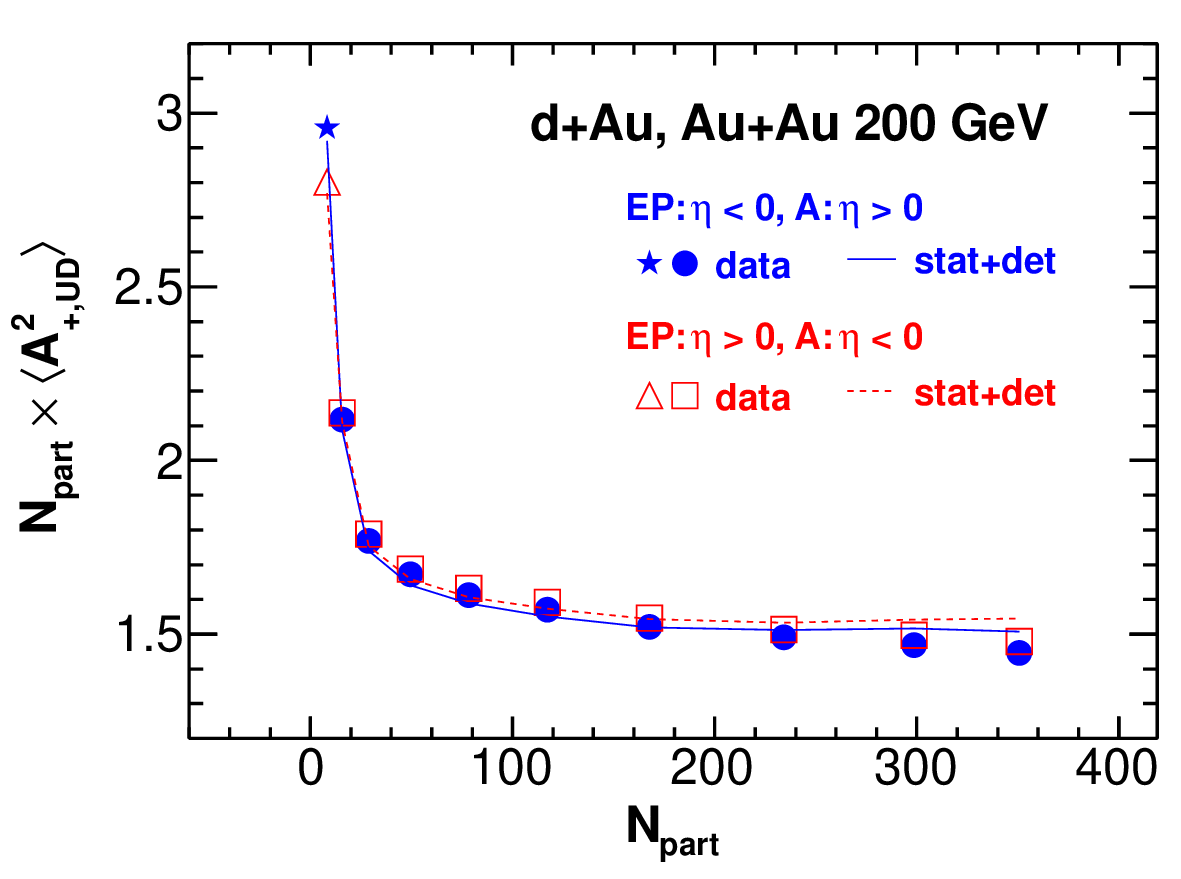}
\includegraphics[width=0.329\textwidth]{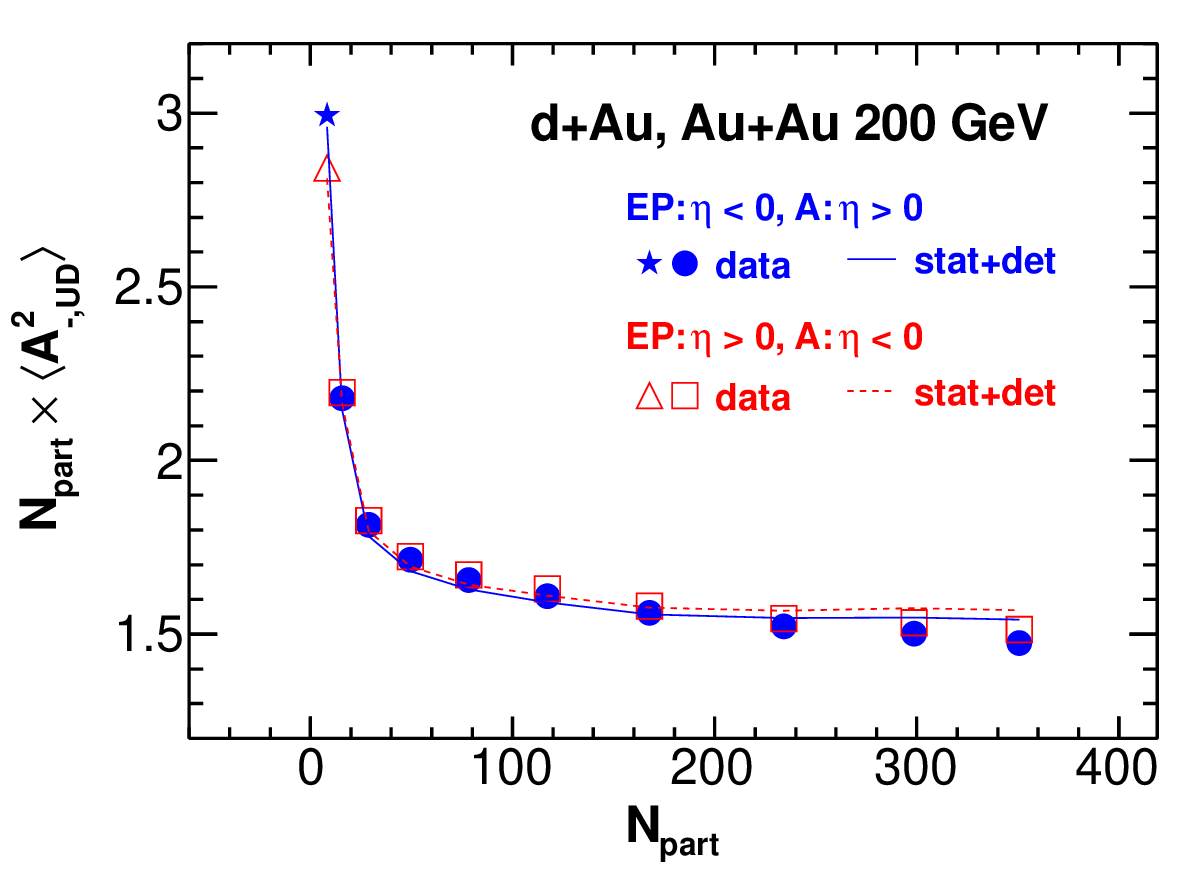}
\includegraphics[width=0.329\textwidth]{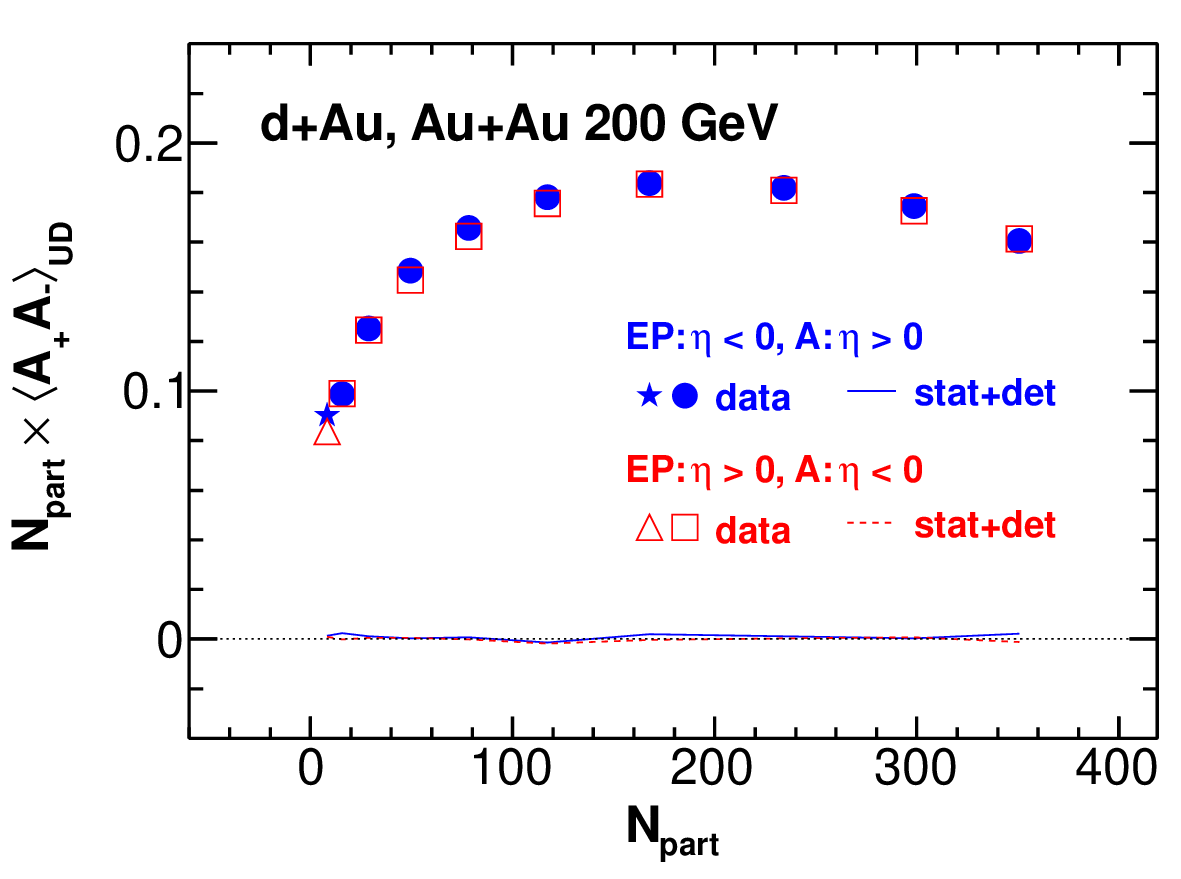}
\end{center}
\caption{(Color online) The asymmetry correlations, $\ApsqUD$ (left panel), $\AmsqUD$ (middle panel), and $\AAUD$ (right panel), multiplied by the number of participants $\Npart$, separately for $\eta>0$ and $\eta<0$. The star and triangle depict the results from $d$+Au collisions. The net effects of the statistical fluctuations and detector non-uniformities are shown as the curves. 
Error bars are statistical only.}
\label{fig:corr_stat}.
\end{figure*}

As seen in Fig.~\ref{fig:phiCorrec} (right panel), before the $\phi$-efficiency correction, the values of $\ApAm$ for the $\eta<0$ region are significantly larger than those in the $\eta>0$ region in central collisions. After the $\phi$-efficiency correction, the values of $\AAUD$ from the $\eta<0$ and $\eta>0$ regions are consistent, as will be shown in Fig.~\ref{fig:corr_stat} (right panel). As for $\Asq$, the corrections for $\ApAm$ for $\eta<0$ are relatively large. In fact, the inefficient sector boundaries seem to have similar effects on the absolute magnitudes of $\Asq$ and $\ApAm$. Relatively, they are more significant on $\ApAm$ than on $\Asq$. The values of $\AALR$ are not shown in Fig.~\ref{fig:phiCorrec} (right panel) but they are similar to $\AAUD$. The curves shown in Fig.~\ref{fig:phiCorrec} will be discussed in Appendix~\ref{app:stat}. 

\subsection{Statistical Fluctuation and Detector Effects\label{app:stat}}

\begin{figure*}[hbt]
\begin{center}
\includegraphics[width=0.329\textwidth]{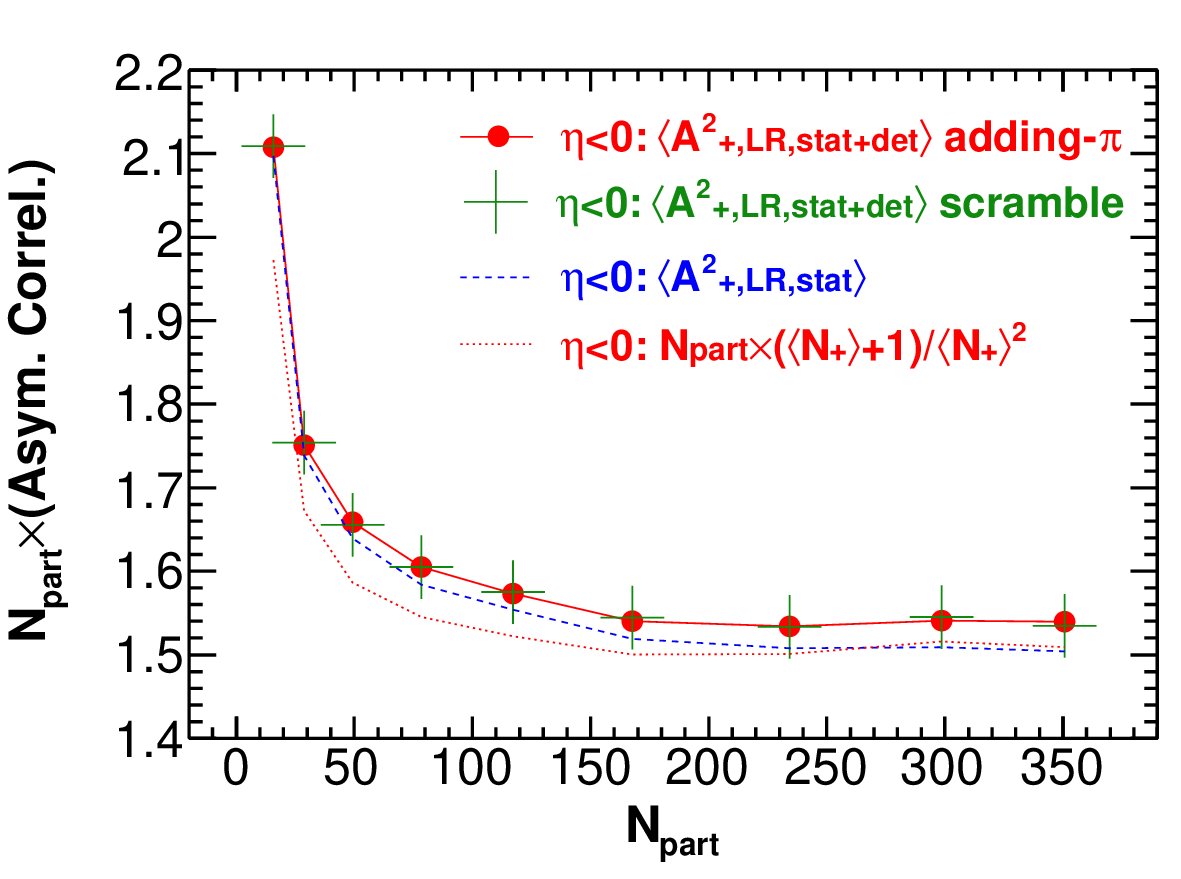}
\includegraphics[width=0.329\textwidth]{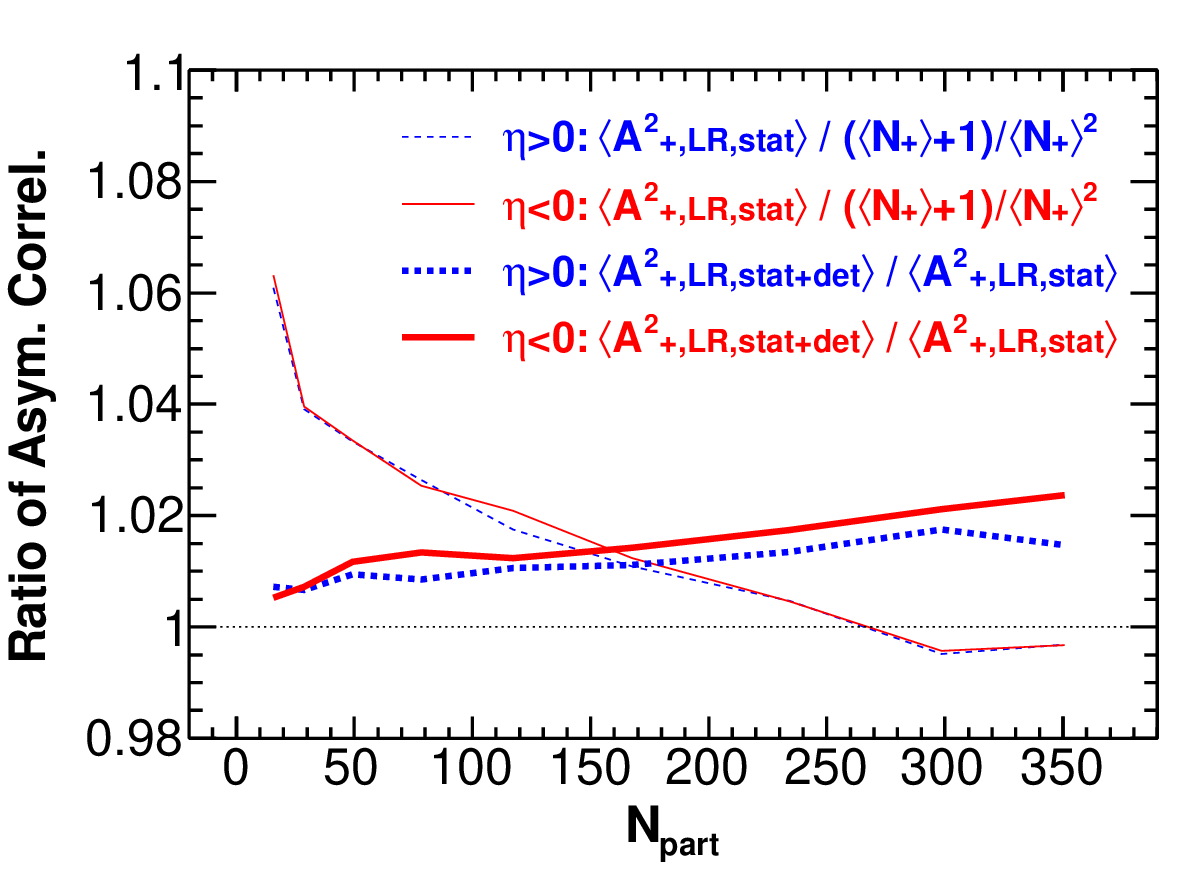}
\includegraphics[width=0.329\textwidth]{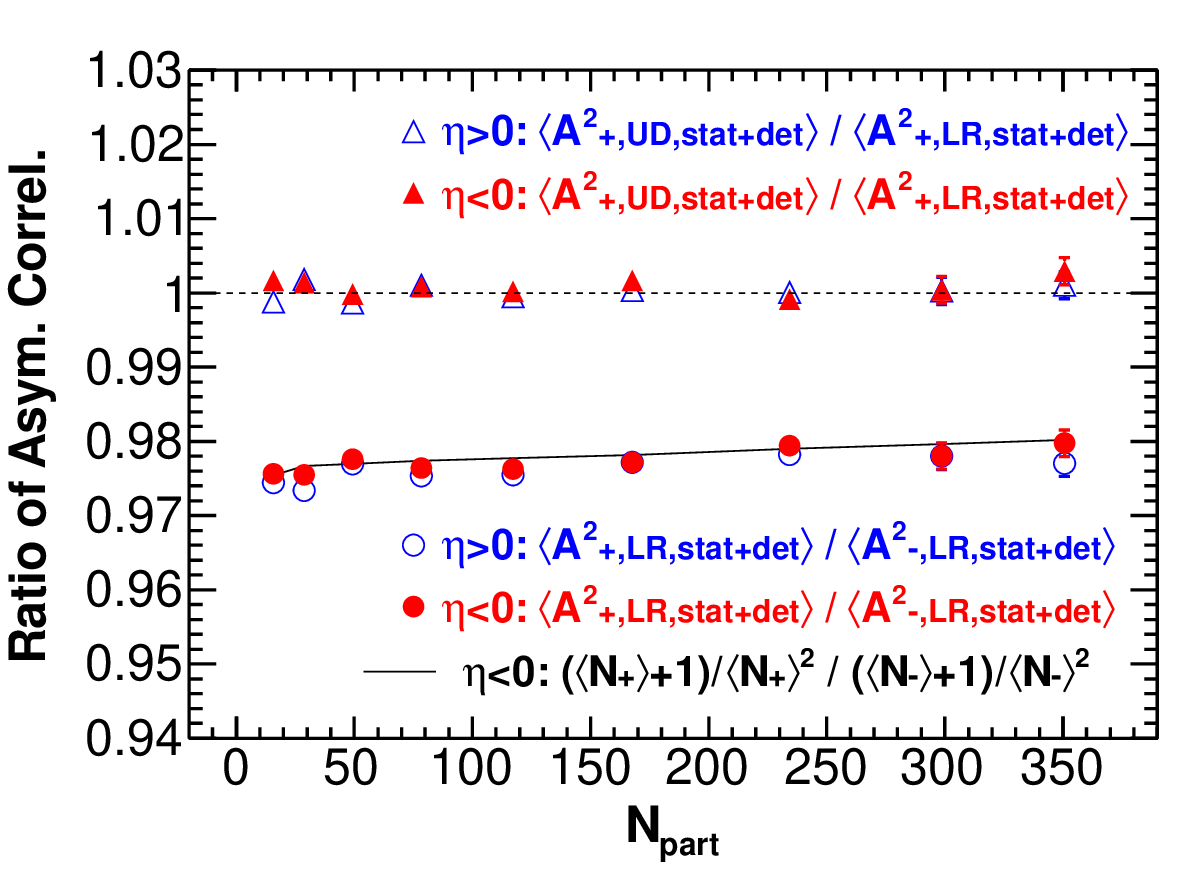}
\end{center}
\caption{(Color online) Left panel: Statistical fluctuation and detector effects in the charge asymmetry variance (multiplied by the number of participants $\Npart$) from the $\eta<0$ region. The dotted curve shows the $1/N$ approximation using Eq.~(\ref{eq:1N}), the dashed curve shows the statistical fluctuations $\AmsqLRst$ obtained by the ``50-50" method (see the text). The solid curve connecting the solid points shows the net effect of the statistical fluctuations and detector non-uniformities, $\AmsqLRrn$, obtained by the adding-$\pi$ method, and the crosses shows the same but obtained from the ``scramble'' method. See the text for details. Middle panel: The statistical fluctuation effects relative to the $1/N$ approximation (thin curves) and the effects due to the imperfect detector (thick curves). The dashed curves are for the region $\eta>0$ and the solid curves are for $\eta<0$. Right panel: The ratios of the statistical fluctuation and detector effects $\ApsqLRrn/\AmsqLRrn$ and $\ApsqUDrn/\ApsqLRrn$, separately for the $\eta>0$ and $\eta<0$ regions.
The $\phi$-acceptance correction was applied for the results in this figure. The error bars are statistical only. 
}
\label{fig:stat}.
\end{figure*}

The variances are non-zero even when no dynamical fluctuations are present. This is due to the trivial effects of statistical fluctuations of the multiplicity. If one takes $\Nu=\mean{\Nu}+\delta\Nu$ and $\Nd=\mean{\Nd}+\delta\Nd$ (where $\Nu$ collectively stands for $\NpU$ and $\NmU$, and $\Nd$ for $\NpD$ and $\NmD$), and assume the fluctuations are Poissonian, the statistical fluctuations can be expanded in $\delta\Npm/\mean{\Npm}$ and approximated by
\begin{widetext}
\be
\ApmsqUDst=\frac{1}{\mean{\Nud}^{2}}\left\langle\left(\frac{\delta\Nu-\delta\Nd}{1+(\delta\Nu+\delta\Nd)/\mean{\Nud}}\right)^{2}\right\rangle\approx\frac{\mean{\Nud}+1}{\mean{\Nud}^{2}}\,,
\label{eq:1N}
\ee
\end{widetext}
A similar formula exists for $\ApmsqLRst$. The multiplicities in Eq.~(\ref{eq:1N}) are the measured multiplicities, not the efficiency-corrected ones. The average efficiency corrections cancel in the present charge asymmetries and do not contribute to the statistical fluctuations. The result by Eq.~(\ref{eq:1N}) for $\ApsqLRst$, as an example, is shown in Fig.~\ref{fig:stat} (left panel) as the dotted curve. This approximation using Eq.~(\ref{eq:1N}) will be referred to as the ``$1/N$'' approximation.

In the data analysis, one can obtain the contributions from statistical fluctuations by assigning each particle randomly into the up or down hemisphere and into the left or right hemisphere (referred to as ``50-50'' method), and then calculating the asymmetry correlation without additional corrections by $1/\epsilon(\phi)$. This would correspond to the case of purely statistical fluctuations with a perfect detector. Figure~\ref{fig:stat} (left panel) shows the values of $\ApsqLRst$ in the region $\eta<0$ obtained by this ``50-50'' method as the dashed curve. As seen from the figure, the $1/N$ approximation of Eq.~(\ref{eq:1N}) underestimates the statistical fluctuations. To quantify the magnitude of the underestimation, Fig.~\ref{fig:stat} (middle panel) shows the ratio of the dashed to the dotted curves from the left panel. Also shown is the corresponding ratio from the $\eta>0$ region. As expected, the statistical fluctuations relative to the $1/N$ approximation are the same between the $\eta>0$ and $\eta<0$ regions. The underestimation is the most severe in peripheral collisions, and is of the order of a few percent. This is presumably due to the non-Poissonian nature of the multiplicity distributions in the peripheral data and the approximations made in Eq.~(\ref{eq:1N}). In central collisions, both the Poissonian multiplicity distribution assumption and the large $N$ approximation should be valid. As can be seen, the statistical fluctuations in central collisions can be well described by Eq.~(\ref{eq:1N}).

The STAR TPC is not precisely uniform in azimuth. There are sector boundaries and different detection efficiencies due to variations in the electronics performance. As mentioned earlier, these non-uniformities are corrected for on average by the $\phi$-dependent correction factor $1/\epsilon(\phi)$ separated by particle charge signs, magnetic field polarities, and positive and negative $\eta$ regions, centrality bins, and $\pt$ bins. However, the event-by-event fluctuations in the efficiencies cannot be corrected. These fluctuations in detector performance introduce a dynamical effect, which are referred to as ``detector effects'' in this paper. These detector effects were quantified in the following way. For each particle, one either adds $\pi$ to the measured azimuthal angle $\phi$ or does nothing. The probability to add $\pi$ is determined by the relative efficiency at $\phi$ and $\phi+\pi$, namely, $\epsilon(\phi+\pi)/(\epsilon(\phi)+\epsilon(\phi+\pi))$. In this way, the physics correlation amongst particles is destroyed, but the detector non-uniformities and elliptic flow correlations of the particles are preserved. The average $\phi$-dependent efficiency is then corrected for each particle depending on its new $\phi$ values, and the charge asymmetry correlation were then calculated. The resultant asymmetry correlation from the region $\eta<0$ for positive particles, $\ApsqLRrn$, is shown in Fig.~\ref{fig:stat} (left panel) by the solid curve connecting the dots. The text ``stat+det'' is used to stand for the net effect of the statistical fluctuations and the detector non-uniformities. The solid curves in Fig.~\ref{fig:stat} (middle panel) show the ratio of $\ApsqLRrn/\ApsqLRst$. This ratio measures the magnitude of the detector effects relative to the pure statistical fluctuations. As seen from the figure, the ratio is approximately unity in peripheral collisions and increases to a few percent above unity in central collisions. This is expected because the detector non-uniformity is the strongest in the most central collisions~\footnote{The present track reconstruction algorithm jumps over gaps between sectors to search for the next hits when reconstructing tracks across the sector boundaries. For track trajectories nearly parallel to a sector boundary, the empty distance over the gap between the adjacent sectors that the algorithm has to bridge is relatively large. The search window for hits in the next sector is made proportionally wider. In central collisions, where the TPC hit occupancy is high, the confusion in the hit finding greatly increases. This results in a lower efficiency in reconstructing those tracks nearly parallel to the sector boundaries in more central collisions.}. The same ratio for the region $\eta>0$ is also shown. 

An alternative approach was also pursued. For each particle, a $\phi$ angle was randomly generated according to the efficiency $\epsilon(\phi)$. The charge asymmetry correlations for this ``scrambled" data were then calculated in the same way as was done 
for the real data. This should be equivalent to the adding-$\pi$ method above, except that it also destroys the contribution from elliptic flow. The results obtained by this ``scramble'' method are shown by the crosses in Fig.~\ref{fig:stat} (left panel) and are consistent with those obtained by the ``adding-$\pi$'' method (solid circles). 

Eq.~(\ref{eq:1N}) indicates that, if there are no dynamical fluctuations, the self-correlation variable is approximately equal to the inverse of the particle multiplicity. Figure~\ref{fig:stat} (right panel) shows the ratio of the positive to negative charge asymmetry correlations from the adding-$\pi$ method. Indeed, the $\ApsqRn$ and $\AmsqRn$ differ slightly. The values of $\ApsqRn$ are smaller than those of $\AmsqRn$ by about 2\%. This difference is consistent with the slight (about 2\%) excess of positive particles over negative particles~\cite{Levente}. 

The statistical fluctuation and the detector effects should not depend on the reaction plane orientation and they should be equal between \ud\ and \lr. It has been verified that $\ApsqUDrn=\ApsqLRrn$ and $\AmsqUDrn=\AmsqLRrn$. Figure~\ref{fig:stat} (right panel) shows as examples the ratios of $\ApsqUDrn/\ApsqLRrn$ from the adding-$\pi$ method for the $\eta>0$ and $\eta<0$ regions. They are consistent with unity. The same is observed for the negative particle asymmetry correlations.

\subsection{Consistency Checks\label{app:check}}

As shown in Fig.~\ref{fig:phiCorrec}, the charge asymmetry variances and covariances without applying the $\phi$-acceptance corrections differ from those with the corrections. This section checks the corresponding statistical fluctuation and detector effects with and without the $\phi$-acceptance corrections. The adding-$\pi$ method was used to calculate the statistical fluctuation and detector effects. The results are superimposed as the curves of the corresponding colors in Fig.~\ref{fig:phiCorrec} for the $\phi$-efficiency corrected and uncorrected asymmetry correlation data. 
As seen, the statistical fluctuation and detector effects for $\ApAm$ without the $\phi$-efficiency corrections are no longer zero, especially for the $\eta<0$ region. 
However, the differences in the data points with and without the $\phi$-efficiency corrections appear to be the same as those between the corresponding curves for the statistical fluctuation and detector effects. This indicates that the differences observed in the data from different $\eta$ regions are due to the statistical fluctuation and detector effects. This is demonstrated more quantitatively by the dynamical asymmetry correlations after subtracting the statistical fluctuation and detector effects.

Figure~\ref{fig:corr_stat} (left and middle panels) shows the $\phi$-efficiency corrected charge asymmetry variances $\ApsqUD$ and $\AmsqUD$ from the $\eta>0$ and $\eta<0$ regions separately. Superimposed are the combined statistical fluctuation and detector effects $\ApsqUDrn$ and $\AmsqUDrn$. The $\AsqLR$ are similar to the $\AsqUD$. Figure~\ref{fig:corr_stat} (right panel) shows the $\phi$-efficiency corrected covariances, $\AAUD$. The statistical fluctuation and detector effects in the covariances are zero, as indicated by the superimposed curves (obtained from the adding-$\pi$ method). The $\AALR$ is similar to the $\AAUD$. In Fig.~\ref{fig:corr_stat}, the minimum-bias d+Au data are also shown. The analysis procedure for d+Au is the same as for Au+Au collisions. The d+Au data follow the trend of the peripheral Au+Au data.

The asymmetry correlation results from $\eta<0$ and $\eta>0$ should ideally be equal in Au+Au collisions because of the collision symmetry about mid-rapidity. This is approximately the case for the covariances, and both are consistent with zero. It is, however, not true for the variances. The corrected variances in $\eta<0$ are larger than the corresponding ones in $\eta>0$ by 1-3\%. This is partially due to the fact that the variance magnitudes are dominated by the statistical fluctuations of multiplicities, which are approximately inversely proportional to the average multiplicity (see below). The 10\% inefficiency in the $11\pi/6<\phi<2\pi$ region for $\eta<0$ results in a 2\% larger magnitude for the variances compared to $\eta>0$. This is indeed shown by the two differing curves which are the corresponding statistical and detector effects.


The above effects are cancelled in the difference between \ud\ and \lr\ asymmetry correlations, which is the main observable sensitive to the local parity violation. This is because the $\phi$-dependent acceptance corrections and the effect of statistical fluctuation of the multiplicities are identical for \ud\ and \lr\ asymmetries.

\begin{figure*}[hbt]
\begin{center}
\includegraphics[width=0.329\textwidth]{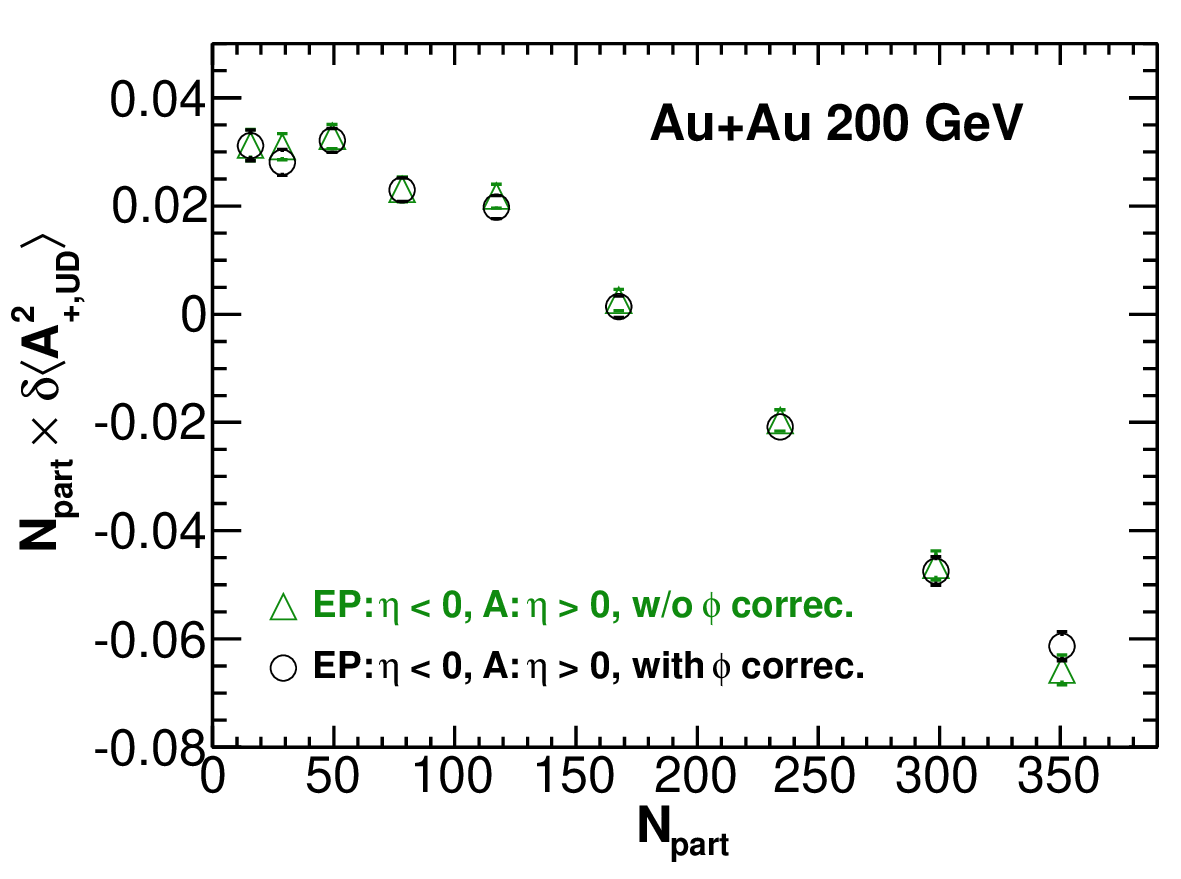}
\includegraphics[width=0.329\textwidth]{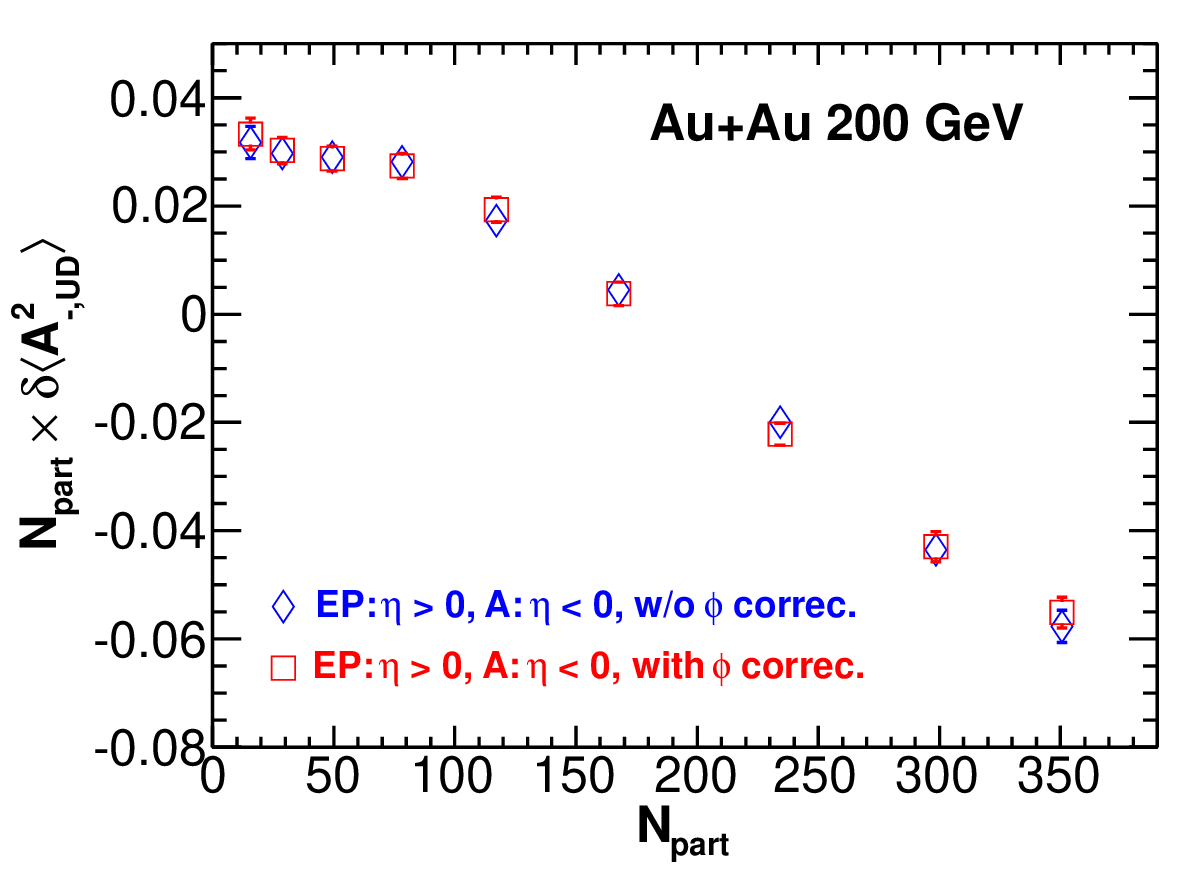}
\includegraphics[width=0.329\textwidth]{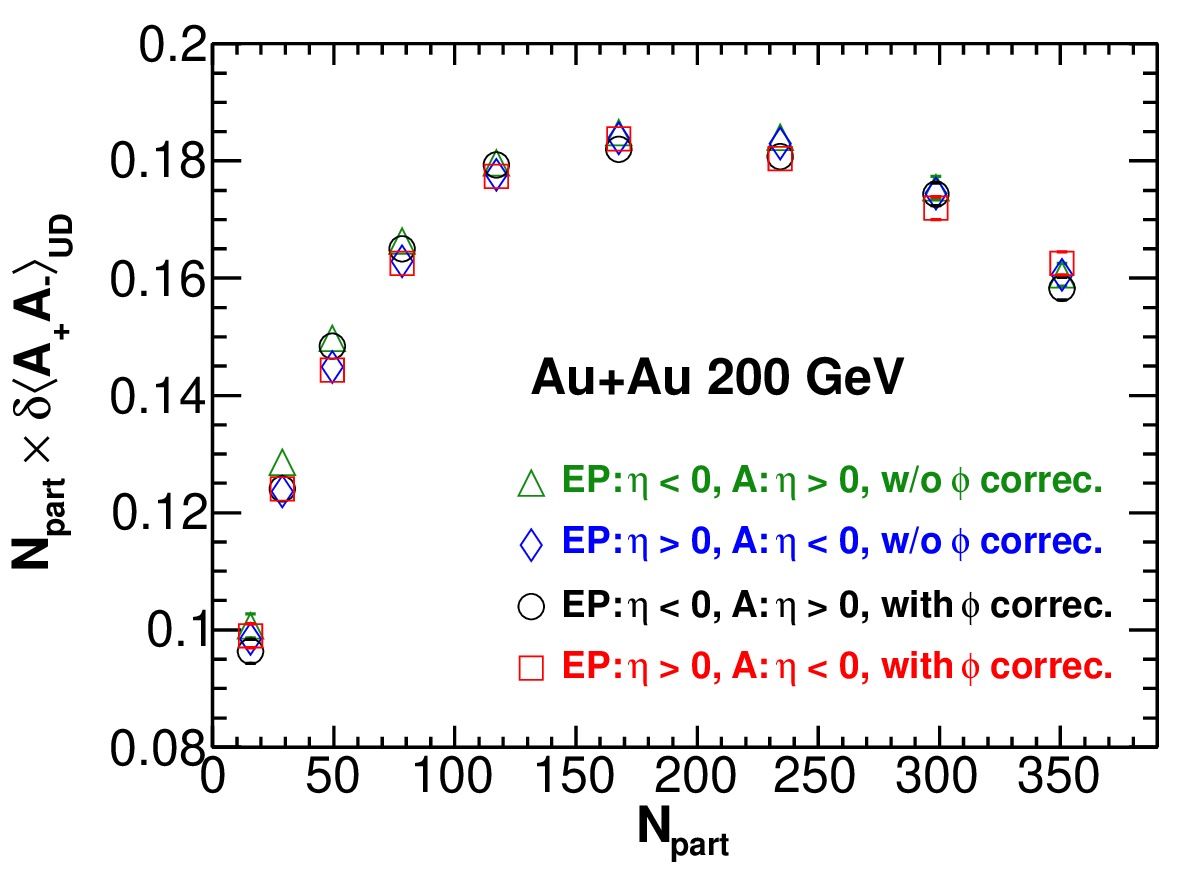}
\end{center}
\caption{(Color online) The statistical and detector effect subtracted asymmetry correlations, $\delta\ApsqLR$ (left panel), $\delta\AmsqLR$ (middle panel), and $\delta\AALR$ (right panel), multiplied by the number of participants $\Npart$, before and after the corrections for the $\phi$-dependent acceptance $\times$ efficiency. This figure corresponds to Fig.~\ref{fig:phiCorrec}. The error bars are statistical only.}
\label{fig:phiCorrec2}.
\end{figure*}

The statistical fluctuation and detector effects $\ApmsqRn$ (as obtained by the adding $\pi$ method) were subtracted from $\Apmsq$ to obtain the dynamical asymmetry correlations: $\delta\Apmsq=\Apmsq-\ApmsqSt$. Figure~\ref{fig:phiCorrec2} (left and middle panels) show, respectively, the dynamical $\delta\ApsqUD$ and $\delta\AmsqUD$ from the $\eta>0$ and $\eta<0$ regions, by taking the differences between the data points and the corresponding curves in Fig.~\ref{fig:phiCorrec} (left and middle panels). Similarly, Fig.~\ref{fig:phiCorrec2} (right panel) shows the dynamical $\daa\AAUD$. 
The corresponding statistical fluctuation and detector effects were subtracted from the charge asymmetry variances and covariances, calculated both with and without applying the $\phi$-acceptance corrections. 
The resultant dynamical results are consistent with and without the $\phi$-acceptance corrections, as shown in Fig.~\ref{fig:phiCorrec2}.
This indicates that no residual effects of the $\phi$-acceptance remains after the corrections for the statistical fluctuations and detector effects. Nonetheless, the $\phi$-acceptance corrections were applied in the present analysis. Unless specified, all results in this paper have the $\phi$-acceptance corrections applied.

A test was also performed where an acceptance hole was articifically created within a restricted $\phi$ region by randomly discarding 50\% of the particles in the $\phi$ region and at the same time reducing the acceptance $\times$ efficiency $\epsilon(\phi)$ by a factor of 2. The analysis was repeated with the remaining particles to calculate the charge asymmetry variances and the statistical fluctuation and detector effects to the dynamical variances. The dynamical variances were consistent with the results previously obtained without the artificial acceptance hole. This is also true for the charge asymmetry covariances.

\begin{figure}[hbt]
\begin{center}
\includegraphics[width=0.4\textwidth]{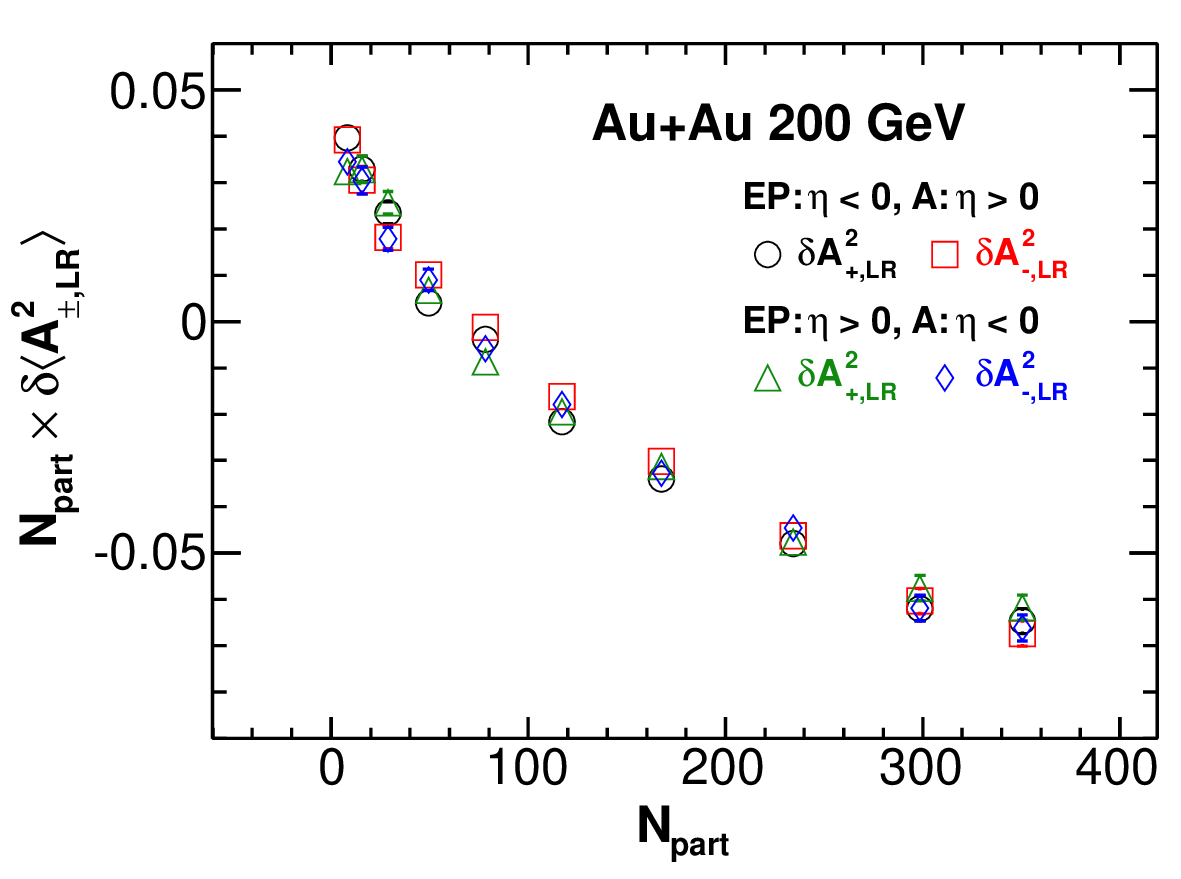}
\end{center}
\caption{(Color online) Dynamical charge asymmetry variance after removing the statistical and detector effects, $\delta\ApmsqLR=\ApmsqLR-\ApmsqLRrn$, scaled by the number of participants $\Npart$. The results are consistent between positive and negative $\eta$ regions and between positive and negative charges. 
Error bars are statistical only.}
\label{fig:etaRegion}.
\end{figure}

Figure~\ref{fig:etaRegion} shows the dynamical $\delta\ApmsqLR$ values from the $\eta>0$ and $\eta<0$ regions with the $\phi$-acceptance corrections. 
This plot demonstrates that the results from the two $\eta$ regions are consistent, and the results are the same for positive and negative charge asymmetries. This is also true for \ud\ asymmetries. 

Since the dynamical charge asymmetry variances from the $\eta<0$ and $\eta>0$ regions are consistent, and those of the positive and negative charge asymmetry dynamical variances are equal within the statistical uncertainties, as shown in Fig.~\ref{fig:etaRegion}, the average of the results from the two $\eta$ regions is used. Also, the average between the positive and negative charge asymmetry variances: $\delta\AsqUD=(\delta\ApsqUD+\delta\AmsqUD)/2$ and $\delta\AsqLR=(\delta\ApsqLR+\delta\AmsqLR)/2$, is used. Since the covariances from the two $\eta$ regions are also the same (see Fig.~\ref{fig:corr_stat} (right panel) and Fig.~\ref{fig:phiCorrec2} (right panel)), the average between the two regions is also used. The average variances and covariances are reported.


\subsection{Effect of the Event-Plane Resolution\label{app:EPres}}

The constructed event-plane is not as same as the reaction plane. The inaccuracy, or event-plane resolution, is due to the finite multiplicity of particles used for the event-plane reconstruction. 
The observed differences between the \ud\ and \lr\ charge asymmetry correlations are affected by the event-plane resolution. The magnitudes of the measured asymmetry correlation differences are reduced from their true values ({\it i.e.} with respect to the real reaction plane) due to the finite event-plane resolution. 

\begin{figure}[hbt]
\begin{center}
\includegraphics[width=0.4\textwidth]{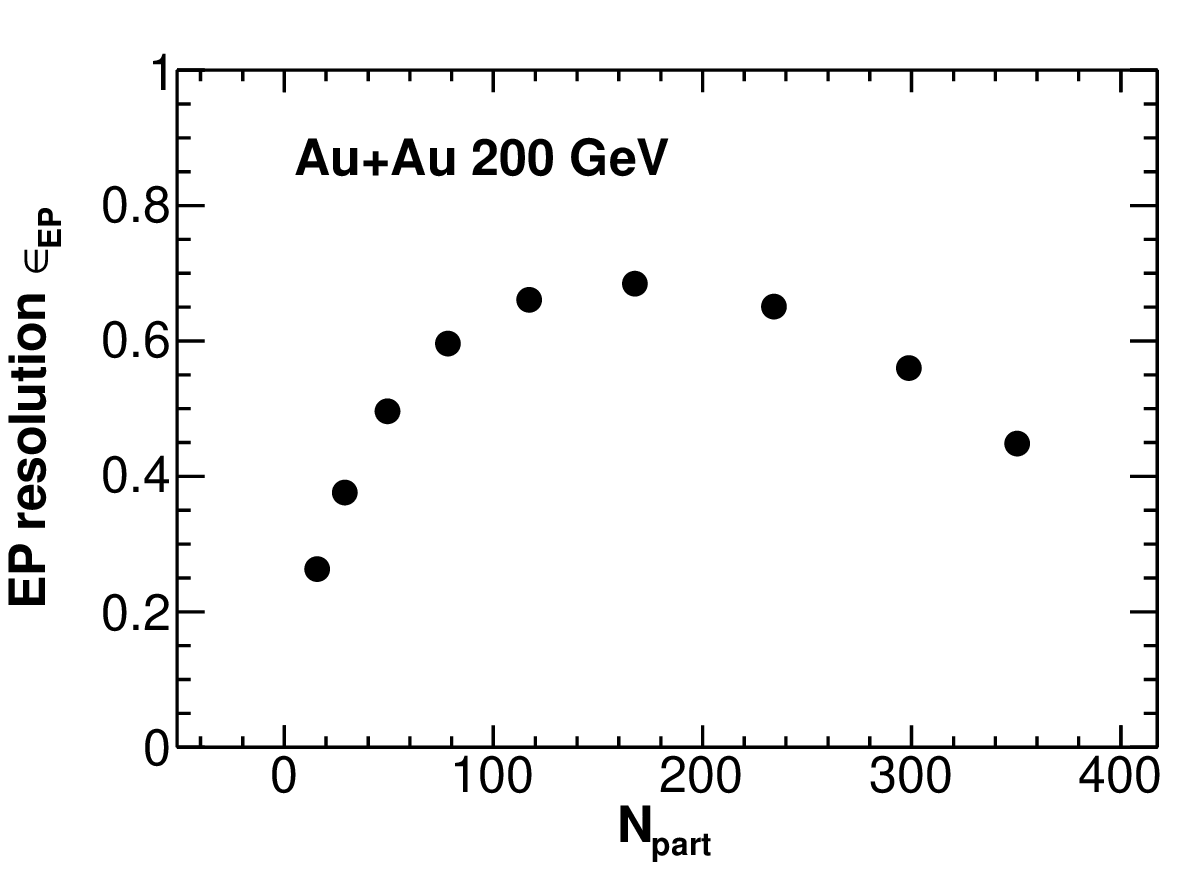}
\end{center}
\caption{The second-harmonic event-plane resolution of sub-events ($\eta<0$ and $\eta>0$) as a function of the number of participants. The statistical errors are smaller than the symbol sizes.}
\label{fig:EPres}
\end{figure}

The event-plane resolution can be calculated, approximately, by $\EPres=\sqrt{\mean{\cos2(\psi_{{\rm EP},\eta>0}-\psi_{{\rm EP},\eta<0})}}$, where $\psi_{{\rm EP},\eta>0}$ and $\psi_{{\rm EP},\eta<0}$ are the reconstructed event plane azimuths from particles at $\eta>0$ and $\eta<0$, respectively. Note, this event plane resolution is for the event planes constructed from sub-events (half-events), which is most relevant for the present studies because the event planes of the half-events was used. Figure~\ref{fig:EPres} shows the $\EPres$ as a function of centrality. The event-plane resolution is maximal in mid-central collisions and decreases towards peripheral and central collisions.

\begin{figure*}[hbt]
\begin{center}
\includegraphics[width=0.329\textwidth]{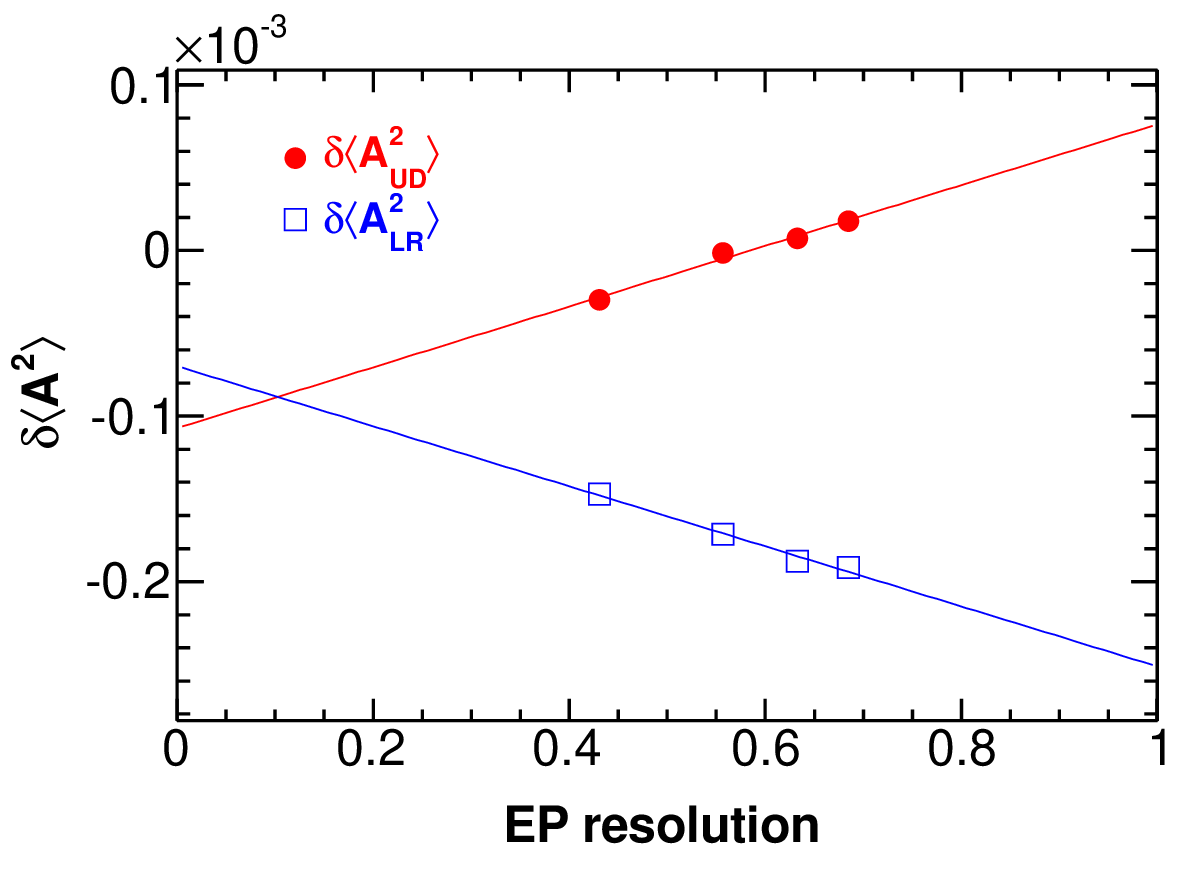}
\includegraphics[width=0.329\textwidth]{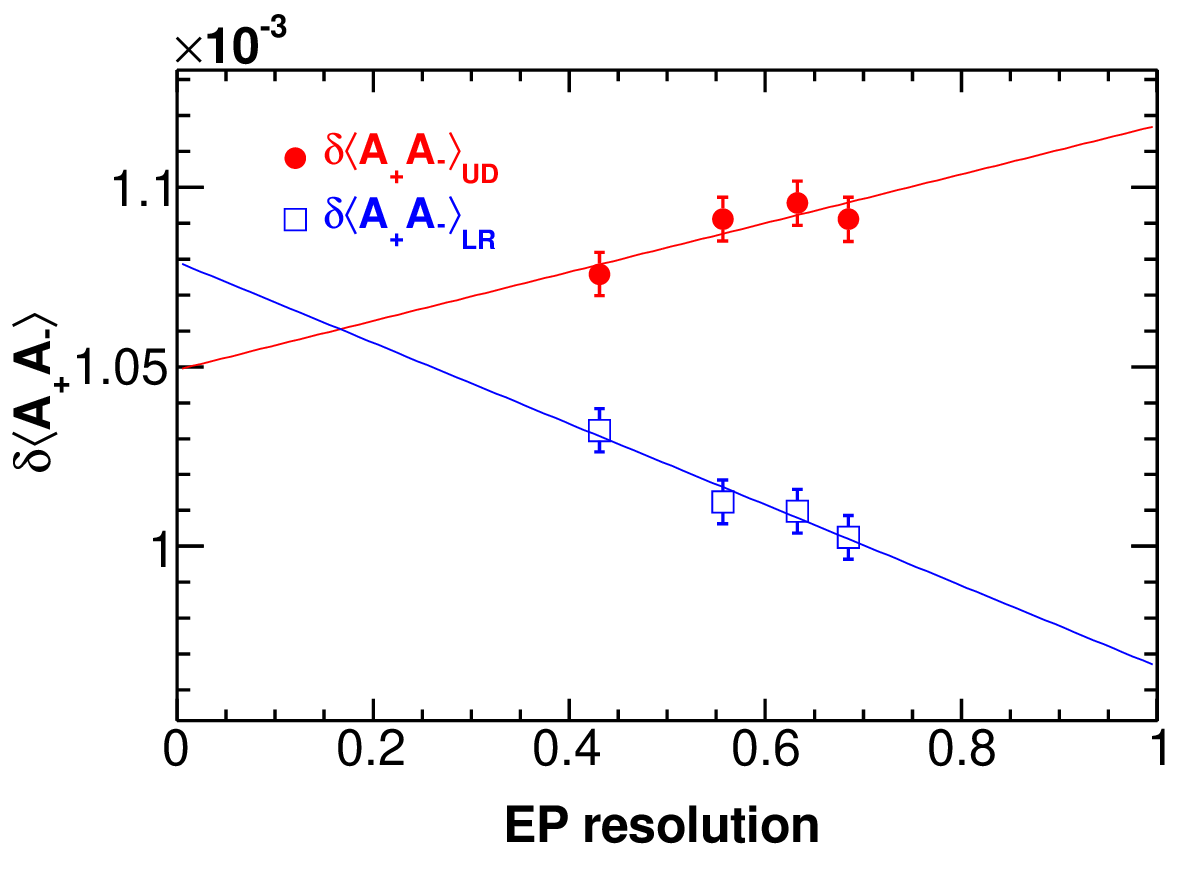}
\includegraphics[width=0.329\textwidth]{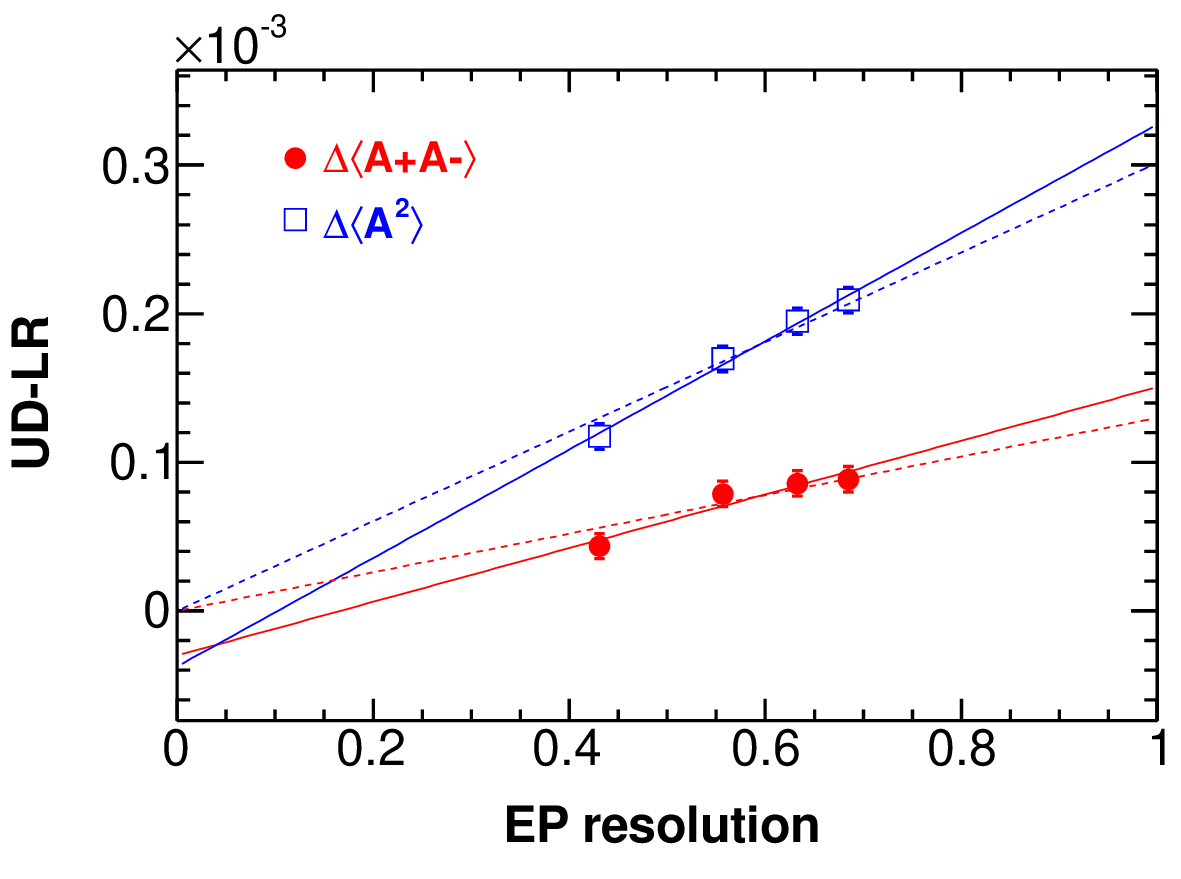}
\end{center}
\caption{(Color online) Charge multiplicity asymmetry correlations, $\daa\Asq$ (left panel) and $\daa\ApAm$ (middle panel), and their differences between \ud\ and \lr\ (right panel) as a function of the event-plane resolution $\EPres(f)=\sqrt{\mean{\cos2(\psi_{{\rm EP},\eta>0}(f)-\psi_{{\rm EP},\eta<0}(f))}}$ in 20-30\% central Au+Au collisions. The error bars are statistical only. The solid lines are free linear fits to the data, while the dashed lines are linear fits with the intercept fixed to zero at $\EPres(f)=0$.}
\label{fig:EPres_ext}
\end{figure*}

To study the dependence of the present results on the event-plane resolution, a certain fraction of particles were randomly discarded ($f$=25\%, 50\%, and 75\%) from the event-plane reconstruction, thereby artificially reducing the event-plane resolution. Figure~\ref{fig:EPres_ext} shows $\delta\Asq$, $\daa\ApAm$ and their differences between \ud\ and \lr\ as a function of $\EPres(f)=\sqrt{\mean{\cos2(\psi_{{\rm EP},\eta>0}(f)-\psi_{{\rm EP},\eta<0}(f))}}$ for 20-30\% central Au+Au collisions. The rightmost data point corresponds to the event-plane resolution of the half-events where no particles were discarded.

The \ud\ and \lr\ asymmetry correlations vary with the \ep\ resolution in opposite directions. 
The differences between \ud\ and \lr, $\Delta\Asq$ and $\Delta\ApAm$, increase with the event-plane resolution as expected. The present results are reported as measured with respect to the reconstructed event-plane, without correcting for the reductions due to the event-plane resolution. The reason this correction is not performed is because it is not known how the asymmetry correlations depend on the \ep\ resolution outside the measured range of \ep\ resolution in Fig.~\ref{fig:EPres_ext}. As an estimate, the asymmetry correlation differences between \ud\ and \lr\ with a perfect \ep\ resolution are described in Sec.~\ref{sec:asym} using a linear extrapolation. The linear extrapolation with fixed zero intercept (dashed lines in Fig.~\ref{fig:EPres_ext} (right panel)) would be correct if the high-order harmonic terms in Eq.~(\ref{eq:expansion}) are negligible.

Despite the difficulty in extrapolating to a perfect \ep\ resolution, the true differences in the correlations between \ud\ and \lr, with ideal event-plane resolution of unity, should be larger than the measured differences reported here. 
The conclusions made using the presently measured correlation differences between \ud\ and \lr\ can, therefore, are only made stronger if the reaction plane could be measured precisely.

\begin{figure*}[hbt]
\begin{center}
\includegraphics[width=0.329\textwidth]{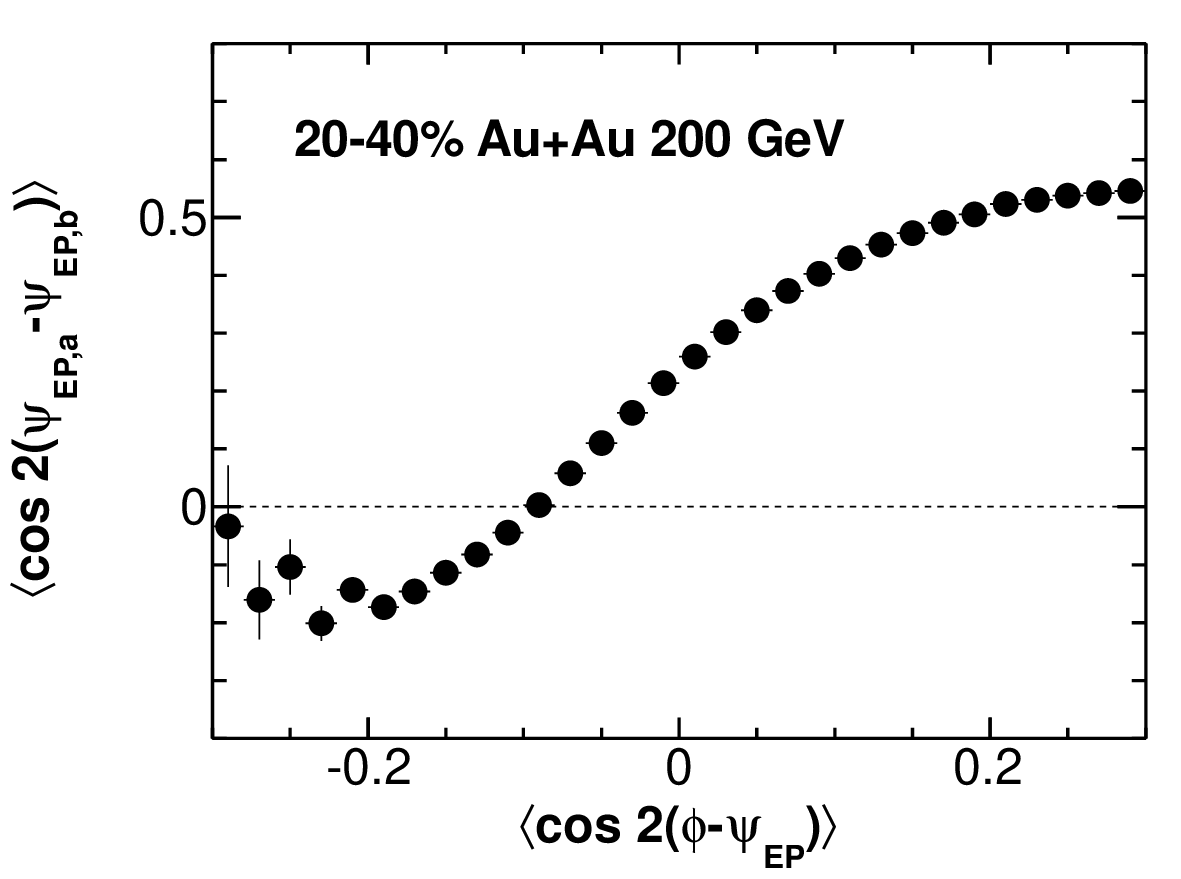}
\includegraphics[width=0.329\textwidth]{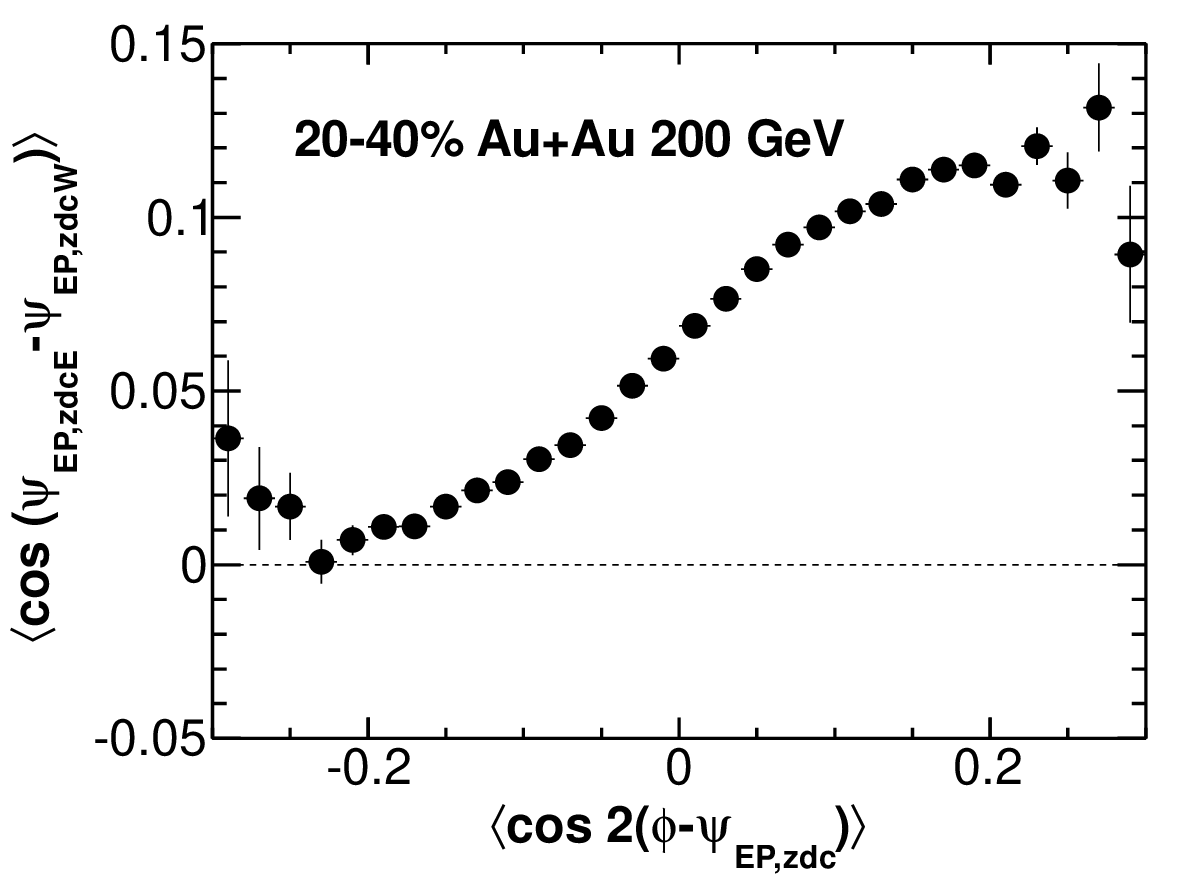}\\
\includegraphics[width=0.329\textwidth]{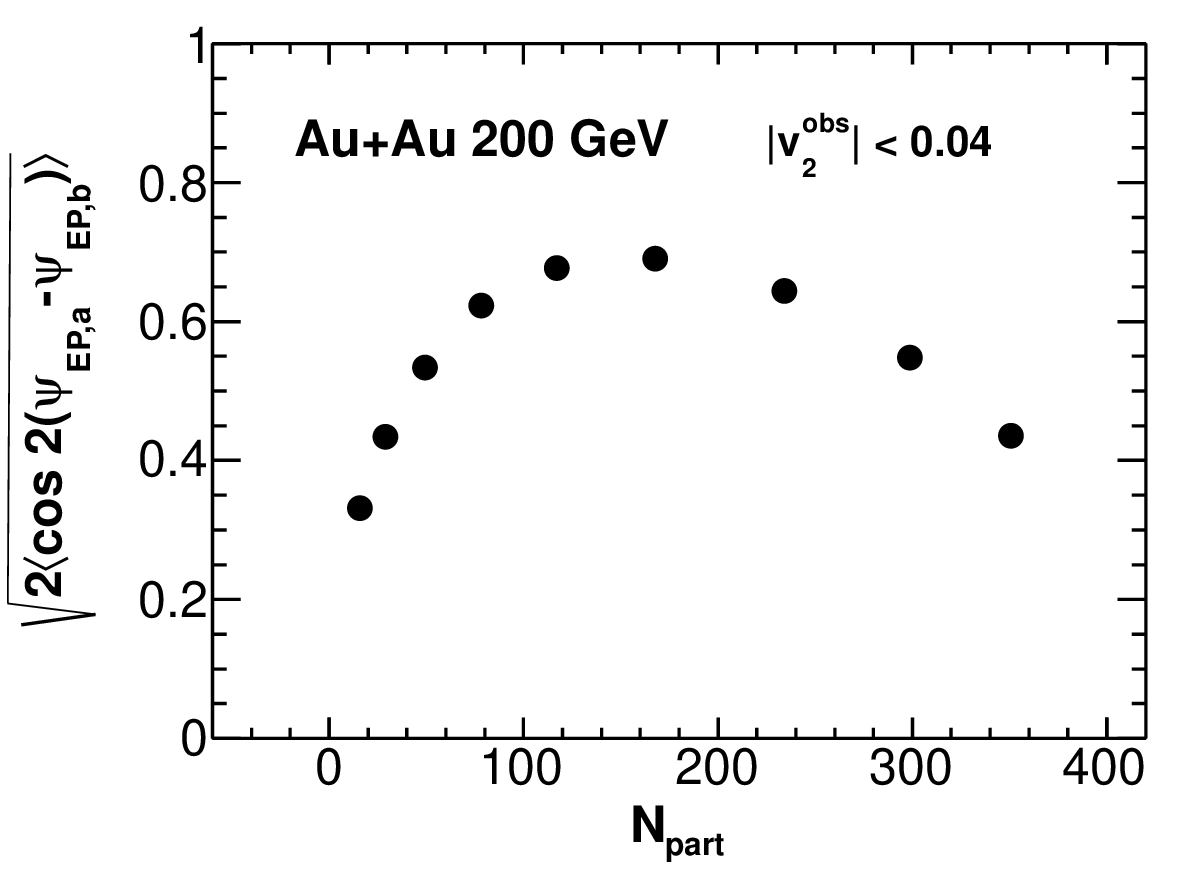}
\includegraphics[width=0.329\textwidth]{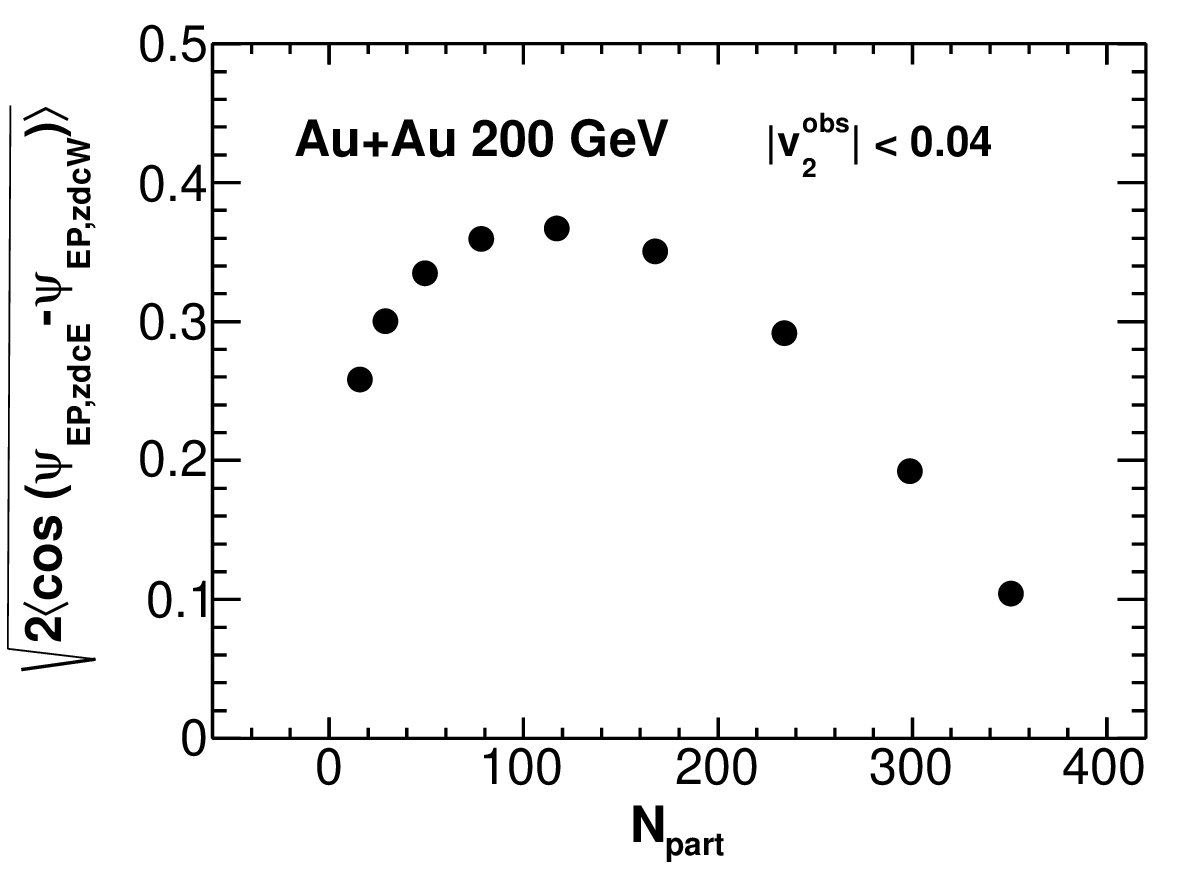}
\end{center}
\caption{The TPC second-harmonic (upper left) and ZDC-SMD first harmonic (upper right) event-plane resolution as a function of $\vlow$ in 20-40\% central Au+Au collisions. The TPC second-harmonic (lower left) and ZDC-SMD first harmonic (lower right) event-plane resolution for events with $|\vlow|<0.04$ as a function of centrality. The error bars are statistical only.}
\label{fig:EPresV2}
\end{figure*}

The particle multiplicity asymmetry correlations were measured as a function of the anisotropy of those particles within one half of the TPC relative to the \ep\ reconstructed from the other half of the TPC. The event-plane resolution varies with the particle anisotropy $\vlow$ even though the two quantities are from different regions of phase space. The half-event used for \ep\ reconstruction was randomly subdivided into two quarters, a and b. The \ep\ of the quarter events was reconstructed. The \ep\ resolution of the half event was assessed via $\mean{\cos2(\psi_{\rm EP,a}-\psi_{\rm EP,b})}$. Figure~\ref{fig:EPresV2} (upper left panel) shows $\mean{\cos2(\psi_{\rm EP,a}-\psi_{\rm EP,b})}$ as a function of $\vlow$. For significantly negative $\vlow$ events, the values of $\mean{\cos2(\psi_{\rm EP,a}-\psi_{\rm EP,b})}$ are negative, suggesting that the reconstructed \ep\ for those events is not the true reaction plane, perhaps even being orthogonal to, rather than aligned with, the reaction plane. This would mean that the \ud\ and \lr\ hemispheres are flipped for the events with significantly negative $\vlow$.

The asymmetry correlations in those events with nearly zero $\vlow$ were also studied. It was found that $\mean{\cos2(\psi_{\rm EP,a}-\psi_{\rm EP,b})}$ is positive for those events for all centralities. Figure~\ref{fig:EPresV2} (lower left panel) shows the estimated event-plane resolution, $\sqrt{2\mean{\cos2(\psi_{\rm EP,a}-\psi_{\rm EP,b})}}$, for events with $|\vlow|<0.04$ as a function of centrality.

The asymmetry correlation dependence on the event-by-event particle distribution anisotropy was also studied with respect to the first harmonic event plane reconstructed from the ZDC-SMD signals. The event-plane resolution was obtained by the correlation between the event planes reconstructed from the east and west ZDC-SMD separately. The corresponding resolutions are shown in Fig.~\ref{fig:EPresV2} (right panels).

\begin{thebibliography}{99}
\bibitem{wpBRAHMS} I.~Arsene {\it et al.} (BRAHMS Collaboration), Nucl.~Phys.~{\bf A757}, 1 (2005).

\bibitem{wpPHOBOS} B.~B.~Back {\it et al.} (PHOBOS Collaboration), Nucl.~Phys.~{\bf A757}, 28 (2005). 

\bibitem{wpSTAR} J.~Adams \etal~(STAR Collaboration), Nucl.~Phys.~{\bf A757}, 102 (2005).

\bibitem{wpPHENIX} K.~Adcox \etal~(PHENIX Collaboration), Nucl.~Phys.~{\bf A757}, 184 (2005).


\bibitem{TDLee1} T.~D.~Lee, Phys.~Rev.~D {\bf 8}, 1226 (1973).

\bibitem{TDLee2} T.~D.~Lee and G.~C.~Wick, Phys.~Rev.~D {\bf 9}, 2291 (1974).

\bibitem{Morley} P.~D.~Morley and I.~A.~Schmidt, Z.~Phys.~{\bf C26}, 627 (1985).

\bibitem{PV} D.~Kharzeev, R.~D.~Pisarski, and M.~H.~G.~Tytgat, Phys.~Rev.~Lett.~{\bf 81}, 512 (1998).

\bibitem{PV1Qsep} D.~Kharzeev, Phys.~Lett.~B {\bf 633}, 260 (2006).

\bibitem{PV2Qsep}  D.~Kharzeev and A.~Zhitnitsky, Nucl.~Phys.~{\bf A797}, 67 (2007).

\bibitem{PV3Qsep} K.~Fukushima, D.~E.~Kharzeev, and H.~J.~Warringa, Phys.~Rev.~D {\bf 78}, 074033 (2008).

\bibitem{PVquench} D.~E.~Kharzeev, L.~D.~McLerran, and H.~J.~Warringa, Nucl.~Phys.~{\bf A803}, 227 (2008).

\bibitem{Voloshin} S.~A.~Voloshin, Phys.~Rev.~C {\bf 70}, 057901 (2004).


\bibitem{CorrelatorPRL} B.~I.~Abelev {\it et al.} (STAR Collaboration), Phys.~Rev.~Lett.~{\bf 103}, 251601 (2009). 

\bibitem{CorrelatorPRC} B.~I.~Abelev {\it et al.} (STAR Collaboration), Phys.~Rev.~C~{\bf 81}, 054908 (2010). 

\bibitem{ALICE} B.~Abelev {\it et al.} (ALICE Collaboration), Phys.~Rev.~Lett.~{\bf 110}, 012301 (2013).


\bibitem{QuanWang} Quan Wang, Ph.D.~thesis, Purdue University, 2012 [http://drupal.star.bnl.gov/STAR/theses/phd/quanwang, arXiv:1205.4638].

\bibitem{STAR} K.~H.~Ackermann {\it et al.} (STAR Collaboration), Nucl.~Instrum.~Meth.~{\bf A499}, 624 (2003).

\bibitem{CTB} F.~S.~Bieser {\it et al.} (STAR Collaboration), Nucl.~Instrum.~Meth.~{\bf A499}, 766 (2003).

\bibitem{ZDC} C.~Adler {\it et al.}, Nucl.~Instrum.~Meth.~{\bf A499}, 433 (2003).

\bibitem{spec200} J.~Adams {\it et al.} (STAR Collaboration), Phys.~Rev.~Lett.~{\bf 92}, 112301 (2004).

\bibitem{Levente} B.~I.~Abelev \etal~(STAR Collaboration), Phys.~Rev.~C~{\bf 79}, 034909 (2009).

\bibitem{TPC1} K.~H.~Ackermann {\it et al.} (STAR Collaboration), Nucl.~Phys.~{\bf A661}, 681 (1999).

\bibitem{TPC2} M.~Anderson {\it et al.}, Nucl.~Instrum.~Meth.~{\bf A499}, 659 (2003).

\bibitem{flowMethod} A.~M.~Poskanzer and S.~A.~Voloshin, Phys.~Rev.~C {\bf 58}, 1671 (1998).


\bibitem{WangG} Gang Wang, Ph.D.~thesis, UCLA, 2005 [http://drupal.star.bnl.gov/STAR/theses/ph-d/gang-wang].

\bibitem{ChenJY} L.~Adamczyk \etal~(STAR Collaboration), Phys.~Rev.~Lett.~{\bf 108}, 202301 (2012).

\bibitem{Wang} F.~Wang, Phys.~Rev.~C {\bf 81}, 064902 (2010). 

\bibitem{Pratt} S.~Pratt, S.~Schlichting, and S.~Gavin, Phys.~Rev.~C {\bf 84} 024909 (2011).

\bibitem{jetspec} J.~Adams \etal~(STAR Collaboration), Phys.~Rev.~Lett.~{\bf 95}, 152301 (2005).

\bibitem{Horner} M.~M.~Aggarwal \etal~(STAR collaboration), Phys.~Rev.~C {\bf 82}, 024912 (2010). 

\bibitem{3part} B.~I.~Abelev \etal~(STAR Collaboration), Phys.~Rev.~Lett.~{\bf 102}, 052302 (2009).

\bibitem{ridge} B.~I.~Abelev \etal~(STAR Collaboration), Phys.~Rev.~C~{\bf 80}, 064912 (2009).

\bibitem{Pawan} B.~I.~Abelev \etal~(STAR Collaboration), Phys.~Rev.~Lett.~{\bf 105}, 022301 (2010). 


\bibitem{Aoqi} H. Agakishiev \etal~(STAR Collaboration), arXiv:1010.0690.

\bibitem{Petersen} H.~Petersen, T.~Renk, and S.A.~Bass, Phys.~Rev.~C {\bf 83}, 014916 (2011). 

\bibitem{Dhevan} L. Adamczyk {\it et al.} (STAR Collaboration), Phys. Rev. C {\bf 88}, 064911 (2013). 

\bibitem{Mueller} M.~Asakawa, A.~Majumder, and B.~M\"{u}ller, Phys.~Rev.~C {\bf 81}, 064912 (2010). 

\bibitem{Koch} A.~Bzdak, V.~Koch, and J.~Liao, Phys.~Rev.~C {\bf 81}, 031901(R) (2010).

\bibitem{Koch2} J.~Liao, V.~Koch, and A.~Bzdak, Phys.~Rev.~C {\bf 82}, 054902 (2010).

\bibitem{GLMa} G.-L.~Ma and B.~Zhang, Phys.~Lett.~B {\bf 700}, 39 (2011). 


\bibitem{Voloshin_UU} S.A.~Voloshin, Phys.~Rev.~Lett.~{\bf 105}, 172301 (2010).


\end{thebibliography}
\end{document}